\documentclass[iop]{emulateapj}

\pdfoutput=1
\usepackage{apjfonts}
\usepackage{natbib}
\usepackage{epsf}
\usepackage{color}
\usepackage{amsmath}
\usepackage{float}
\newcommand{\spitzer}{\textit{Spitzer}}

\newcommand{\jwst}{\textit{JWST}}
\newcommand{\hst}{\textit{HST}}
\newcommand{\hff}{\textit{HFF}}
\newcommand{\clash}{\textit{CLASH}}
\newcommand{\mone}{{M0416}}
\newcommand{\mtwo}{{M0717}}
\newcommand{\mthree}{{M1149}}
\newcommand{\aone}{{A2744}}
\newcommand{\atwo}{{A1063}}
\newcommand{\athree}{{A370}}
\newcommand{\sex}{\hbox{SExtractor}}
\newcommand{\lsim}{\mathrel{\hbox{\rlap{\lower.55ex \hbox{$\sim$}} \kern-.3em \raise.4ex \hbox{$<$}}}}
\newcommand{\gsim}{\mathrel{\hbox{\rlap{\lower.55ex \hbox{$\sim$}} \kern-.3em \raise.4ex \hbox{$>$}}}}

\begin{document}

\submitted{Received --- Accepted --- Published}

\title{HFF-DeepSpace Photometric Catalogs of the Twelve \textit{Hubble} Frontier Fields, Clusters and Parallels:  Photometry, Photometric Redshifts, and Stellar Masses}

\author{\sc Heath V.\ Shipley\altaffilmark{1}, 
Daniel Lange-Vagle\altaffilmark{1,2},
Danilo Marchesini\altaffilmark{1},
Gabriel B.\ Brammer\altaffilmark{2},
Laura Ferrarese\altaffilmark{3},
Mauro Stefanon\altaffilmark{4},
Erin Kado-Fong\altaffilmark{1,5},
Katherine E.\ Whitaker\altaffilmark{6},
Pascal A.\ Oesch\altaffilmark{7},
Adina D.\ Feinstein\altaffilmark{1},
Ivo Labb\'{e}\altaffilmark{4},
Britt Lundgren\altaffilmark{8},
Nicholas Martis\altaffilmark{1},
Adam Muzzin\altaffilmark{9},
Kalina Nedkova\altaffilmark{1},
Rosalind Skelton\altaffilmark{10},
Arjen van der Wel\altaffilmark{11}}
\altaffiltext{1}{Department of Physics \& Astronomy, Tufts University, 574 Boston Avenue Suites 304, Medford, MA 02155, USA; \email{heath.shipley@tufts.edu}}
\altaffiltext{2}{Space Telescope Science Institute, 3700 San Martin Drive, Baltimore, MD 21218, USA}
\altaffiltext{3}{National Research Council of Canada, Herzberg Astronomy and Astrophysics Program, 5071 West Saanich Road, Victoria, BC V9E 2E7, Canada}
\altaffiltext{4}{Leiden Observatory, Leiden University, NL-2300 RA Leiden, Netherlands}
\altaffiltext{5}{Department of Astrophysical Sciences, Princeton University, Peyton Hall, Princeton, NJ 08544, USA}
\altaffiltext{6}{Department of Physics, University of Connecticut, Storrs, CT 06269, USA}
\altaffiltext{7}{Geneva Observatory, University of Geneva, Ch.\ des Maillettes 51, 1290 Versoix, Switzerland}
\altaffiltext{8}{Department of Astronomy, University of Wisconsin, Madison, WI 53706, USA}
\altaffiltext{9}{Department of Physics and Astronomy, York University, 4700 Keele St., Toronto, Ontario, MJ3 1P3, Canada}
\altaffiltext{10}{South African Astronomical Observatory, PO Box 9, Observatory, Cape Town 7935, South Africa}
\altaffiltext{11}{Max-Planck Institut f\"{u}r Astronomie, K\"{o}nigstuhl 17, D-69117, Heidelberg, Germany}

\begin{abstract}

We present \textit{Hubble} multi-wavelength photometric catalogs, including (up to) 17 filters with the Advanced Camera for Surveys and Wide Field Camera 3 from the ultra-violet to near-infrared for the \textit{Hubble} Frontier Fields and associated parallels.  We have constructed homogeneous photometric catalogs for all six clusters and their parallels.  To further expand these data catalogs, we have added ultra-deep $K_{S}$-band imaging at 2.2~\micron\ from the \textit{Very Large Telescope} HAWK-I and \textit{Keck-I} MOSFIRE instruments.  We also add post-cryogenic \spitzer\ imaging at 3.6~\micron\ and 4.5~\micron\ with the Infrared Array Camera (IRAC), as well as archival IRAC 5.8~\micron\ and 8.0~\micron\ imaging when available.  We introduce the public release of the multi-wavelength (0.2--8~\micron) photometric catalogs, and we describe the unique steps applied for the construction of these catalogs.  Particular emphasis is given to the source detection band, the contamination of light from the bright cluster galaxies and intra-cluster light.  In addition to the photometric catalogs, we provide catalogs of photometric redshifts and stellar population properties.  Furthermore, this includes all the images used in the construction of the catalogs, including the combined models of bright cluster galaxies and intra-cluster light, the residual images, segmentation maps and more.  These catalogs are a robust data set of the \textit{Hubble} Frontier Fields and will be an important aide in designing future surveys, as well as planning follow-up programs with current and future observatories to answer key questions remaining about first light, reionization, the assembly of galaxies and many more topics, most notably, by identifying high-redshift sources to target.

\end{abstract}


\section{INTRODUCTION}
\label{intro}

\begin{deluxetable*}{lccccccl}		%%%%%%%%%%%%%  TABLE  %%%%%%%%%%%%%%%
\tablecaption{Hubble Frontier Fields \vspace{-6pt}
\label{fields}}
\tablecolumns{8}
%\tablewidth{0pc}
%\setlength{\tabcolsep}{0pt}
\tablehead{
\colhead{Field} & \colhead{R.A.} & \colhead{Dec.} & \colhead{Cluster} & \colhead{Science Area} & \colhead{$F814W$ Area} & \colhead{$F160W$ Area} & \colhead{Name} \\
 & \colhead{(h m s)} & \colhead{(d m s)} & \colhead{$z_\mathrm{spec}$} & \colhead{(arcmin$^2$)} & \colhead{(arcmin$^2$)} & \colhead{(arcmin$^2$)} &  
}
\startdata
Abell 2744 & 00 14 21.20 & $-$30 23 50.10 & 0.308 & 18.2 & 18.2 & 5.4 & A2744-clu\\
Parallel & 00 13 53.27 & $-$30 22 47.80 & & 11.9 & 11.9 & 5.0 & A2744-par\\
MACS J0416.1-2403 & 04 16  8.38 & $-$24 04 20.80 & 0.396 & 14.1 & 14.1 & 6.2 & M0416-clu\\
Parallel & 04 16 33.40 & $-$24 06 49.10 & & 11.9 & 11.9 & 5.0 & M0416-par\\
MACS J0717.5+3745 & 07 17 34.00 & $+$37 44 49.00 & 0.545 & 15.4 & 15.4 & 6.6 & M0717-clu\\
Parallel & 07 17 32.63 & $+$37 44 59.70 & & 13.0 & 12.9 & 6.5 & M0717-par\\
MACS J1149.5+2223 & 11 49 35.43 & $+$22 23 44.63 & 0.543 & 12.5 & 12.2 & 8.4 & M1149-clu\\
Parallel & 11 49 40.46 & $+$22 18 01.53 & & 14.3 & 14.3 & 5.3 & M1149-par\\
Abell S1063 & 22 48 44.30 & $-$44 31 48.40 & 0.348 & 14.6 & 14.6 & 5.9 & A1063-clu\\
Parallel & 22 49 17.80 & $-$44 32 43.30 & & 12.2 & 11.9 & 6.6 & A1063-par\\
Abell 370 & 02 39 52.80 & $-$1 34 36.00 & 0.375 & 15.1 & 13.8 & 8.3 & A370-clu\\
Parallel & 02 40 13.51 & $-$1 37 34.00 & & 11.9 & 11.9 & 5.0 & A370-par
\enddata
\tablecomments{``Science Area'' refers to the coverage area of the detection band (Section \ref{detection}) in each field.  We refer to the clusters and parallels by the names designated in ``Name'' throughout this work for simplicity.}
\end{deluxetable*}		%%%%%%%%%%%%%  TABLE  %%%%%%%%%%%%%%%

Galaxy formation and evolution remain important topics of research in astronomy with many questions remaining.  Large multi-wavelength photometric surveys have made it possible to study galaxy formation and evolution over most of cosmic time by observing large populations of galaxies.  Recently, many surveys have leveraged ground- and space-based near-IR selected galaxy samples aimed to answer many topics from the build-up of the stellar mass function \citep[e.g.,][]{Marchesini2009, PG2008, Muzzin2013a, Nantais2016, Song2016, Grazian2015, Tomczak2014}, the star formation--mass relation \citep[e.g.][]{Whitaker2012, Duncan2014, Shivaei2015, Ly2015, Salmon2015}, the structural evolution of galaxies \citep[e.g.,][]{Franx2008, Bell2012, Wuyts2012, vanderWel2012, Chang2013}, star-formation histories of galaxies \citep{Papovich2011, Pacifici2016, Tomczak2016, Webb2015, Gonzalez2014}, the formation of clusters \citep{Muzzin2008, Muzzin2013b, Papovich2012, Hatch2016} and the stellar mass--metallicity relation \citep{Tremonti2004, Wuyts2012b, Ly2015, Maier2015, Zahid2014, Yabe2012}.

One recent effort to further our knowledge of galaxy formation and evolution is represented by the \hst\ \textit{Frontier Fields} (\hff) program \citep{Lotz2017}.  The \hff\ program is a multi-cycle Hubble program consisting of 840 orbits of Director's Discretionary (DD) time that imaged six fields centered on strong lensing galaxy clusters in parallel with six blank fields.  Along with \hst, the \textit{Spitzer Space Telescope} has devoted 1000 hours of DD time to image the \hff\ fields at 3.6\micron\ and 4.5\micron\ with IRAC (Capak et al., in prep).  The \hff\ combines the power of \hst\ and \spitzer\ with the natural strong lensing gravitational telescopes of massive galaxy clusters to produce the deepest observations of clusters and their lensed galaxies ever obtained.  We further include ultra-deep $K_S$ imaging from Keck and VLT \citep{Brammer2016} and deep \hst\ UV imaging (Siana et al., in prep) that bridges the UV to near-IR between the \hst/ACS/WFC3 and \spitzer/IRAC imaging surveys.  The \hff\ is further complemented by grism spectroscopy \citep[GLASS][]{Treu2015}, deep far-IR imaging with \textit{Herschel} \citep{Rawle2016}, 1.1 mm continuum detections from ALMA \citep{GL2017, Laporte2017}, Chandra ACIS imaging (archival and additional program, PI C.\ Jones-Forman) and JVLA imaging \citep{vanWeeren2017}, SCUBA-2 lensing cluster survey of radio-detected sub-mm galaxies \citep{Hsu2017} and LMT \citep{Pope2017}, in addition to many ground-based photometric (Subaru and Gemini) and spectroscopic (MUSE, VLT, etc.) programs.

The six clusters $-$ Abell 2744, MACS J0416.1-2403, MACS J0717.5+3745, MACS J1149.5+2223, Abell S1063 and Abell 370 (for simplicity we designate a name for each field in Table \ref{fields}) $-$ were selected based on their lensing strength, sky darkness, Galactic extinction, parallel field suitability, accessibility to ground-based facilities, \hst, \spitzer\ and \jwst\ observability, and pre-existing ancillary data.  The primary science goals of the twelve \hff\ fields are to 1) reveal the population of galaxies at $z = 5-10$ that are $10-50$ times intrinsically fainter than any presently known, 2) solidify our understanding of the stellar masses and star formation histories of faint galaxies, 3) provide the first statistically meaningful morphological characterization of star-forming galaxies at $z > 5$, and 4) find $z > 8$ galaxies magnified by the cluster lensing, with some bright enough to make them accessible to spectroscopic follow-up \citep{Lotz2017}.

The \hff\ poses many challenges akin to previous cluster surveys \citep[e.g.\ CLASH][]{Postman2012} due to the large fraction of light coming from the cluster itself.  How does one preserve the information of the cluster galaxies but gain access to hidden/obscured background or underlying objects in the fields?  The method most preferred is to model out the bright cluster galaxies (bCGs) dominating the majority of light.  We define the term bCG to be ``bright'' cluster galaxy, as different from the traditional ``brightest'' cluster galaxy (BCG) terminology used in the literature, and hereafter refer to them as bCGs.  There are various methods to accomplish this using GALFIT \citep{Peng2010}, IRAF and others \citep[e.g.,][]{Connor2017, Merlin2016} to measure the light profiles of the bCGs and then subtract off the resulting model without destroying the background/underlying objects that are the reason for using the galaxy clusters as lenses.  Specifically, the \hff\ are densely packed massive clusters from $0.3 < z < 0.6$ with dozens of bCGs that require modeling.  Furthermore, the intra-cluster light (ICL) and bCGs light are entangled and need to be modeled together to appropriately remove the light they contribute to each image, which varies from band to band in each field \citep[e.g., see][for a study of the ICL]{Montes2014}.  For a few fields in the \hff\ (e.g.\ \mone\ cluster), this is further complicated by nearby bright galaxies that also must be modeled, if possible.  Below, we discuss fully our approach and solutions to these challenges posed by the \hff\ observations.

We provide catalogs of photometric redshifts and stellar population properties for each field in the \hff, in addition to the photometric catalogs \citep[similar to the ASTRODEEP collaboration][but utilizing different methodology]{Merlin2016,Castellano2016,DC2017}.  Furthermore, the public release is accompanied by all the images used in the construction of the catalogs, including the combined models of the bCGs and ICL, the residual images after bCG modeling, segmentation maps and more.\footnote{see \url{http://cosmos.phy.tufts.edu/~danilo/HFF/Download.html} for catalogs and data products}  The outline of the paper is as follows. In Section~\ref{data}, we describe the datasets and data reduction steps performed.  In Section~\ref{modeling and catalogs}, we describe our photometric methods, catalog format, flags and completeness, including a detailed description of our process for modeling out the bCGs (Section~\ref{bcg modeling}).  In Section~\ref{photo check}, we verify the quality and consistency of the catalogs.  In Section~\ref{derived properties}, we describe the photometric redshift, rest-frame color, stellar population parameter fits to the SEDs and derived lensing magnifications.  In Section~\ref{summary}, we summarize our data products and catalogs that have been generated of the \hff\ survey.  We use the AB magnitude system throughout \citep{Oke1971} and if necessary, a $\Lambda$CDM cosmology with $\Omega_M$ = 0.3, $\Omega_{\Lambda}$ = 0.7 and $H_0 = 70~$km s$^{-1}$ Mpc$^{-1}$.

\section{DATA SETS}
\label{data}

\begin{deluxetable*}{lllll}		%%%%%%%%%%%%%  TABLE  %%%%%%%%%%%%%%%
\vspace{-60pt}
\tablecaption{Image Sources \vspace{-6pt}
\label{image sources}}
\tablecolumns{5}
\tabletypesize{\footnotesize}
\tablewidth{0pc}
\setlength{\tabcolsep}{0pt}
\tablehead{
\colhead{Field} & \colhead{Filters} & \colhead{Telescope/Instrument} & \colhead{Survey} & \colhead{Reference} \\
}
\startdata
\aone -clu & $F275W$, $F336W$ & \hst/UVIS & PID: 14209 & PI: B.\ Siana \\
 & $F435W$, $F606W$, $F814W$ & \hst/ACS & \hff & \citet{Lotz2017} \\
 & $F105W$**, $F125W$, $F140W$**, $F160W$ & \hst/WFC3 & \hff & \citet{Lotz2017} \\
 & $K_S$ & VLT/HAWK-I & KIFF & \citet{Brammer2016} \\
 & 3.6\micron, 4.5\micron, 5.8\micron, 8.0\micron & \spitzer/IRAC & & see Section \ref{irac} for details \\
\noalign{\smallskip}
\hline
\noalign{\smallskip}
\aone -par & $F435W$, $F606W$, $F814W$ & \hst/ACS & \hff & \citet{Lotz2017} \\
 & $F105W$**, $F125W$, $F140W$, $F160W$ & \hst/WFC3 & \hff & \citet{Lotz2017} \\
 & $K_S$ & VLT/HAWK-I & KIFF & \citet{Brammer2016} \\
 & 3.6\micron, 4.5\micron & \spitzer/IRAC & & see Section \ref{irac} for details \\
\noalign{\smallskip}
\hline
\noalign{\smallskip}
\mone -clu & $F225W$, $F390W$ & \hst/UVIS & CLASH & \citet{Postman2012}  \\
 & $F275W$, $F336W$ & \hst/UVIS & PID: 14209 & PI: B.\ Siana \\
 & $F435W$, $F606W$, $F814W$ & \hst/ACS & \hff & \citet{Lotz2017} \\
 & $F475W$**, $F625W$**, $F775W$** & \hst/ACS & PID: 12459 & PI: M.\ Postman \\
 & $F850LP$ & \hst/ACS & CLASH & \citet{Postman2012} \\
 & $F105W$**, $F125W$**, $F140W$**, $F160W$** & \hst/WFC3 & \hff & \citet{Lotz2017} \\
 & $F110W$** & \hst/WFC3 & CLASH & \citet{Postman2012} \\
 & $K_S$ & VLT/HAWK-I & KIFF & \citet{Brammer2016} \\
 & 3.6\micron, 4.5\micron & \spitzer/IRAC & & see Section \ref{irac} for details \\
\noalign{\smallskip}
\hline
\noalign{\smallskip}
\mone -par & $F435W$, $F606W$, $F814W$ & \hst/ACS & \hff & \citet{Lotz2017} \\
 & $F775W$**, $F850LP$** & \hst/ACS & PID: 12459 & PI: M.\ Postman \\
 & $F105W$**, $F125W$, $F140W$, $F160W$ & \hst/WFC3 & \hff & \citet{Lotz2017} \\
 & $K_S$ & VLT/HAWK-I & KIFF & \citet{Brammer2016} \\
 & 3.6\micron, 4.5\micron & \spitzer/IRAC & & see Section \ref{irac} for details \\
\noalign{\smallskip}
\hline
\noalign{\smallskip}
\mtwo -clu & $F225W$, $F390W$ & \hst/UVIS & CLASH & \citet{Postman2012}  \\
 & $F275W$, $F336W$ & \hst/UVIS & PID: 14209 & PI: B.\ Siana \\
 & $F435W$, $F606W$, $F814W$ & \hst/ACS & \hff & \citet{Lotz2017} \\
 & $F475W$**, $F625W$**, $F775W$**, $F850LP$** & \hst/ACS & PID: 12103 & PI: M.\ Postman \\
 & $F555W$ & \hst/ACS & CLASH & \citet{Postman2012} \\
 & $F105W$, $F125W$, $F140W$, $F160W$ & \hst/WFC3 & \hff & \citet{Lotz2017} \\
 & $F110W$ & \hst/WFC3 & CLASH & \citet{Postman2012} \\
 & $K_S$ & Keck/MOSFIRE & KIFF & \citet{Brammer2016} \\
 & 3.6\micron, 4.5\micron & \spitzer/IRAC & & see Section \ref{irac} for details \\
\noalign{\smallskip}
\hline
\noalign{\smallskip}
\mtwo -par & $F435W$, $F606W$, $F814W$ & \hst/ACS & \hff & \citet{Lotz2017} \\
 & $F105W$, $F125W$, $F140W$, $F160W$ & \hst/WFC3 & \hff & \citet{Lotz2017} \\
 & $K_S$ & Keck/MOSFIRE & KIFF & \citet{Brammer2016} \\
 & 3.6\micron, 4.5\micron & \spitzer/IRAC & & see Section \ref{irac} for details \\
\noalign{\smallskip}
\hline
\noalign{\smallskip}
\mthree -clu & $F225W$, $F390W$ & \hst/UVIS & CLASH & \citet{Postman2012}  \\
 & $F275W$, $F336W$ & \hst/UVIS & PID: 14209 & PI: B.\ Siana \\
 & $F435W$, $F606W$, $F814W$ & \hst/ACS & \hff & \citet{Lotz2017} \\
 & $F475W$, $F555W$, $F625W$, $F775W$, $F850LP$ & \hst/ACS & CLASH & \citet{Postman2012} \\
 & $F105W$, $F125W$, $F140W$, $F160W$ & \hst/WFC3 & \hff & \citet{Lotz2017} \\
 & $F110W$ & \hst/WFC3 & CLASH & \citet{Postman2012} \\
 & $K_S$ & Keck/MOSFIRE & KIFF & \citet{Brammer2016} \\
 & 3.6\micron, 4.5\micron & \spitzer/IRAC & & see Section \ref{irac} for details \\
\noalign{\smallskip}
\hline
\noalign{\smallskip}
\mthree -par & $F435W$, $F606W$, $F814W$ & \hst/ACS & \hff & \citet{Lotz2017} \\
 & $F105W$, $F125W$, $F140W$, $F160W$ & \hst/WFC3 & \hff & \citet{Lotz2017} \\
 & $K_S$ & Keck/MOSFIRE & KIFF & \citet{Brammer2016} \\
 & 3.6\micron, 4.5\micron & \spitzer/IRAC & & see Section \ref{irac} for details \\
\noalign{\smallskip}
\hline
\noalign{\smallskip}
\atwo -clu & $F225W$, $F390W$ & \hst/UVIS & CLASH & \citet{Postman2012}  \\
 & $F275W$, $F336W$ & \hst/UVIS & PID: 14209 & PI: B.\ Siana \\
 & $F435W$, $F606W$, $F814W$ & \hst/ACS & \hff & \citet{Lotz2017} \\
 & $F475W$, $F625W$, $F775W$, $F850LP$ & \hst/ACS & CLASH & \citet{Postman2012} \\
 & $F105W$, $F125W$, $F140W$, $F160W$ & \hst/WFC3 & \hff & \citet{Lotz2017} \\
 & $F110W$** & \hst/WFC3 & PID: 12458 & PI: M.\ Postman \\
 & $K_S$ & VLT/HAWK-I & KIFF & \citet{Brammer2016} \\
 & 3.6\micron, 4.5\micron, 5.8\micron, 8.0\micron & \spitzer/IRAC & & see Section \ref{irac} for details \\
\noalign{\smallskip}
\hline
\noalign{\smallskip}
\atwo -par & $F435W$, $F606W$, $F814W$ & \hst/ACS & \hff & \citet{Lotz2017} \\
 & $F105W$, $F125W$, $F140W$, $F160W$ & \hst/WFC3 & \hff & \citet{Lotz2017} \\
 & $K_S$ & VLT/HAWK-I & KIFF & \citet{Brammer2016} \\
 & 3.6\micron, 4.5\micron & \spitzer/IRAC & & see Section \ref{irac} for details \\
\noalign{\smallskip}
\hline
\noalign{\smallskip}
\athree -clu & $F275W$, $F336W$ & \hst/UVIS & PID: 14209 & PI: B.\ Siana \\
 & $F435W$, $F606W$, $F814W$ & \hst/ACS & \hff & \citet{Lotz2017} \\
 & $F475W$**, $F625W$** & \hst/ACS & PID: 11507 & PI: K.\ Noll \\
 & $F475W$** & \hst/ACS & PID: 11582 & PI: A.\ Blain \\
 & $F625W$** & \hst/ACS & PID: 13790 & PI: S.\ Rodney \\
 & $F105W$, $F125W$, $F140W$, $F160W$ & \hst/WFC3 & \hff & \citet{Lotz2017} \\
 & $F110W$** & \hst/WFC3 & PID: 11591 & PI: J.P.\ Kneib \\
 &  &  & PID: 13790 & PI: S.\ Rodney \\
 & $K_S$ & VLT/HAWK-I & KIFF & \citet{Brammer2016} \\
 & 3.6\micron, 4.5\micron, 5.8\micron, 8.0\micron & \spitzer/IRAC & & see Section \ref{irac} for details \\
\noalign{\smallskip}
\hline
\noalign{\smallskip}
\athree -par & $F435W$, $F606W$, $F814W$ & \hst/ACS & \hff & \citet{Lotz2017} \\
 & $F105W$, $F125W$, $F140W$, $F160W$ & \hst/WFC3 & \hff & \citet{Lotz2017} \\
 & $K_S$ & VLT/HAWK-I & KIFF & \citet{Brammer2016} \\
 & 3.6\micron, 4.5\micron & \spitzer/IRAC & & see Section \ref{irac} for details \\
\enddata
\vspace{-6pt}
\tablecomments{\hst/ACS and \hst/WFC3-IR bands marked by (**) are processed internally by our group to improve and/or include any additional data that is available (see Section \ref{redux}). }
\end{deluxetable*}		%%%%%%%%%%%%%  TABLE  %%%%%%%%%%%%%%%

The twelve \textit{Hubble Frontier Fields} (\hff) have been observed with \hst/WFC3, \hst/ACS \citep{Lotz2017}, \spitzer\ and two ground-based observatories (\textit{VLT} and \textit{Keck-I}) for added ultra-deep $K_{S}$-band imaging \citep{Brammer2016}.  In each field, the data consist of the ACS $F435W$, $F606W$, $F814W$ and WFC3 $F105W$, $F125W$, $F140W$, $F160W$ images obtained from the \hff\ Program.  In this section, we describe our data reduction steps and summarize all other space- and ground-based data that are used to construct the catalogs.

The photometric catalogs make use of 22 filters (see Table~\ref{image sources}) and corresponding image mosaics from not only the \hff\ program but previous programs that have observed the \hff\ fields \citep[e.g.\ CLASH][]{Postman2012}.  We projected all $Hubble$ data onto the astrometric grid and pixel scale defined in the data released products for the \hff\ Program, specifically the $F160W$ filter, but allowing for larger coverage areas from the additional data (i.e.\ Abell 2744 cluster, hereafter \aone -clu, see Table~\ref{fields}).

\subsection{Hubble Frontier Fields Imaging}

\subsubsection{Sources of Data}

\begin{figure*}[htpb]	%%%%%%%%%%%%%%%  PLOT  %%%%%%%%%%%%%%%%%
\epsscale{1.}
\plottwo{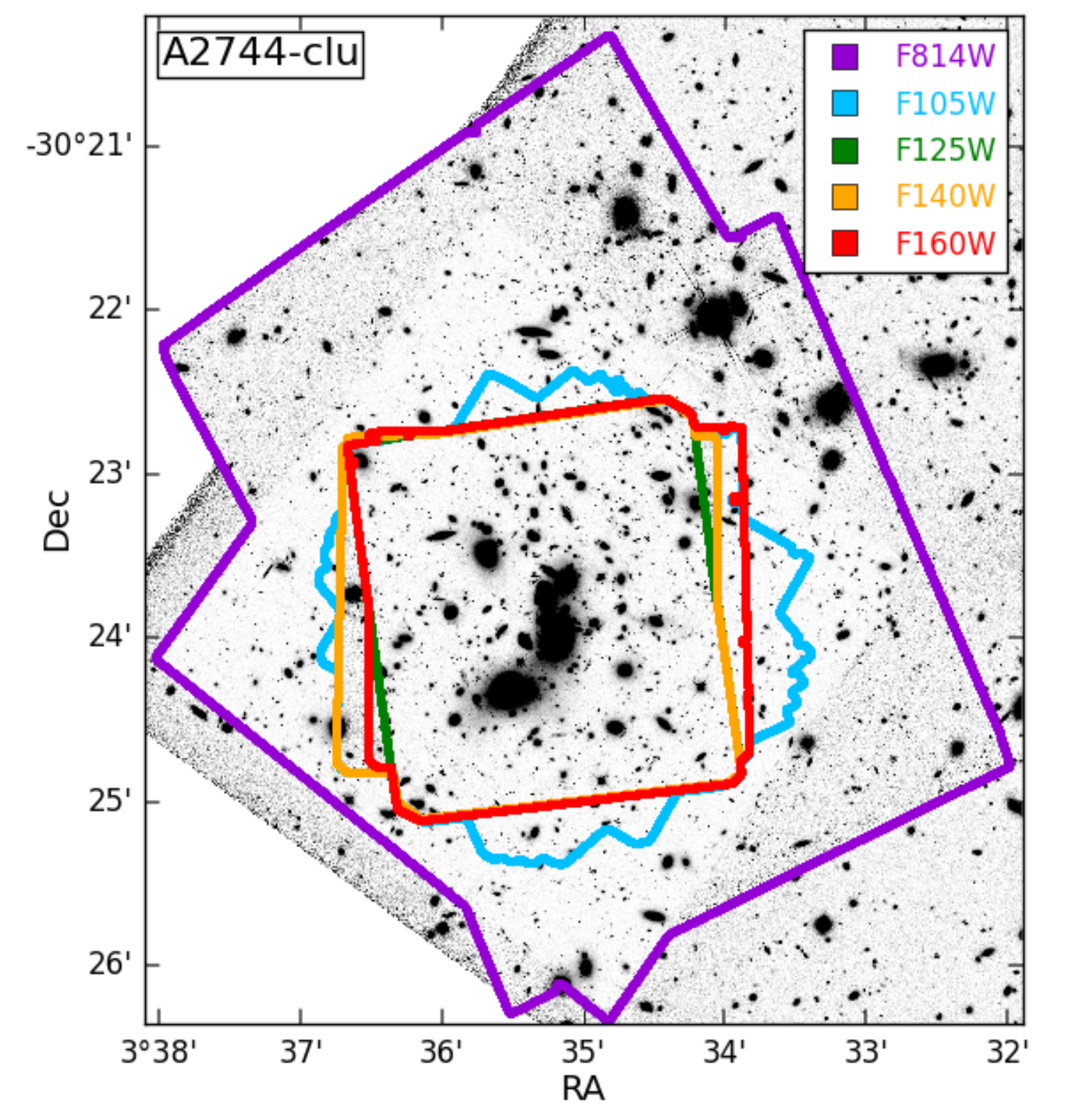}{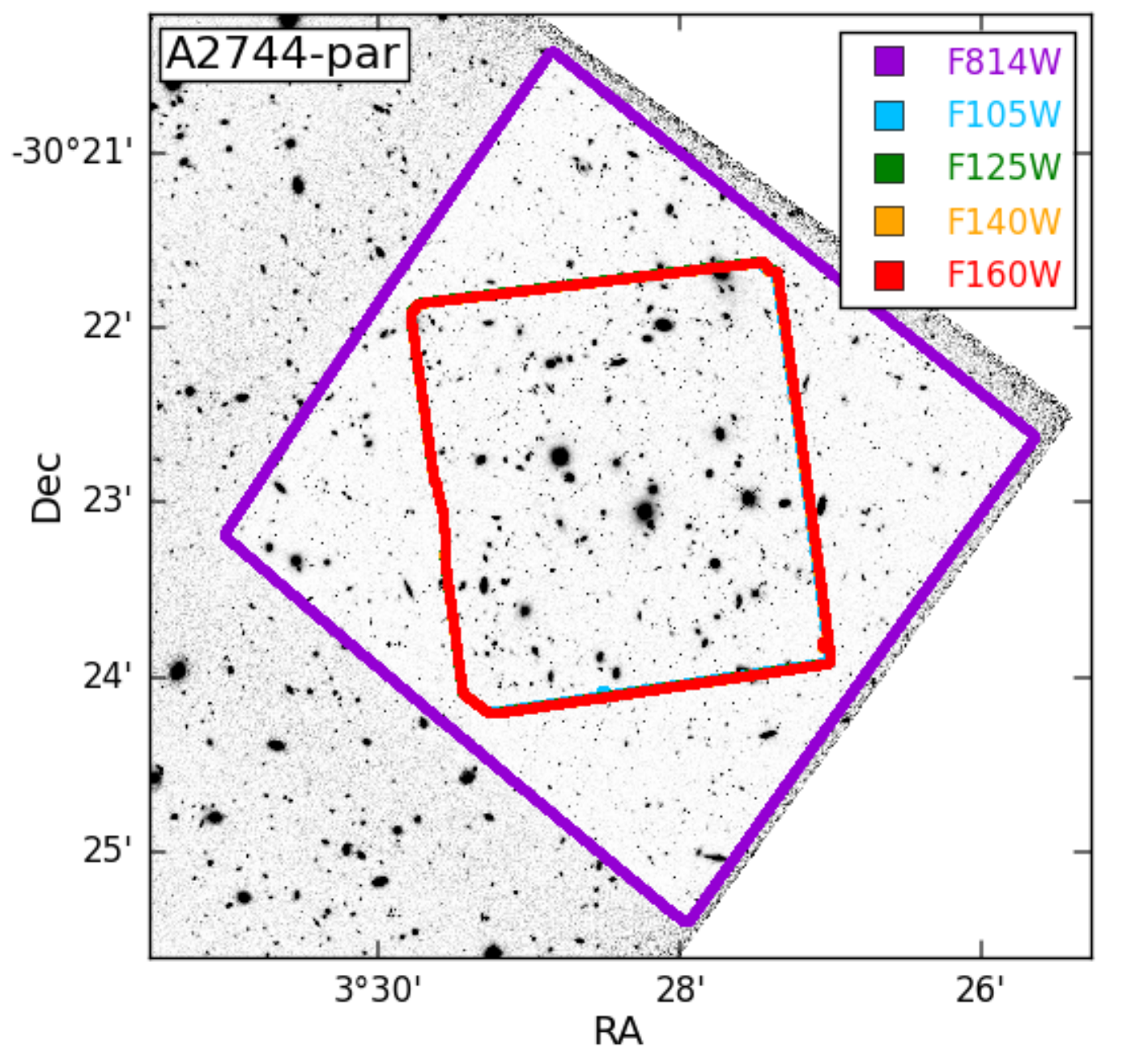}
\plottwo{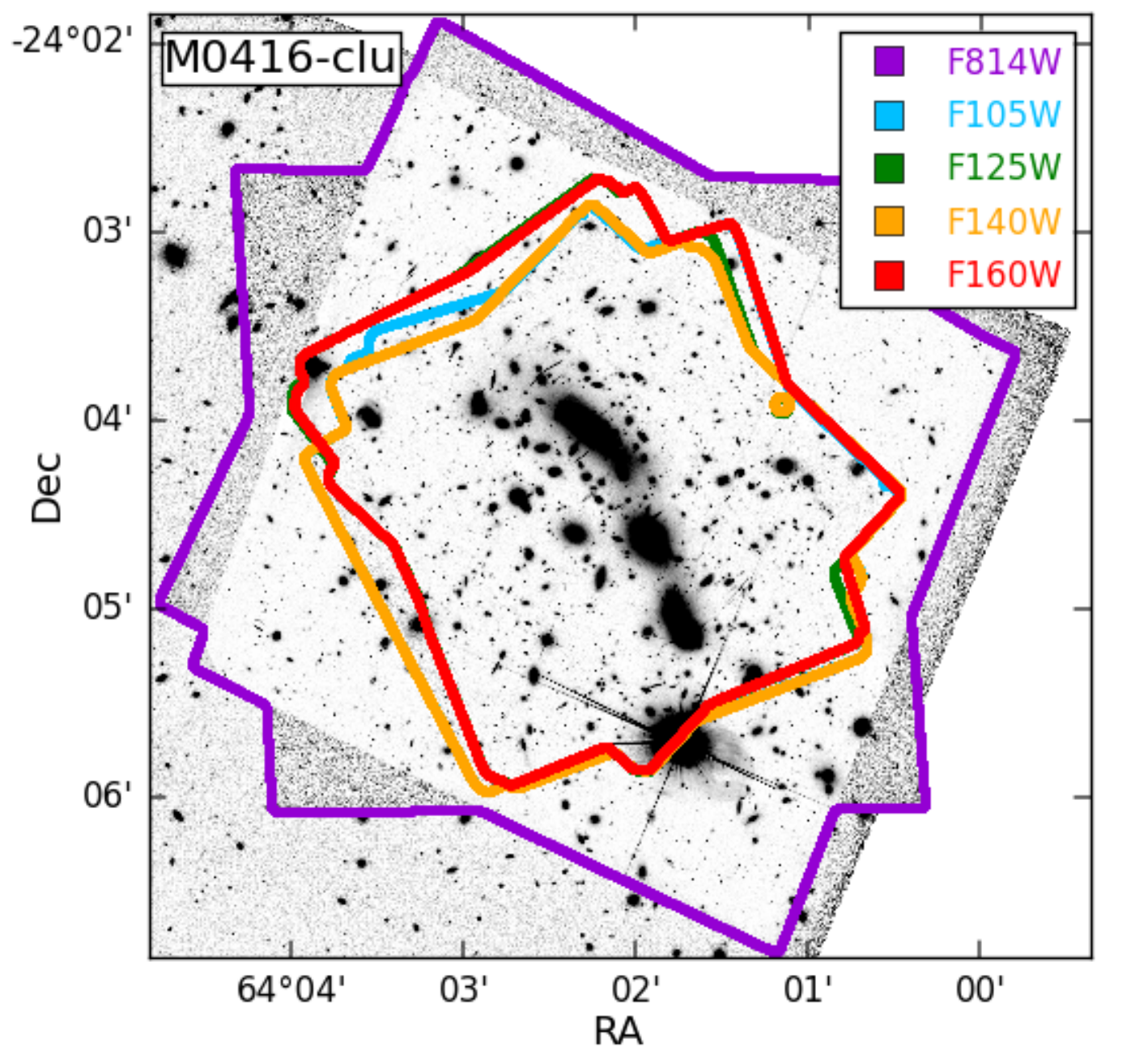}{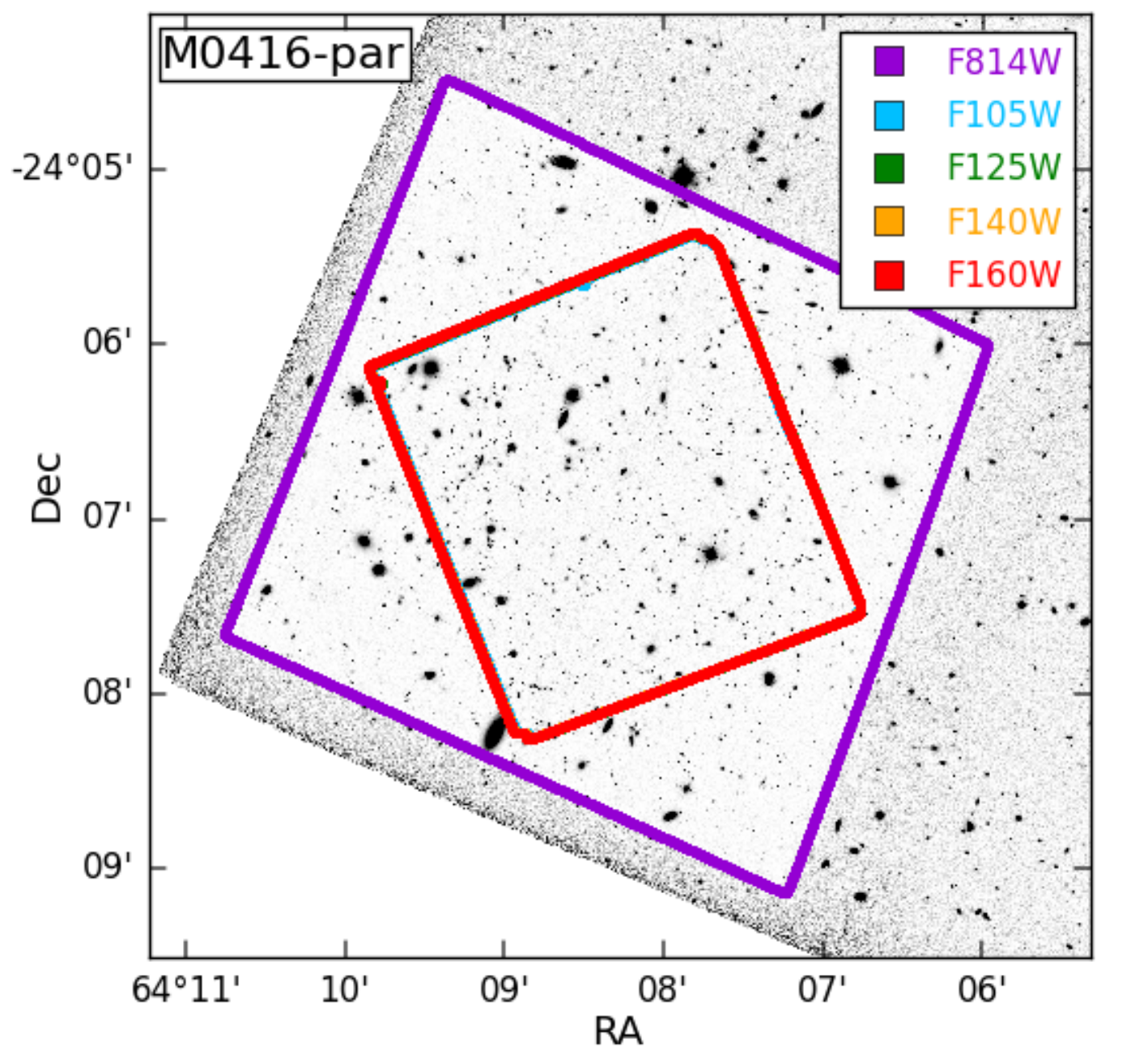}
\plottwo{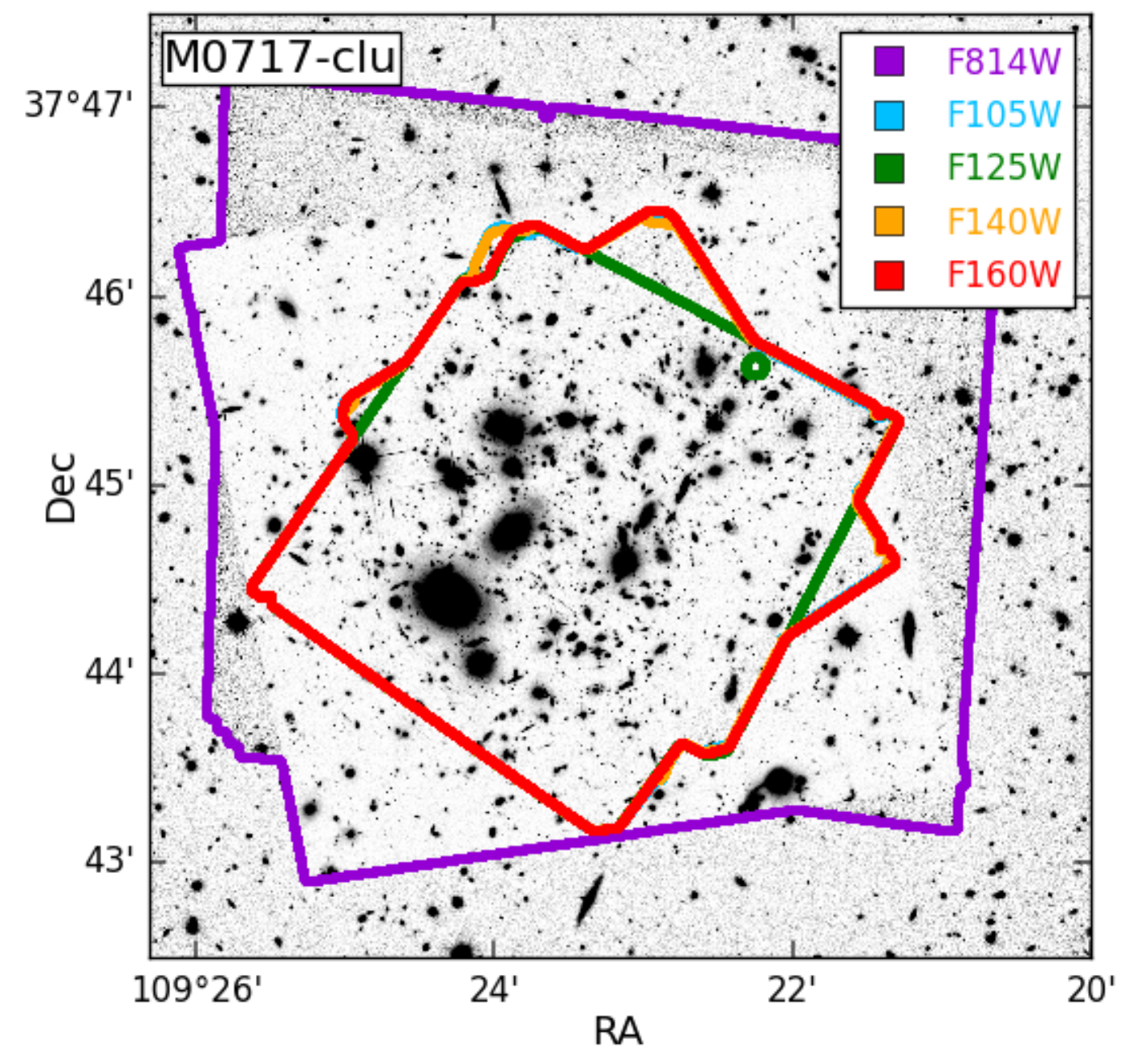}{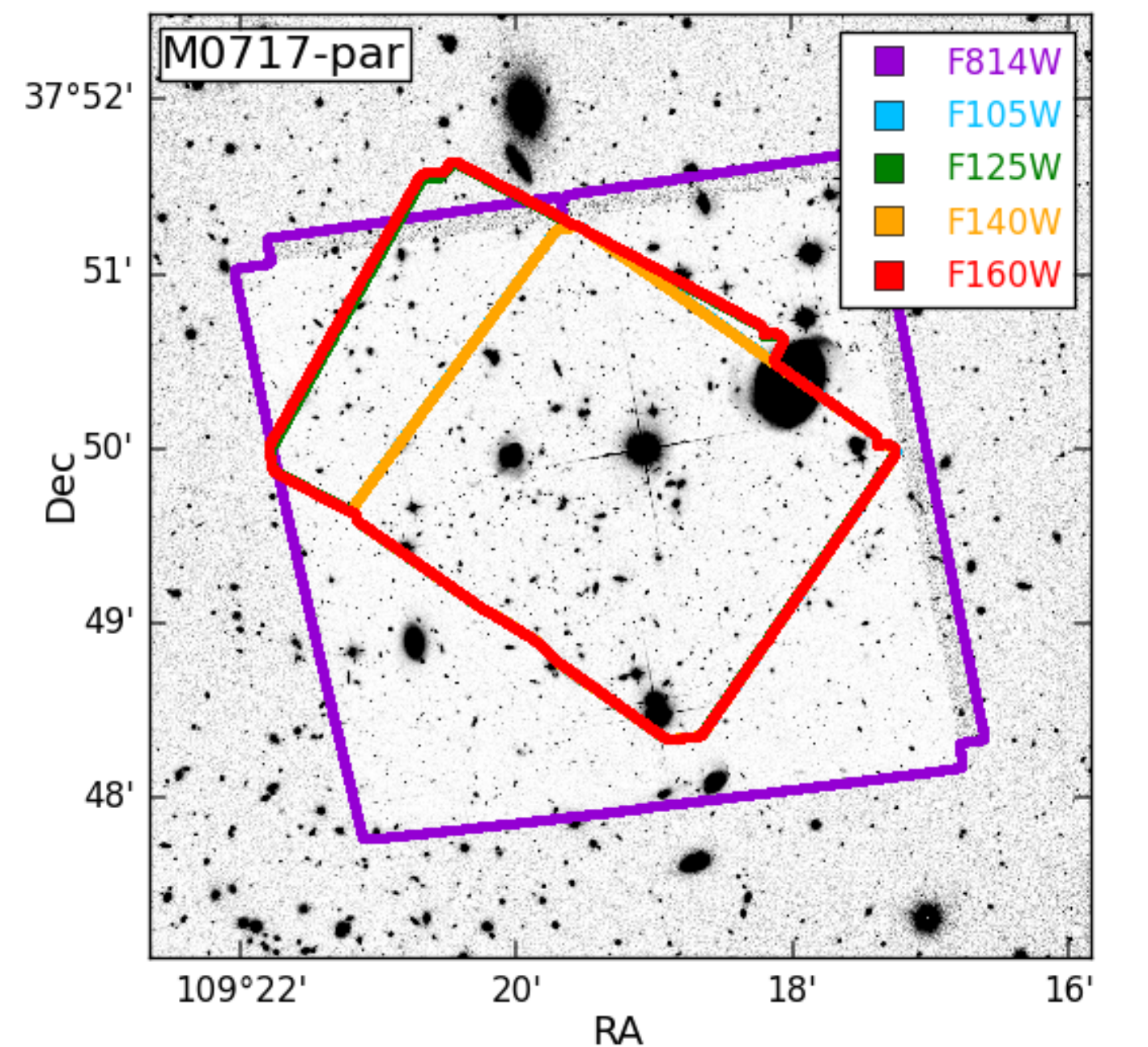}
\caption{Layout of the Hubble observations used.  The catalogs presented here cover the entire area encompassed by the five bands ($F814W$, $F105W$, $F125W$, $F140W$, $F160W$; i.e.\ the detection band, see Section \ref{detection}).  The imaging is of the $F814W$ band inside its border and the $K_S$ band outside of it.  North is up and East is to the left.}
\label{hff layout 1}
\end{figure*}		%%%%%%%%%%%%%%%  PLOT  %%%%%%%%%%%%%%%%%

\begin{figure*}[htbp]	%%%%%%%%%%%%%%%  PLOT  %%%%%%%%%%%%%%%%%
\epsscale{1.}
\plottwo{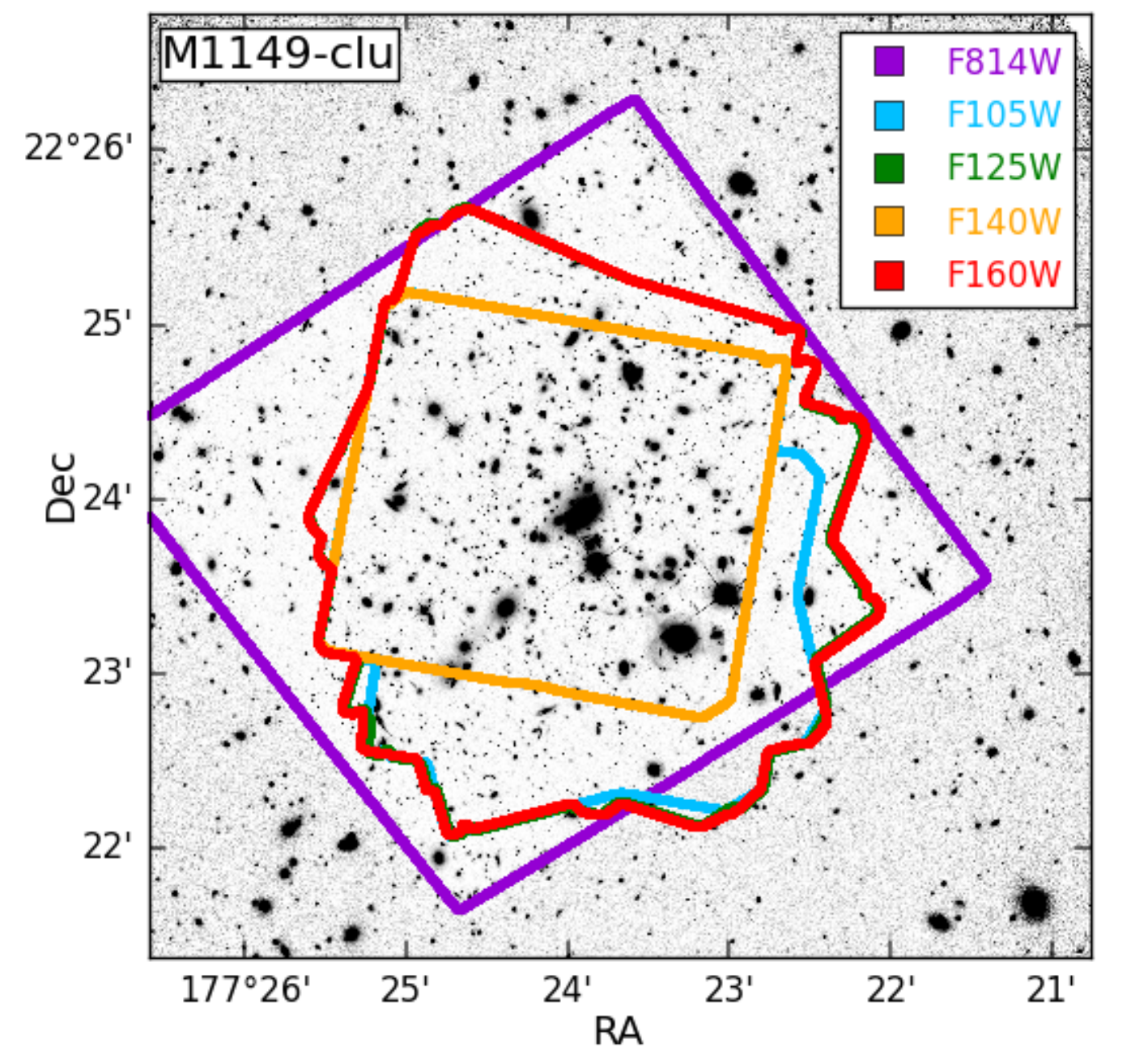}{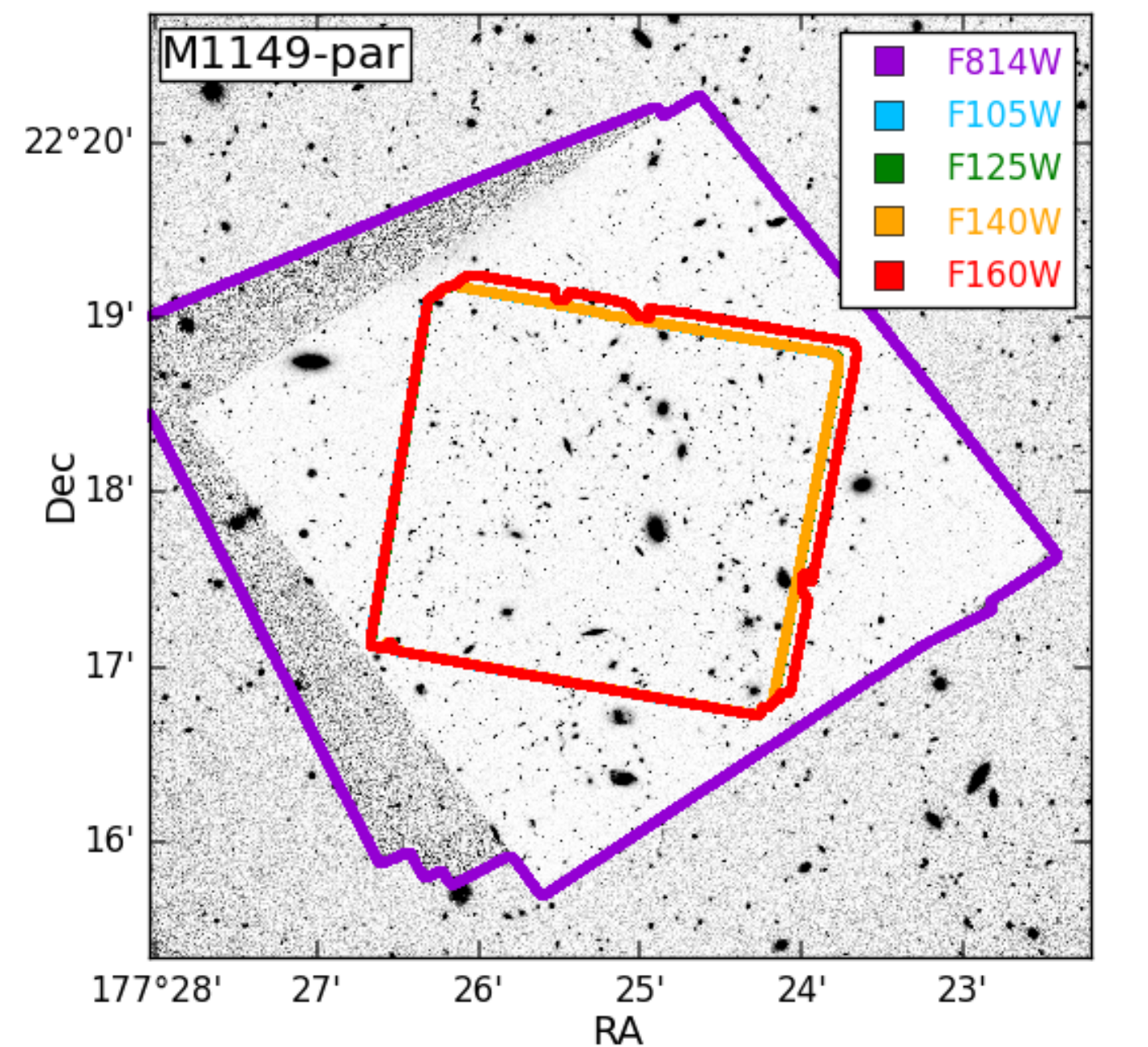}
\plottwo{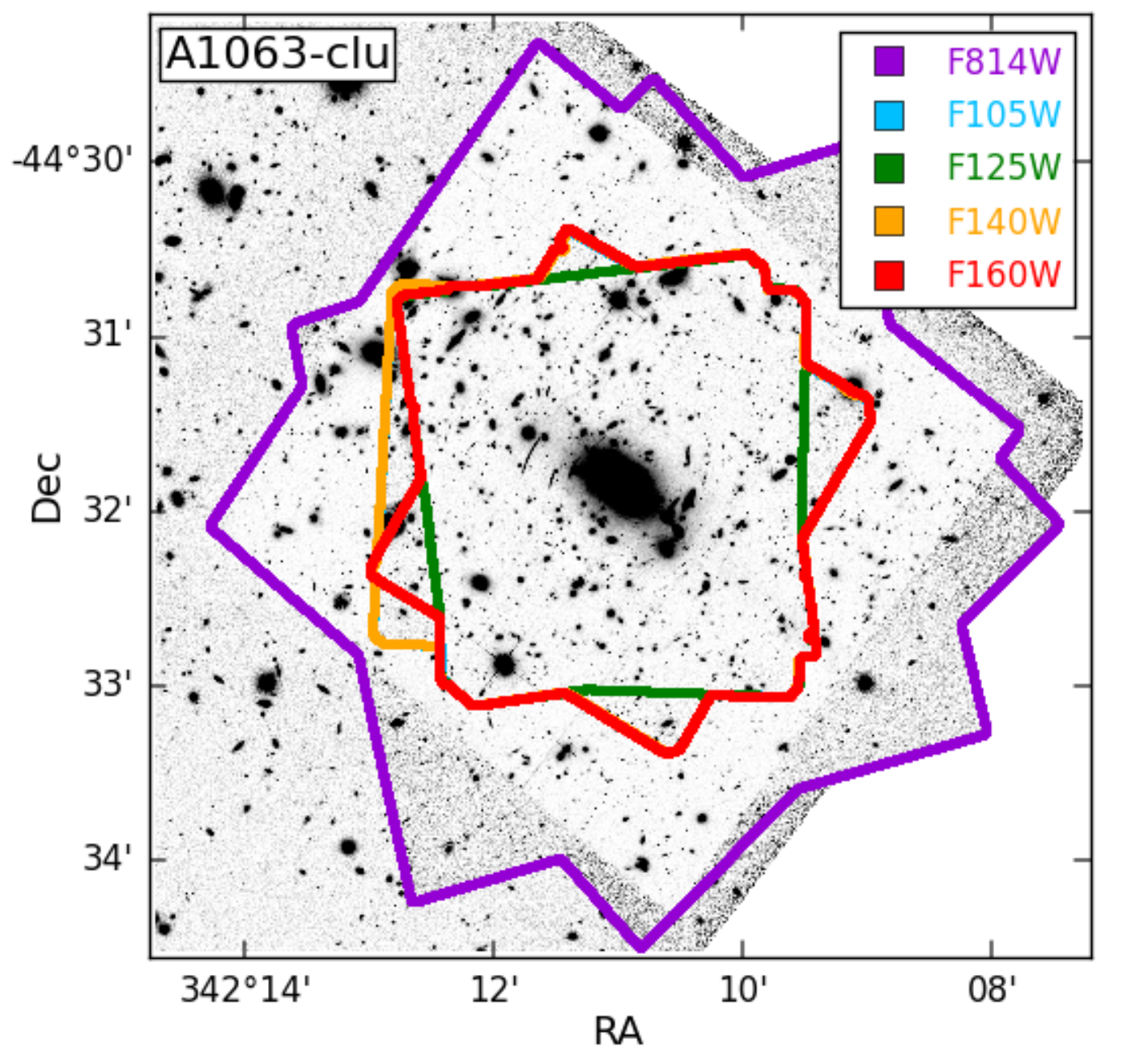}{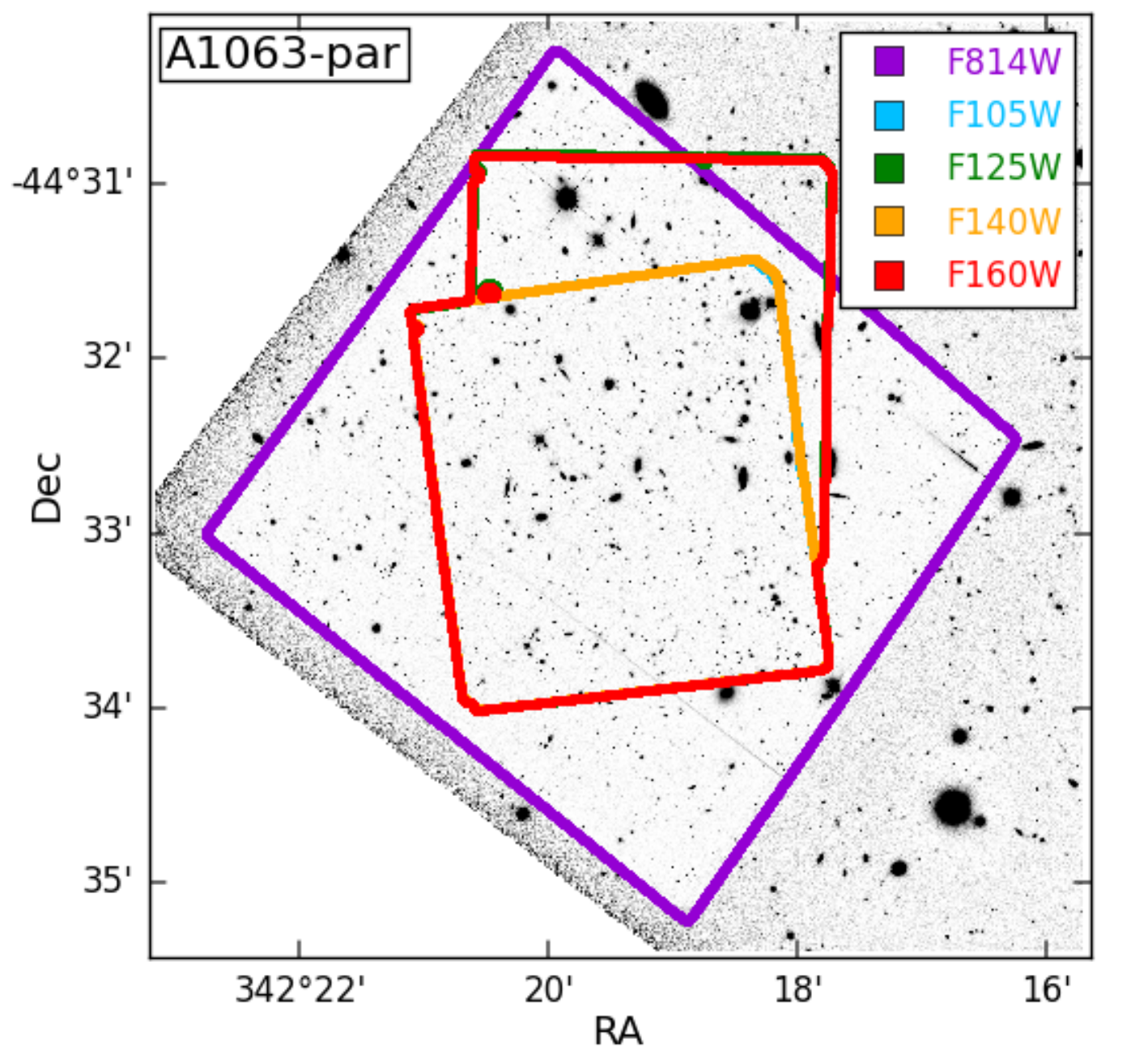}
\plottwo{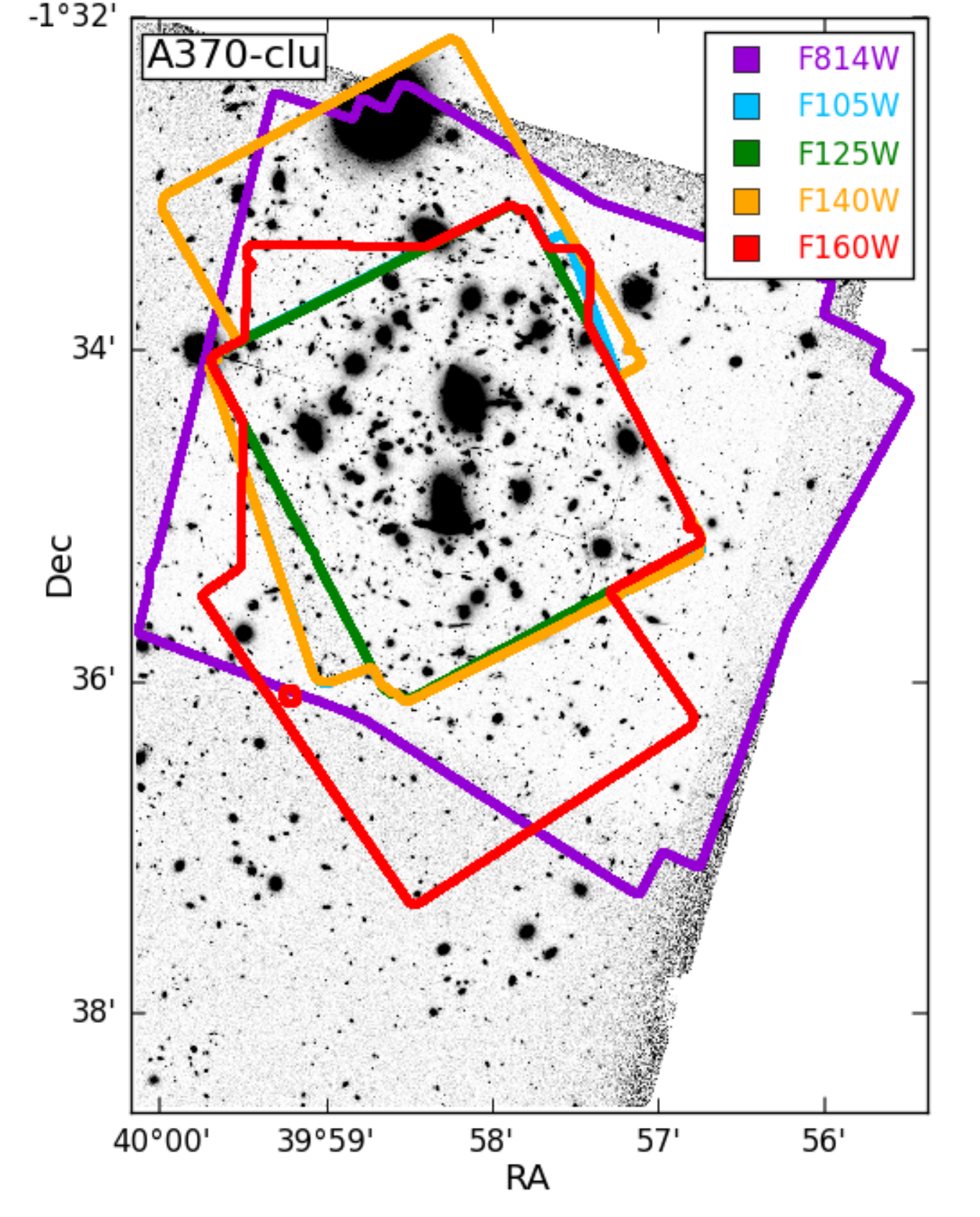}{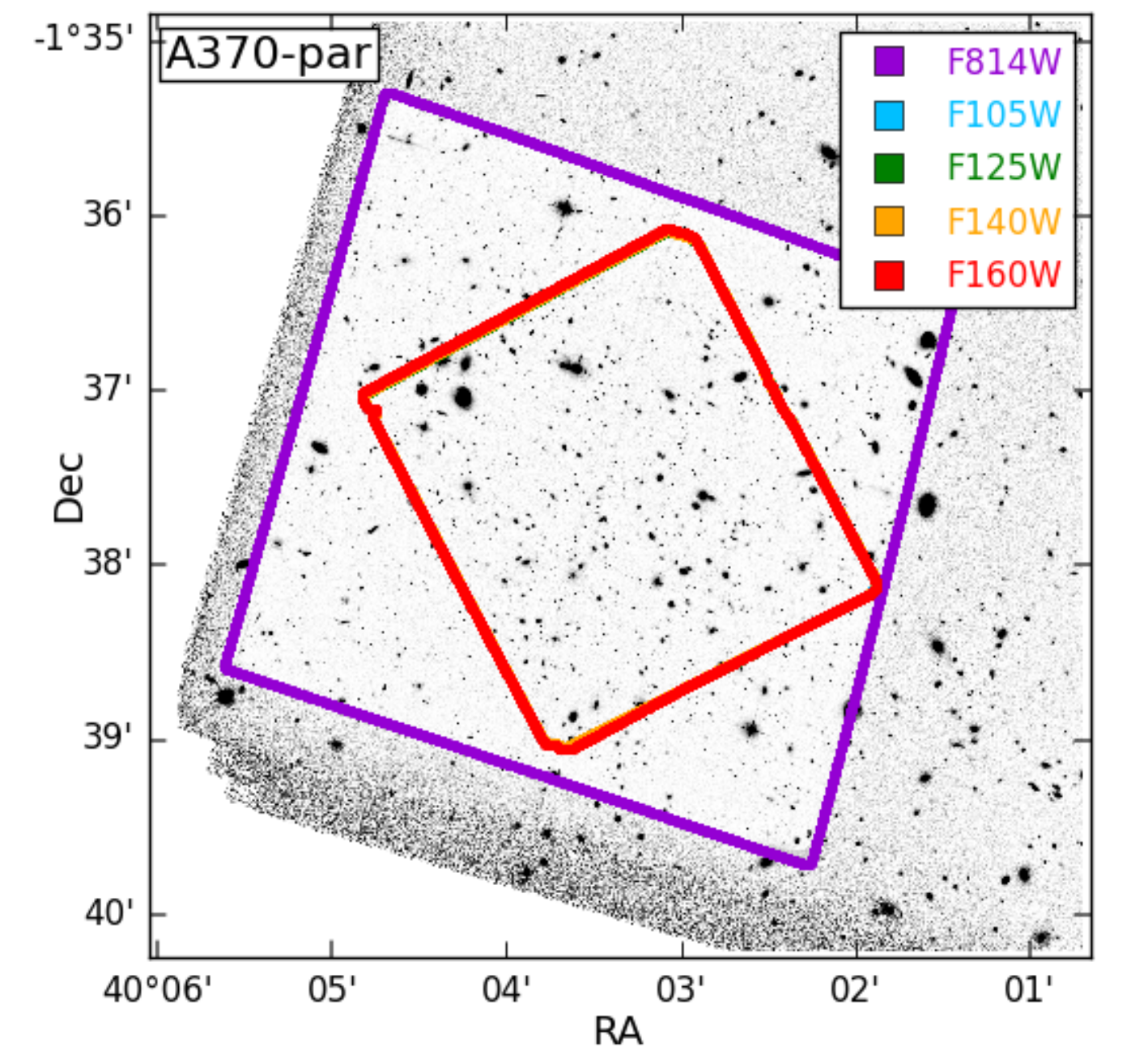}
\caption{Same as Figure \ref{hff layout 1} of the last 3 cluster and parallel fields observed (labeled in plot).}
\label{hff layout 2}
\end{figure*}		%%%%%%%%%%%%%%%  PLOT  %%%%%%%%%%%%%%%%%

To maximize the depth and coverage of the \textit{Hubble Frontier Fields}, we collected imaging from any previous \hst\ observations utilizing the ACS and WFC3 instruments for any of the 17 filters in our catalogs.  The coordinates and coverage areas of all twelve fields' catalogs are given in Table~\ref{fields}.  The ``Science Area'' column indicates the region covered by the $F814W$, $F105W$, $F125W$, $F140W$ and $F160W$ bands (i.e.\ the detection band, see Section~\ref{detection}).  Other \hst\ programs have carried out observations of the \hff\ and we have incorporated these additional data sets into our mosaics for each field, where available, to increase the depth and area coverage of the catalogs (see Table~\ref{image sources}).  Furthermore, \clash\ and other smaller surveys of the \hff\ have added filters beyond those observed with the \hff\ Program albeit to shallower depths.

All near-IR \hst\ observations are obtained using the Wide Field Camera 3 IR detector (WFC3/IR), which has a $1024\times1024$ HgCdTe array.  The usable portion of the detector is $1014\times1014$ pixels, covering a region of $136\arcsec\times123\arcsec$ across with a native pixel scale of $0\farcs128$ pixel$^{-1}$ (at the central reference pixel).  The \hff\ observations are done in four wide filters:  $F105W$, $F125W$, $F140W$ and $F160W$, which cover the wavelength ranges of $\sim0.9\micron - 1.2\micron$, $\sim1.1\micron - 1.4\micron$, $\sim1.2\micron - 1.6 \micron$ and $\sim1.4\micron - 1.7 \micron$, respectively.  The standard designations for the four filters are $Y_{F105W}$, $J_{F125W}$, $JH_{F140W}$ and $H_{F160W}$, however we will refer to them by the \hst\ filter name to avoid confusion with ground-based bandpasses.\footnote{see ``WFC3 Instrument Handbook'' for additional information}  The available UV data are obtained using the WFC3/UVIS detector which has two $2051\times4096$ UV optimized e2v CCDs.  The usable portion of the detector is rhomboidal, covering a region of $162\arcsec\times162\arcsec$ across with a native pixel scale of $0\farcs04$ pixel$^{-1}$ (at the central reference pixel).

All visible \hst\ observations are obtained using the Advanced Camera for Surveys WFC detector (ACS/WFC), which has two $2048\times4096$ SITe CCDs array.  The usable portion of the detector is $4040\times4040$ pixels, covering a region of $202\arcsec\times202\arcsec$ across with a native pixel scale of $0\farcs049$ pixel$^{-1}$ (at the central reference pixel).  The \hff\ observations are done in three wide filters:  $F435W$, $F606W$ and $F814W$, which cover the wavelength ranges of $\sim0.35\micron - 0.5\micron$, $\sim0.5\micron - 0.7\micron$ and $\sim0.7\micron - 0.95 \micron$, respectively.  The standard designations for the three filters are $B_{F435W}$, $V_{F606W}$ and $I_{F814W}$, however we will refer to them by the \hst\ filter name to avoid confusion.\footnote{see ``ACS Instrument Handbook'' for additional information}

\subsubsection{Data Reduction}
\label{redux}

We downloaded the \hst\ science and weight images for each available filter\footnote{\hff\ images are downloaded from:  \url{http://www.stsci.edu/hst/campaigns/frontier-fields/FF-Data}.} of the \hff\ fields following the observational scheduled defined by the program\footnote{see \url{http://www.stsci.edu/hst/campaigns/frontier-fields/HST-Survey}}.  The \hff\ images downloaded are the latest version available (v1.0 in all cases; Koekemoer et al., in prep).  Other \hst\ images covering the \hff\ fields are downloaded from the MAST archive\footnote{These images are downloaded and processed internally from the MAST archive.  See \url{https://archive.stsci.edu/hst/search.php}}.  In a few cases, this required that we process some of the \hff\ filter images internally when additional data is available.  We designate these bands in Table \ref{image sources}.  We downloaded \clash\ archival science and weight images during February 2015\footnote{see \url{https://archive.stsci.edu/prepds/clash/}}.  All data images used are constructed from the best available data pipelines at the time.  Before modeling out the bCGs in each field (see Section \ref{bcg modeling}), data reduction steps are performed to prepare the science and weight images for modeling.  We describe these steps in the following paragraphs.

The final mosaics in each filter for the \hff\ Program release are stacked and drizzled image products at 30 and 60 mas pixel scales, with major artifacts removed.  All images are aligned to the same astrometric grid based on previous \hst\ and ground-based catalogs \citep[see][for further information]{Lotz2017}.  We use the 60 mas pixel scale ($0\farcs06$/pix) images in our analysis for catalog construction as this was the most reasonable for all accompanying data products.  The \clash\  image products are produced similarly but at 30 and 65 mas pixel scales.  We chose the 30 mas images and use the IRAF tool WREGISTER to match the \clash\ images to the 60 mas pixel scale \hff\ images.  In a few cases, we process some of the \clash\ filter images internally when additional data is available or to improve the mosaics (designated in Table~\ref{image sources}).

For additional \hst\ data images that have not been through the \hff\ or \clash\ data release pipelines, these images are produced with \textit{AstroDrizzle} to create the science and weight images from the FLT and ASN files from the MAST archive.  In order to exactly match the pixel scale ($0\farcs06$), we use the $F160W$ filter image from the \hff\ Program as a reference image for \textit{AstroDrizzle} in each field.  Deeper WFC3/UVIS $F275W$ and $F336W$ data have been collected by the \hst\ observing program PID: 14209 (PI: B.\ Siana). We reduce these data internally and the produced mosaics are processed further similarly to the other \hst\ bands.  At this point, we have produced data images for all \hst\ filters, of each field, that we include in the final catalogs and match the released \hff\ images at a pixel scale of $0\farcs06$.  The remaining data reduction and analysis steps are the same for all \hst\ and $K_S$ band images (see Section \ref{kband imaging}).

A background subtraction is performed on each science image for each field using a Gaussian interpolation to smooth out the mosaic and remove sky background.  The Gaussian interpolation is performed by sampling the mosaics in small regions (size of region defined arbitrarily based on results of interpolation) and setting a limiting magnitude and threshold of sources that can contribute to the overall background of the image.  For the \hst\ images and $K_S$ band, we found \sex\ AUTO background subtraction runs best with the following parameters:  mesh size of 64, limiting magnitude of 15 and maximum threshold of 0.01.  If there are multiple epochs for the same field and filter, we combine the background-subtracted mosaics using a weighted mean and simply add the weight images together.

Next, we improve the mosaics for modeling and catalog construction by detecting and cleaning cosmic rays that remain after the initial \textit{AstroDrizzle} combination.  It is important to remove cosmic rays so that they are not detected as sources and to not affect the nearby pixels once we have point-spread function (PSF)-matched images.  We remove any cosmic rays either by hand using a DS9 region mask or by running the image through, L.A.\ Cosmic \citep{vD2001}.  This step improves the image quality and reduces the number of false detections.

The final step we perform, before we model out the bCGs, is to remove any data that would been seen as bad data during the modeling process.  This is accomplished by creating a weight mask for each filter of each field.  We mask any pixel whose value in the weight image is very small compared to the median weight of the image.  The value of the weight pixels should never be negative and ones that have very small values typically have shorter exposure times and have unacceptable science data quality necessary for analysis.  These background subtracted, cosmic ray cleaned, and weight masked images are now ready to be used for the modeling of the bCGs (see Section \ref{bcg modeling}).

\subsection{Additional Data}

For better and more complete photometric catalogs of the \hff, we collect additional data available from ground-based sources ($K_{S}$-band imaging, 2.2\micron) and \spitzer/IRAC (3.6\micron\ and 4.5\micron\ imaging, also 5.8\micron\ and 8.0\micron\ imaging is available for the three Abell clusters) to extend the coverage of these fields into the IR.  The sources of the additional data are described in the following sections.  The raw $K_S$-band images were drizzled to $0\farcs06$ pixel scale to match the \hst\ image grid and the \spitzer/IRAC imaging pixel scale is $5\times$ larger ($0\farcs3$).  We describe the modeling and analysis of these data sets in Section \ref{low res modeling}.

\subsubsection{$K_{S}$-band Imaging}
\label{kband imaging}

Ultra-deep $K_{S}$ imaging of all of the \hff\ clusters and parallels were carried out for the ``$K$-band Imaging of the Frontier Fields'' (``KIFF''\footnote{see \url{http://www.eso.org/sci/observing/phase3/news.html\#kiff}}) project \citep{Brammer2016}.  These observations have been observed with the VLT/HAWK-I and Keck/MOSFIRE instruments for the six clusters and six parallel fields.  The VLT/HAWK-I integrations of the \aone, \mone, \atwo\ and \athree\ clusters and parallels reach 5$\sigma$ limiting depths of $K_{S} \sim 26.0$ (AB, point sources) and have excellent image quality (FWHM $\sim 0\farcs4$). Shorter Keck/MOSFIRE integrations of the \mtwo\ and \mthree\ clusters and parallels reach limiting depths $K_{S} = 25.5$ and 25.1 with seeing FWHM$\sim 0\farcs4$ and $0\farcs5$, respectively.  In all cases, the $K_{S}$-band mosaics cover the primary cluster and parallel \hff\ fields entirely with small exceptions (see Figures \ref{hff layout 1} and \ref{hff layout 2}).  The total area of the $K_{S}$-band imaging is 490 arcmin$^2$.  These observations (at 2.2\micron) fill a crucial gap between the space-based observations of the \hff\ (reddest \hst\ filter, 1.6 \micron) and \spitzer/IRAC (bluest 3.6 \micron).  While not as deep as the space-based observations, these deep $K_{S}$-band images provide important constraints in determining galaxy properties from galaxy modeling that are improved greatly from this extra coverage \citep[see][for more detail]{Brammer2016}.

\subsubsection{$IRAC$ Imaging}
\label{irac}

The multi-wavelength photometric catalogs presented in this work include photometry in the \spitzer/IRAC 3.6 \micron\ and 4.5 \micron\ bands based on the full-depth mosaics assembled by our group.  These data probe rest-frame wavelengths redder than the Balmer Break up to $z\sim 8-10$, and therefore provide important constraints for the derived photometric redshift and stellar population parameters.

The IRAC 3.6 \micron\ and 4.5 \micron\ mosaics combine all the \spitzer/IRAC data available to December 2016.  Specifically, \aone\ and its parallel are combined data from PID 83 (PI: Rieke) and PID 90257 (PI: Soifer); \mone\ and its parallel from PID 90258 (PI: Soifer) and PID 80168 (ICLASH - PI: Bouwens); \mtwo\ and its parallel from PID 90259 (PI: Soifer), PID 60034 (PI: Egami) and PID 90009 (SURFS-UP - PI: Bradac); \mthree\ and its parallel from PID 90260 (PI: Soifer), PID 60034 (PI: Egami) and PID 90009 (SURFS-UP - PI: Bradac); \atwo\ and its parallel from PID 10170 (PI: Soifer), PID 83 (PI: Rieke), and PID 60034 (PI: Egami); finally, \athree\ and its parallel from PID 10171 (PI: Soifer), PID 64 (PI: Fazio), PID 137 (PI: Fazio) and PID 60034 (PI: Egami).

Notably, \aone, \atwo\ and \athree\ clusters benefit from observations of the IRAC 5.8 \micron\ and 8.0 \micron\ bands during the cryogenic mission (PIDs 83, 64 and 137).  Mosaics in these bands are built using the same procedures adopted for the IRAC 3.6 \micron\ and 4.5 \micron\ bands.  Below, we introduce briefly the steps adopted to assemble the IRAC mosaics, referring the reader to \citet{Labbe2015} for a more detailed description of the process.

The reduction of the IRAC data is carried out using the pipeline developed by \citet{Labbe2015}, using the corrected Basic Calibrated Data (cBCD) generated by the Spitzer Science Center (SSC) calibration pipeline. The full process is organized in two passes. During the first pass, each cBCD frame is corrected for background and persistence from very bright stars and other artifacts. Then the frames of each Astronomical Observation Request (AOR) are registered to the reference frame (the \hff\ detection image) and median combined. During the second pass, the pipeline removes cosmic rays, improves the background subtraction and carefully aligns the frames to the reference image, before the final co-addition of the frames.  The resulting mosaics have a pixel scale of $0\farcs3$ and the same tangential point of the \hff\ detection image. The average exposure in the 3.6 \micron\ and 4.5 \micron\ is $\sim 50 h$, corresponding to an AB magnitude depth of $\sim25$ ($5 \sigma$, aperture $3\farcs0$ diameter).  In all cases, the IRAC imaging for the 3.6 \micron\ and 4.5 \micron\ bands cover the primary cluster and parallel fields entirely of the \hff.

Accurate PSFs are key for robust photometry.  For each mosaic, the pipeline generates a spatially varying empirical PSF.  At each position in a grid across the mosaic, a high signal-to-noise (S/N) template PSF, obtained from observations of $\sim 200$ stars, is rotated and weighted according to the rotation angles and exposure time map of each AOR at the specific position on the grid. The final PSF is constructed combining the set of rotated, weighted templates.

\section{Photometry}
\label{modeling and catalogs}

\begin{figure}[ht!]	%%%%%%%%%%%%%%%  PLOT  %%%%%%%%%%%%%%%%%
\epsscale{1.25}
\plotone{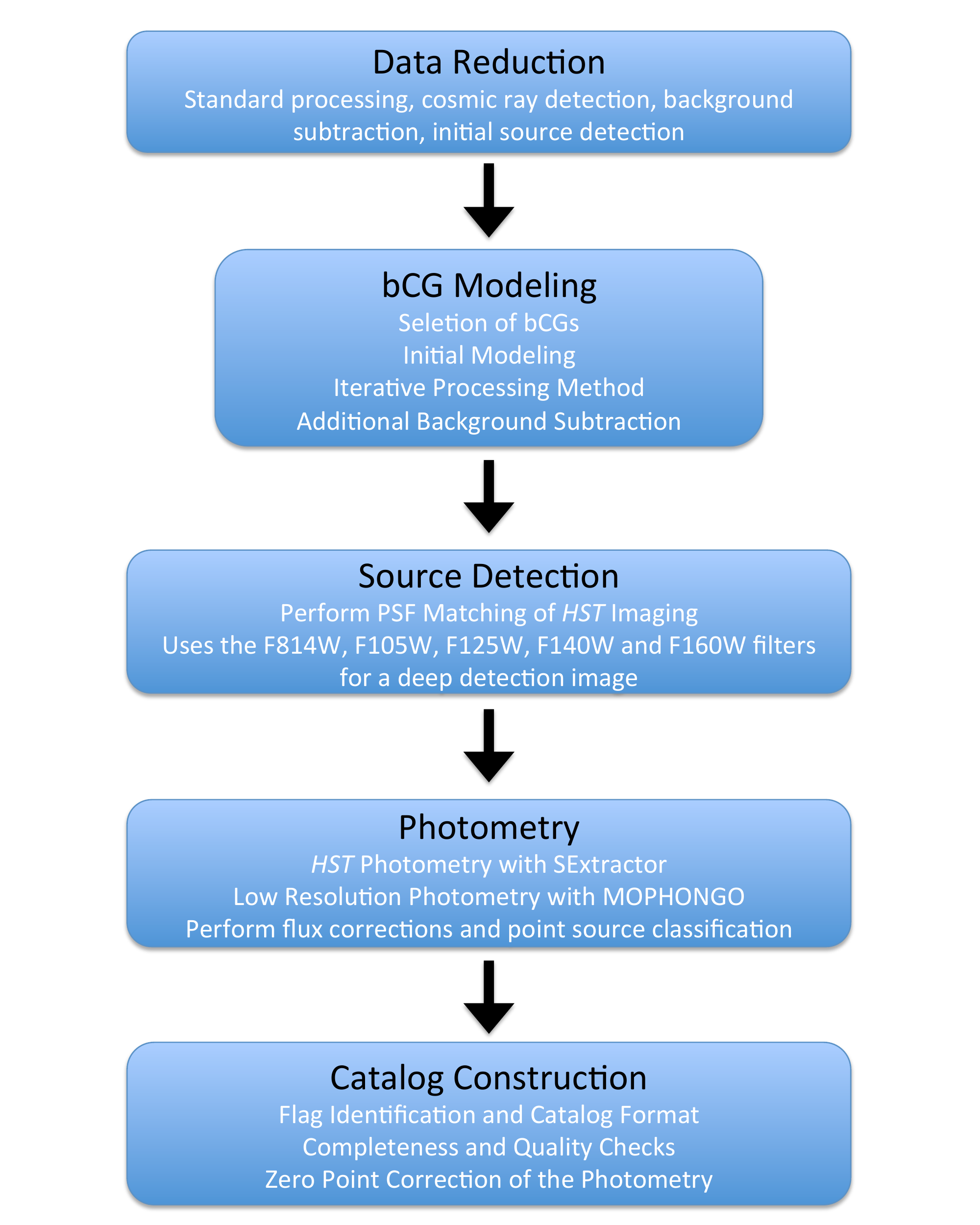}
\caption{Illustration of the main steps performed from data reduction to final catalog construction for all the data presented here of the \hff\ cluster and parallel fields. }
\vspace{6pt}
\label{procedure diagram}
\end{figure}		%%%%%%%%%%%%%%%  PLOT  %%%%%%%%%%%%%%%%%

Here, we describe the procedure for producing the photometric catalogs of each field.  We start by following the standard pipeline for image processing, performing background subtraction on each image before combining multiple epochs (if already not performed by the \hff\ data core team) and cleaning the images for remaining artifacts and cosmic rays (Section \ref{redux}).  Next, we model out bCGs from each field that contribute significant light and perform an additional background subtraction on the resulting bCGs out mosaics (Section \ref{bcg modeling}).  We then PSF match the shorter wavelength bands to the WFC3/$F160W$ band and perform a source detection with \sex\ for each field using a detection image created from the $F814W$, $F105W$, $F125W$, $F140W$ and $F160W$ bands (Section \ref{detection}).  Finally, fluxes are estimated for each band of each field with \sex\ and error analysis is performed (Section \ref{hst photometry}).  We show a diagram of the procedure in Figure \ref{procedure diagram}.

\subsection{Modeling Out of bCGs}
\label{bcg modeling}

One of our main science goals for these catalogs is to identify sources magnified by the gravitational potential of the cluster galaxies and the cluster itself.  To accomplish this, we need to model out the light from the galaxy cluster members or at least the brightest members that contribute the most light to the cluster and ICL.  We adopt a method that measures the isophotal parameters of a galaxy and removes the resulting model as described by \citet{Ferrarese2006}.  We summarize the procedure and additions necessary for our modeling purposes.  In the following sections, we describe our selection of bCGs that contribute significantly to the light of the cluster.  We summarize the procedure that creates a galaxy model for a bCG.  Finally, we describe our iterative process that improves on the initial models to produce a final cluster model.

\subsubsection{Selection of bCGs}
\label{selection of bcgs}

\begin{figure*}[ht!]	%%%%%%%%%%%%%%%  PLOT  %%%%%%%%%%%%%%%%%
\epsscale{0.55}
\plotone{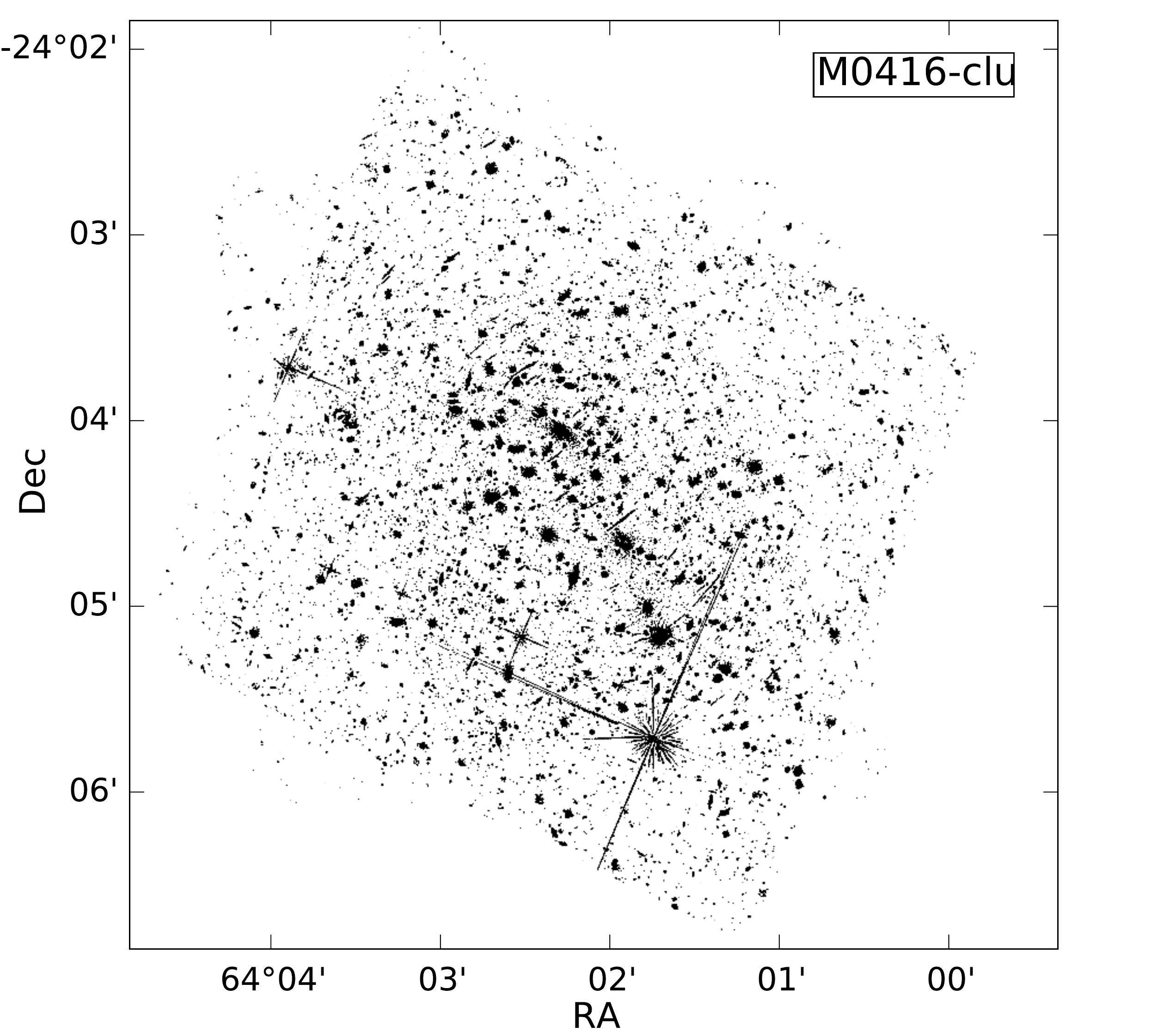}
\epsscale{0.55}
\plotone{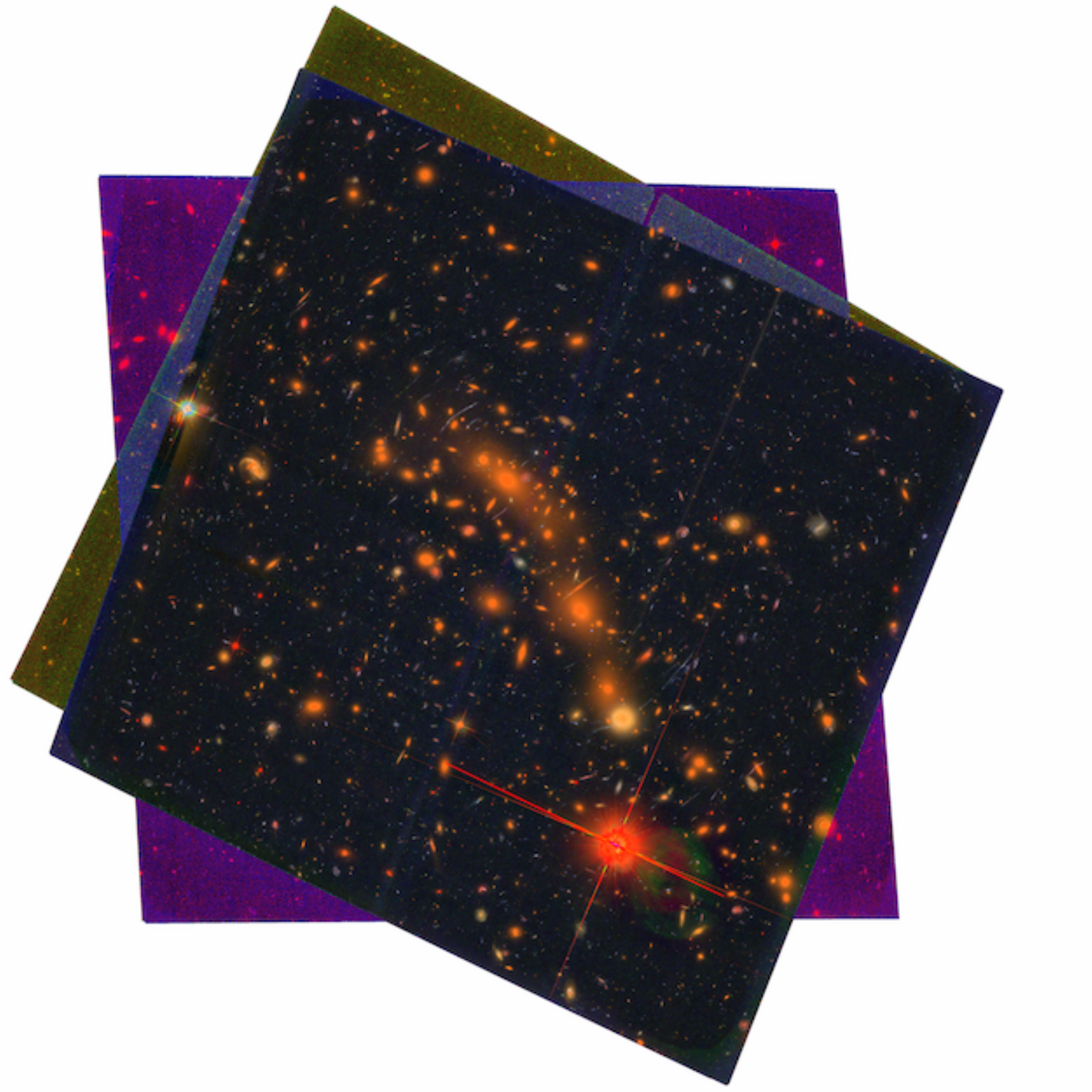}
\caption{(Left panel) Initial segmentation map of \mone\ cluster using a heavy background subtraction with \sex\ for identifying bCGs to be modeled out (all other fields can be found in the Appendix).  North is up and East is to the left.  (Right panel) A false color RGB image of the cluster made with the $F435W$, $F606W$ and $F814W$ bands to better identify cluster members (reddish-orange colored galaxies; refer to Section \ref{selection of bcgs} for further details).}
\label{initial maps}
\end{figure*}

As a first pass, we identify bCGs to be modeled out using an over-subtracted background detection image to produce a segmentation map and associated initial catalog.  This detection image is constructed in the same manner that is performed for our final catalogs (Section \ref{detection}).  This is accomplished using \sex\ with an aggressive background subtraction to identify the centers of all sources and have a complete as possible initial catalog (Figure \ref{initial maps}, left panel).

		%%%%%%%%%%%%%%%  PLOT  %%%%%%%%%%%%%%%%%

We take great care to identify cluster members by color with RGB mosaics of each field (cluster members appear as reddish-orange galaxies, right panel of Figure \ref{initial maps}).  We create the RGB mosaics of each field using the $F435W$, $F606W$ and $F814W$ bands.  These bands are chosen to limit the light from the brightest galaxies that would wash out the detail to identify smaller contributing cluster members affecting our ability to identify background sources.  Ultimately, the selection of cluster members to be modeled out is done in a somewhat arbitrary manner but guided by the principals that these galaxies are bright and/or affecting nearby background sources and appear in many bands for better modeling.  For these reasons, we are more aggressive in our selection of bCGs to model out that fall within the WFC3 footprint and less aggressive outside the WFC3 footprint (i.e.\ the ACS).  Also, we choose to model out fainter cluster members that have lensed sources nearby that affect their photometry.

Furthermore, due to the limitations of the modeling code to handle nearby resolved spiral galaxies, we choose not to model them out (even if the source contributes significantly to the light in the field) as the resulting residual and model are undesirable.  However, we do model out nearby bright elliptical galaxies when possible (e.g.\ \mone\ and \mtwo\ clusters), but this results in only a few galaxies for all fields.  Also, we limit our selection to not include edge-on disk galaxies of cluster members due to these limitations \citep[see][Section 3.2 for specifics]{Ferrarese2006}.  However, we do note a few edge-on galaxies are selected, where the benefit of modeling out the galaxy improves the detection of background sources.

\subsubsection{Method for Modeling a bCG}
\label{model a bcg}

We summarize here the method used to model a galaxy's light of our selected sources \citep[we refer the reader to][for more detailed information on the modeling procedure]{Ferrarese2006} and describe changes to this code that are necessary for the \hff\ data.  Again, we note that this code is designed originally to model elliptical galaxies and has some shortcomings for spiral galaxies.  However, our improvements using an iterative process have made these shortcomings mostly negligible (see Section \ref{iteration method}).  Furthermore, the adopted method is superior compared to other modeling codes, e.g.\ GALFIT, especially for elliptical galaxies with significant isophotal twisting, which are the predominant type of bright galaxies in the cluster environment and those limiting the full exploitation of the \hff\ cluster data depth.

The IRAF task ELLIPSE is used to measure the isophotal parameters for each modeled galaxy.  The best fitting parameters are determined by minimizing the sum of the squares of the residuals between the data and the ellipse model.  First, a mask is created that masks all sources but the galaxy to be modeled.  This is done using \sex\ to identify all possible sources in the mosaic of the band.  Next, all objects near the center of the bCG are unmasked and an ELLIPSE run is performed with a fixed center.  \sex\ is run again on the residual image using a weight image (which prevents it from picking up noisy areas and residuals) to create a mask of objects near the bCG.  The final mask is built from the first mask outside a region determined by the ELLIPSE run and the new mask inside.  Finally, the central region of the bCG is unmasked and then the mask is blurred, by a Gaussian profile, to minimize pixels that may have been missed during this process.

The next step is to create the model itself.  This is accomplished by using the mask created and performing another ELLIPSE run with all parameters allowed to vary (including the center within 2 pixels\footnote{In every case, the centers determined by ELLIPSE are essentially the same as our centers ($< 1$ pixel offsets) from the selection method (see Section \ref{selection of bcgs}), which are more reliable.}).  The surface brightness parameters are found out to a radius we set arbitrarily, but large enough to measure all the light of the bCG, and this can include ICL.  However, ELLIPSE fails to converge well before this condition is met.  When this happens, the mean values for the five outermost fitted isophotes are calculated and ELLIPSE is run with $\theta$, $\epsilon$ and the isophotal center fixed to these values.  The parameters that are returned from this procedure are given to the IRAF task BMODEL to create the model from the isophotal parameters.  However, BMODEL can have problems getting the interpolation correct, especially at large radii, with spurious results.  This is fixed by splining and interpolating the parameters from the ELLIPSE run that is used for BMODEL.  Furthermore, a local background, for the extent of the bCG model, is estimated and added to produce the final model for the bCG.  This results in a more accurate residual and a smoother profile at larger radii.

Finally, the curve of growth is measured from the largest radii isophote inwards to determine when the model surface brightness falls below the measured sky background for the image.  This is done to help eliminate extra light being modeled that is attributed to the sky.  The resulting built model for the bCG is then subtracted from the mosaic.  We create an input list of all the galaxies that we have selected to be modeled and do an initial run for each galaxy.  This is done in succession for each galaxy to be modeled creating a new mosaic with the galaxy removed.  We manually check the final result after all the bCGs have been modeled out to see if manual input is required.

We make an addition to the galaxy modeling code by creating a ``master mask'' from the original mosaic to be used for each bCG modeled out.  We make the master mask in the same manner as described previously in this section, but all sources are masked.  Then for each bCG modeled out, we use the master mask and substitute in a small portion (a box) of the mask created for the bCG being modeled out.  We substitute mask sizes of $3\farcs6 - 24''$ along a side as determined by the size of the bCG and density of nearby sources.  We add this step because the mask created for the bCG can be affected by residuals from poor modeling of previous bCGs, negatively impacting nearby sources and subsequent modeling.  We discuss the importance of this step further in the iterative process (Section \ref{iteration method}).

We find the procedure for the modeling of a bCG works quite well in an automated way.  But, one aspect that has significant impact requiring manual input for some bCGs is to edit the mask manually, usually masking more area around other nearby sources contributing to the fit.  This is accomplished using the IRAF task IMEDIT.  When this occurs, the mask is saved and used for all future runs as explained further in the iterative process.

\subsubsection{Iterative Processing Method of bCGs}
\label{iteration method}

\begin{figure*}[ht!]	%%%%%%%%%%%%%%%  PLOT  %%%%%%%%%%%%%%%%%
\epsscale{2.25}
\plottwo{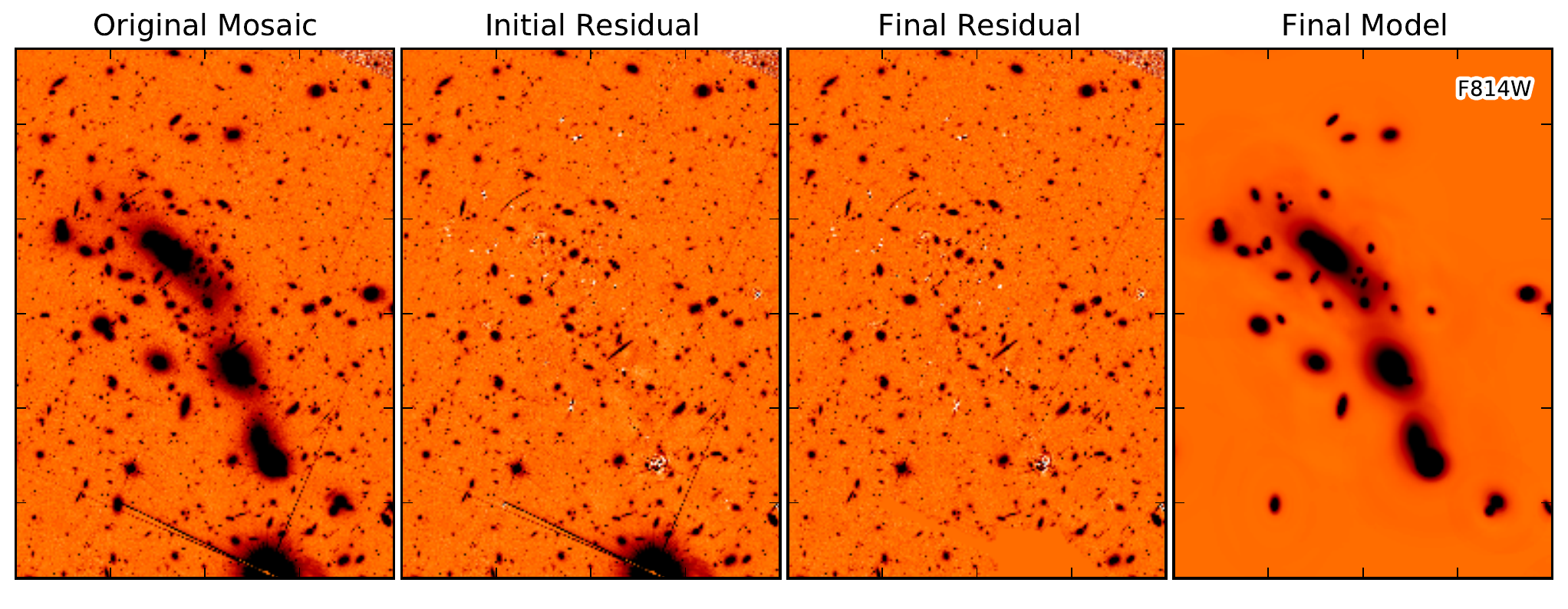}{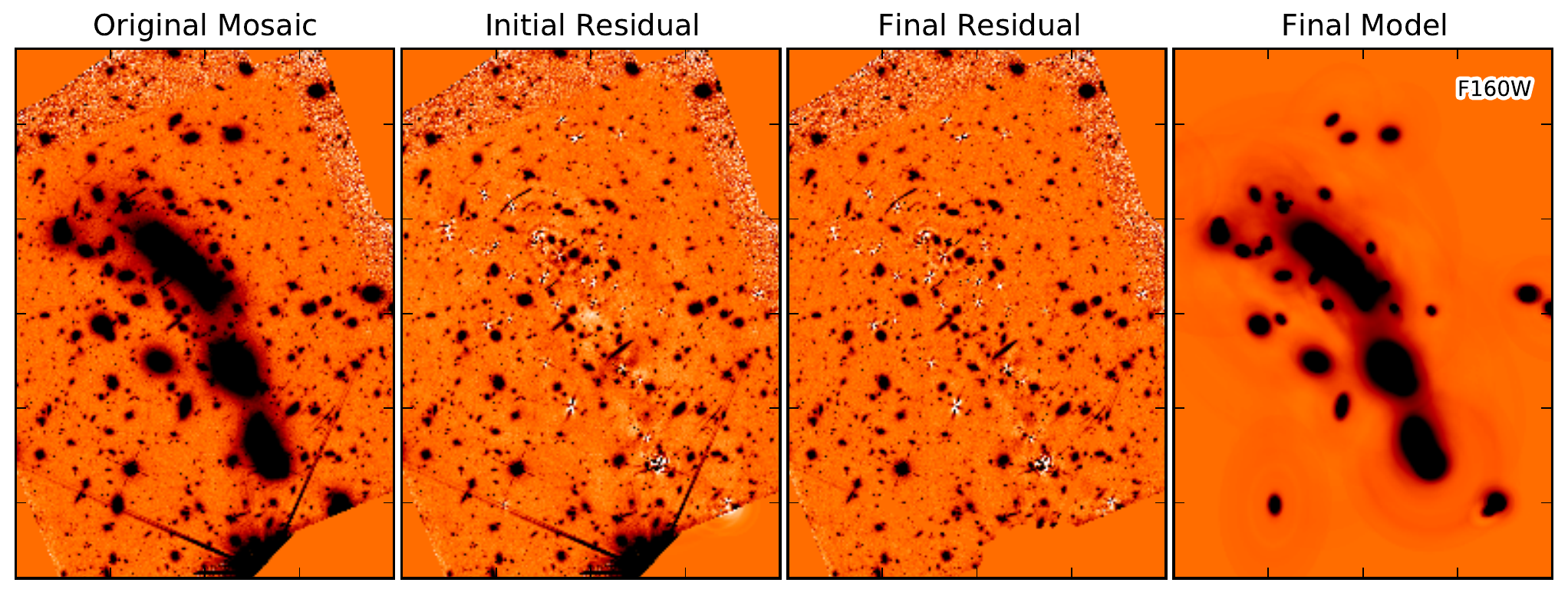}
\caption{Results from the modeling procedure on the \mone\ cluster $F814W$ (top) and $F160W$ (bottom) bands (all other fields can be found in the Appendix).  \textit{Left to right}:  The original image (defined in Section \ref{iteration method}), residual mosaic after the initial run (Section \ref{model a bcg}), final mosaic after the additional sky subtraction (Section \ref{additional skysub}) and model of the cluster after iterative processing method (Section \ref{iteration method}).  All images show the same scale and region of the cluster.  The final residual mosaic is used to extract the photometry of all detected sources except for the modeled out bCGs, whose photometry is extracted from the final model image.}
\label{modeling results}
\end{figure*}

The initial model of the cluster for each band (sum of all modeled bCGs that includes ICL; see Table~\ref{bcgs modeled} for number of bCGs modeled in each field) is a useful result but not very accurate for precise photometry of the remaining sources or reliable photometry of the bCGs themselves (see panels second from left in Figure \ref{modeling results}).  To improve the models themselves and thus improve the photometry, we developed an iteration method that can be run on the resulting models to improve them.  For clarity, we define the term ``original mosaic'' as the mosaic created after the data reduction steps discussed earlier but before any bCG modeling has been performed (including the initial run).

\begin{deluxetable}{lcc}		%%%%%%%%%%%%%  TABLE  %%%%%%%%%%%%%%%
\tablecaption{bCGs Modeled for Each Field \vspace{-6pt}
\label{bcgs modeled}}
\tablecolumns{3}
%\tablewidth{0pc}
%\setlength{\tabcolsep}{12pt}
\tablehead{
\colhead{Field} & \colhead{Cluster} & \colhead{Parallel} \\
 & \colhead{(\# Galaxies)} & \colhead{(\# Galaxies)}
}
\startdata
A2744 & 79 & 27 \\
M0416 & 49 & 12 \\
M0717 & 35 & 7 \\
M1149 & 63 & 9 \\
A1063 & 90 & 22 \\
A370 & 75 & 13
\enddata
\tablecomments{The number of bCGs is for the $F814W$ filter and includes all bCGs that were modeled for that field.  The same amount or less were modeled out for each of the other filters from the same set of bCGs.}
\end{deluxetable}		%%%%%%%%%%%%%  TABLE  %%%%%%%%%%%%%%%

		%%%%%%%%%%%%%%%  PLOT  %%%%%%%%%%%%%%%%%

After the initial run (described above), the code runs through 10 more iterations of each galaxy in the input list of bCGs for the specific field and band\footnote{As described in Section \ref{selection of bcgs}, some selected galaxies fall outside the WFC3 footprint and are not included for those bands.  This varies depending on the specific field and band as each band can have different orientations and coverages from all the included data.} (11 total iterations).  For the first iteration (modeling the bCGs for the second time) we start with the residual image after all the galaxies have been modeled out (i.e.\ the resulting mosaic after the initial run).  Then, in succession, we add back each bCG modeled out one at a time to this residual image (in effect creating a new mosaic with only that bCG included) and re-run the modeling of it.  We then subtract off the new model from this image where the previous model was added back into it.  The result of this improves the model and the residual for that bCG without having contamination from all the surrounding bCGs that hindered the initial models.  This is done for all the galaxies in the input list until completed.

Once all the bCGs have finished creating new models in this manner, we sum and subtract off the new cluster model from the original mosaic and use that to begin the process again for the next iteration.  We find that this method reliably converges after a few iterations and achieves optimal results within 10 iterations.  Also to eliminate further issues from bad fits (as mentioned earlier), we allow for certain bCGs, usually the brightest and/or heavily crowded regions, to create new masks on each iteration and substitute into the master mask (described in Section \ref{model a bcg}).  For the most part, isolated bCGs do not benefit from this (and rarely can result in unsatisfactory models) as nearby galaxies are well masked initially.

In an effort to create the best overall model of the cluster light, we use a high-low mean combine of four iterations from the 10 iterations after the initial run.  We use the IRAF task IMCOMBINE to accomplish this by setting the following parameters (combine=``average'', reject=``minmax'', nlow=``4'', nhigh=``2'') for the cluster models.  The ``nlow'' parameter rejects the four lowest value pixels and the ``nhigh'' parameter rejects the two highest value pixels.  We set the ``nlow'' parameter to reject the models that do not model out enough light at larger radii, which is more of a concern in the final result than the ``nhigh'' parameter.  The ``nhigh'' parameter is set to remove the models with too much light subtracted out in the core, where the models leave residual patterns that are unavoidable (Figure \ref{modeling results}).  This process gave the best results for not including poor models and the smallest residuals leaving a smooth accurate mosaic for each band of each field.

These adjustments make the biggest impacts in allowing the galaxy modeling code to be able to work out poor fits that the IRAF tasks ELLIPSE and BMODEL sometimes return.  We note a few issues still remain, i.e.\ some models have negative flux values in the outermost regions when allowing for large radii isophotes.  This seems to be the ELLIPSE task response to another brighter galaxy being modeled out first that subtracted off too much light.  The ELLIPSE task tries to compensate for this by adding back in light in the outer regions of nearby smaller galaxies currently being modeled (producing negative values in the models).  An example is when the local background (see Section \ref{model a bcg}) is measured and added to (in this case subtracted from) the model.  However, we stress that these issues are minor and the final summed model of the cluster is very accurate (uncertainties $< 1$\% from an estimation of the bCGs measured fluxes).  

The fluxes and uncertainties are measured for the modeled bCGs in the same manner as the sources in the final residual mosaics (described in detail in the following Sections \ref{hst photometry} and \ref{flux corr}) but using the final cluster model for each field and band.  The modeled out bCGs are given an identifier (id) 20000 and above and ``bandtotal'' reference of ``bcg'' (see Section~\ref{cat format}).  The patterns left by the modeled bCGs (primarily in the core) are masked to measure the remaining flux in the final residual mosaics and added to the uncertainties given in the catalogs for each bCG.

\subsubsection{bCG Modeling of the Low-Resolution Data}
\label{low res modeling}

For the ultra-deep $K_{S}$-band mosaics from \citet{Brammer2016}, we are able to use our iterative processing method to model out the bCGs the same way as the \hst\ bands.  This is possible because the pixel scale is equivalent ($0\farcs06$) to the \hst\ bands and the resolution is sufficient to produce an accurate cluster model of the bCGs.  All steps for the $K_S$ band data follow the modeling of the \hst\ bands, including the additional sky subtraction.

For the IRAC mosaics, a different approach needs to be adopted because of the larger pixel scale ($0\farcs3$) of the IRAC mosaics, which is not compatible with the fitting routine used for the bCG modeling.  The approach (to satisfactory results) took advantage of the fact, we have models produced for these bCGs in the shorter wavelength bands.  We use the $F160W$ and $F814W$ models to PSF match and scale them to the IRAC bands (3.6 and 4.5 \micron\ bands for all fields; 5.8 and 8.0 \micron\ bands for the Abell clusters).  $F814W$ models are used only where the $F160W$ mosaic does not cover the bCG models.  Although the $K_{S}$-band models would be preferable due to the closer matching wavelength band, they produce inferior IRAC models of the bCGs because of the differences in the sky background subtraction during data reduction for ground- and space-based observations.

To match the $F160W$ and $F814W$ models appropriately to the IRAC bands, the original mosaic for the $F160W$ and $F814W$ is scaled, registered to the same pixel scale (accomplished with the IRAF task WREGISTER) and PSF matched (see Section \ref{psf matching} for method) to each IRAC band to measure the flux scaling necessary for each model.  We measure the flux in $0\farcs6$ apertures for each $F160W$ model ($F814W$ model, where necessary) to determine the scaling factor for each model.  The $0\farcs6$ aperture is chosen as the best solution as this contained a significant amount of the flux for each bCG model without being contaminated by surrounding galaxies when using the original mosaics for the scaling.

Then, we create the cluster model for the IRAC bands from the $F160W$ and $F814W$ models using these scaling factors.  The models are registered to the pixel scale of the IRAC bands and then PSF matched before applying the scaling.  The IRAC models are summed to create the cluster model and subtracted from the original mosaic for each IRAC band.  While too much light is still subtracted off from the cores of the bCGs, this reflects the same issue with the \hst\ bands at longer wavelengths (Section \ref{iteration method} and see Figure \ref{modeling results}).  This effect is minimal and does not impact the photometry of the IRAC bands.  This method allows for the bCGs to be modeled out of the IRAC bands efficiently without significantly altering the remaining sources.  We follow the same procedure as the \hst\ and $K_S$ bands to measure the fluxes and uncertainties of these IRAC bCG models (see Section~\ref{iteration method}).  This allows each modeled bCG's flux and uncertainty to be measured in a consistent way for all bands in the catalogs.

\subsection{Additional Background Subtraction}
\label{additional skysub}

Once we have the final mosaic with the bCGs modeled out (from the mean of the four best runs), we do an additional sky subtraction.  This is to remove any excess light previously missed during the initial sky subtraction and modeling of the bCGs.  The sky subtraction is performed the same way as earlier for the data reduction process (see Section \ref{redux}) with a Gaussian interpolation of the background.  The result of this sky subtraction is minimal (usually on the order of a few hundredths of a percent for each pixel affected) but improves the background near the borders of the mosaic and the outer regions of the subtracted cluster model (sum of the bCGs modeled out).

\subsection{Source Detection}
\label{detection}

\begin{figure*}[htpb]	%%%%%%%%%%%%%%%  PLOT  %%%%%%%%%%%%%%%%%
\epsscale{2.25}
\plottwo{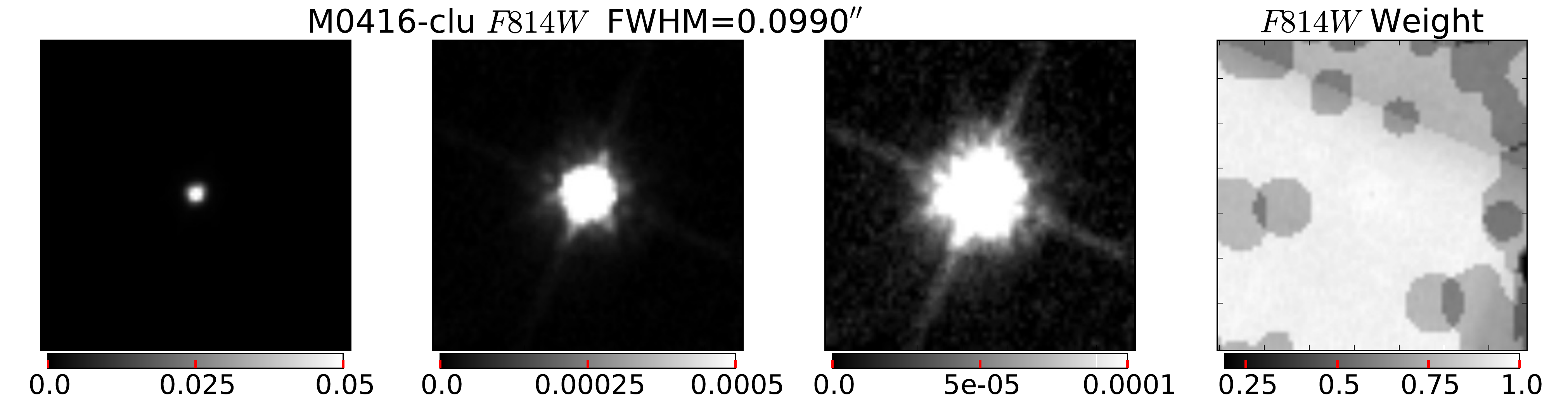}{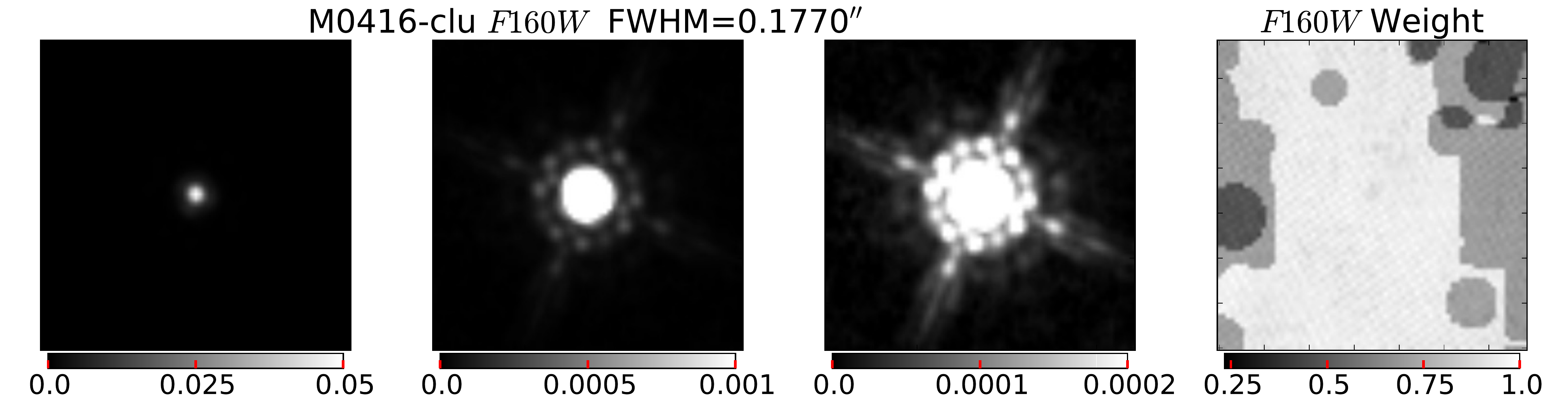}
\caption{Point-spread functions (PSFs) for the ACS/$F814W$ band the WFC3/$F160W$ band in the \mone\ cluster (all other fields can be found in the Appendix). The construction of the PSFs is described in Section \ref{psf matching}. For each filter we show three scales (top panels for $F814W$ and bottom for $F160W$ as labeled) to illustrate the structure of the PSF (from \textit{left to right}:  the core, the first Airy ring and the diffraction spikes).  The images are normalized to a maximum value of one.  The grayscale bars show the scale for each panel. These are different for the ACS and WFC3 as a result of the different FWHMs (listed above the images). We also show the combined weight images for each PSF.  The weight is largest in the center and lower at larger radii and not consistent as shown due to masking of neighboring objects (this is the reason for darker circles appearing).}
\label{psfs}
\end{figure*}		%%%%%%%%%%%%%%%  PLOT  %%%%%%%%%%%%%%%%%

For each field, we create a deep detection image from the bCGs modeled out residual images (see Figure \ref{modeling results}, second panels from the right) of the $F814W$, $F105W$, $F125W$, $F140W$ and $F160W$ bands.  Before we combine the bands to create the detection image, we perform a separate background subtraction on the five mosaics.  This is a separate step, independent from the photometry additional background subtraction (Section \ref{additional skysub}) of each individual band mosaic and is used only for creating the detection image.  This background subtraction utilizes a spline interpolation to better smooth and normalize the background to zero improving our detection of sources when the bands are combined.  We mask all the residuals from the bCGs and any ICL or contaminant (cosmic ray, bad pixel, etc.) that was missed previously.  Then, the images are PSF-matched to the $F160W$ image.  We combine these images together to produce a weighted mean mosaic, using the corresponding error images (obtained from the inverse variance maps) to properly weight the images.  We divide the weighted mean mosaic by its error image to noise-equalize the weighted mean mosaic.  This forms a deep detection image of the central field and larger coverage with the $F814W$ band.  Since the variable weight from each band is taken into account using this method, we do not input a weight map to \sex\ during source detection.

As each cluster and parallel field is significantly different, we allow the detection and analysis thresholds to vary slightly from field to field.  The detection and analysis thresholds are set in the range of $3-6$ depending on the field's specific noise properties (same value for both thresholds).  We require a minimum area of 4 pixels for detection. The de-blending threshold is set to 32, with a minimum contrast parameter of $5\times10^{-6}$ for all fields.  A Gaussian filter of 4 pixels is used to smooth the images before detection.  The detection parameters are chosen as a compromise between de-blending neighboring galaxies and splitting large objects into multiple components \citep[following a similar approach to][]{Skelton2014}.  After an initial run, we check the detection image with the sources found to ensure ICL and residuals did not get identified as sources that are not apparent in the individual images but detectable in the deep detection image.  For these instances, we mask the detected ICL and residuals and re-run \sex\ with the same parameters as defined previously.  This procedure results in the best overall detected sample of sources.

\subsection{PSF Matching of the HST Imaging}
\label{psf matching}

We PSF-match all the \hst\ ACS and WFC3 mosaics to the $F160W$ mosaic, which has the largest PSF FWHM of the \hst\ filters, before performing aperture photometry using the procedure discussed in \citet{Skelton2014}.  Below, we summarize and discuss our results for the \hst\ filters.

\begin{figure}[ht!]	%%%%%%%%%%%%%%%  PLOT  %%%%%%%%%%%%%%%%%
\epsscale{1.15}
\plotone{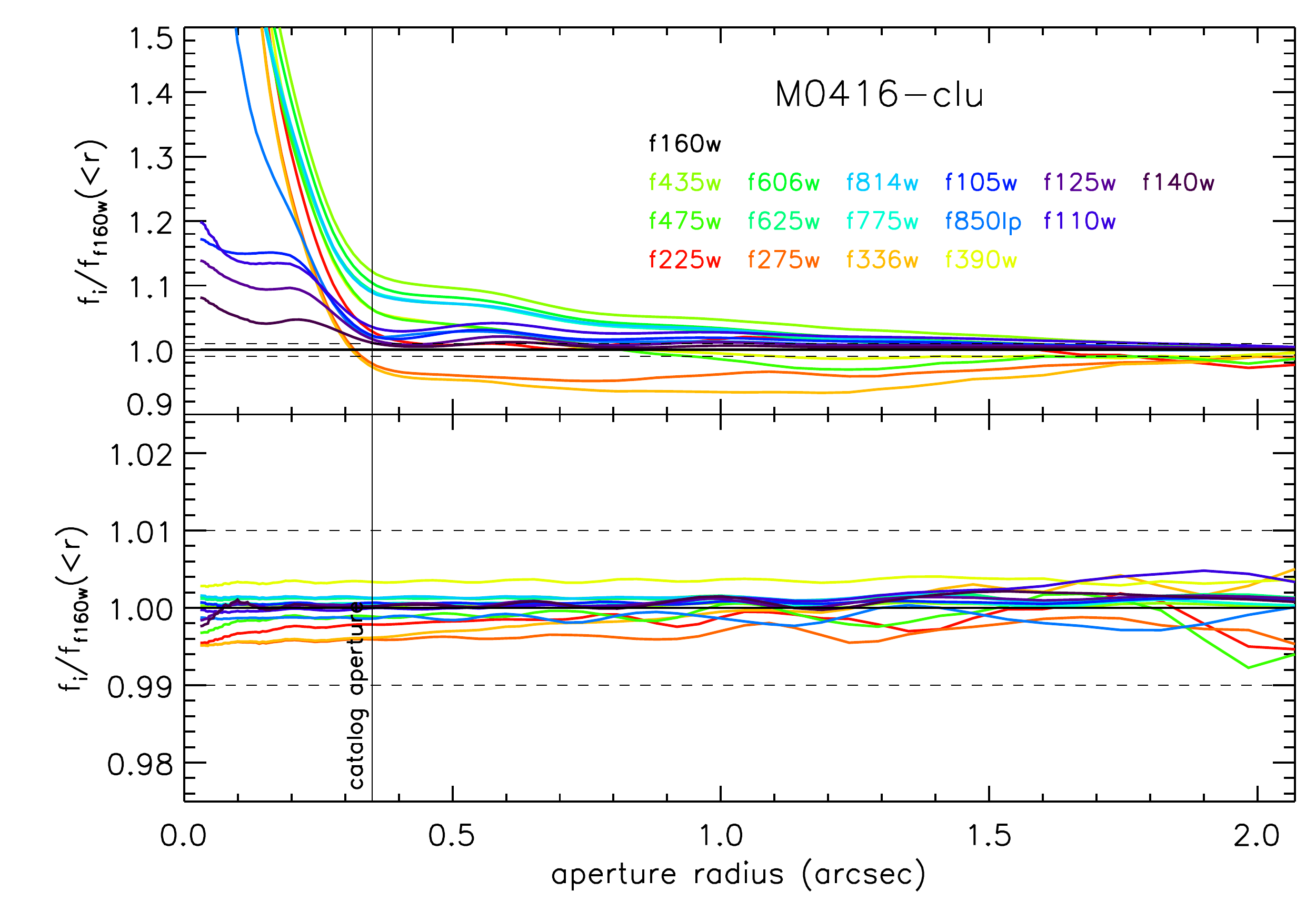}
\caption{Growth curves showing the fraction of light enclosed as a function of radius for each \hst\ filter relative to the $F160W$ growth curve in the \mone\ cluster (all other fields can be found in the Appendix).  The upper and lower panels show the growth curves before and after convolution to match the $F160W$ PSF, respectively.  Note the change in scale between the upper and lower panels. The dashed line in each panel represents a 1\% difference from the $F160W$ PSF.  After PSF-matching, the resulting growth curves in all bands are consistent with the $F160W$ PSF to well within 1\%.}
\label{growth curves}
\end{figure}

We create an empirical PSF for each \hst\ mosaic by stacking isolated unsaturated stars.  This selection is performed by measuring the ratio of flux within a small aperture to a large aperture to correctly identify appropriate stars, adjusting the criteria as necessary for each band.  The number of stars vary for each field and band but the selection results with at least a few stars (3 or more) to tens of stars in each band of the ACS and WFC3 bands.  The UVIS bands present more of a challenge as there are not many sources in these bands.  However, we are able to make use of at least two or more point-like sources in each band of each field that produces satisfactory results (discussed later in this section and demonstrated by the growth curves in Figure \ref{growth curves}).  We make postage stamp cut-outs of these stars following the same parameters detailed in \citet{Skelton2014} with a couple of adjustments.  Since we do not have dozens of stars to choose from in our fields, we allow for large shifts during the re-centering and normalizing process.  Since these are densely packed fields, we do a visual inspection of the PSFs after they are created to check for any contaminants and, if necessary, re-perform the process after additional masking.

\begin{figure}[ht!]	%%%%%%%%%%%%%%%  PLOT  %%%%%%%%%%%%%%%%%
\epsscale{1.15}
\plotone{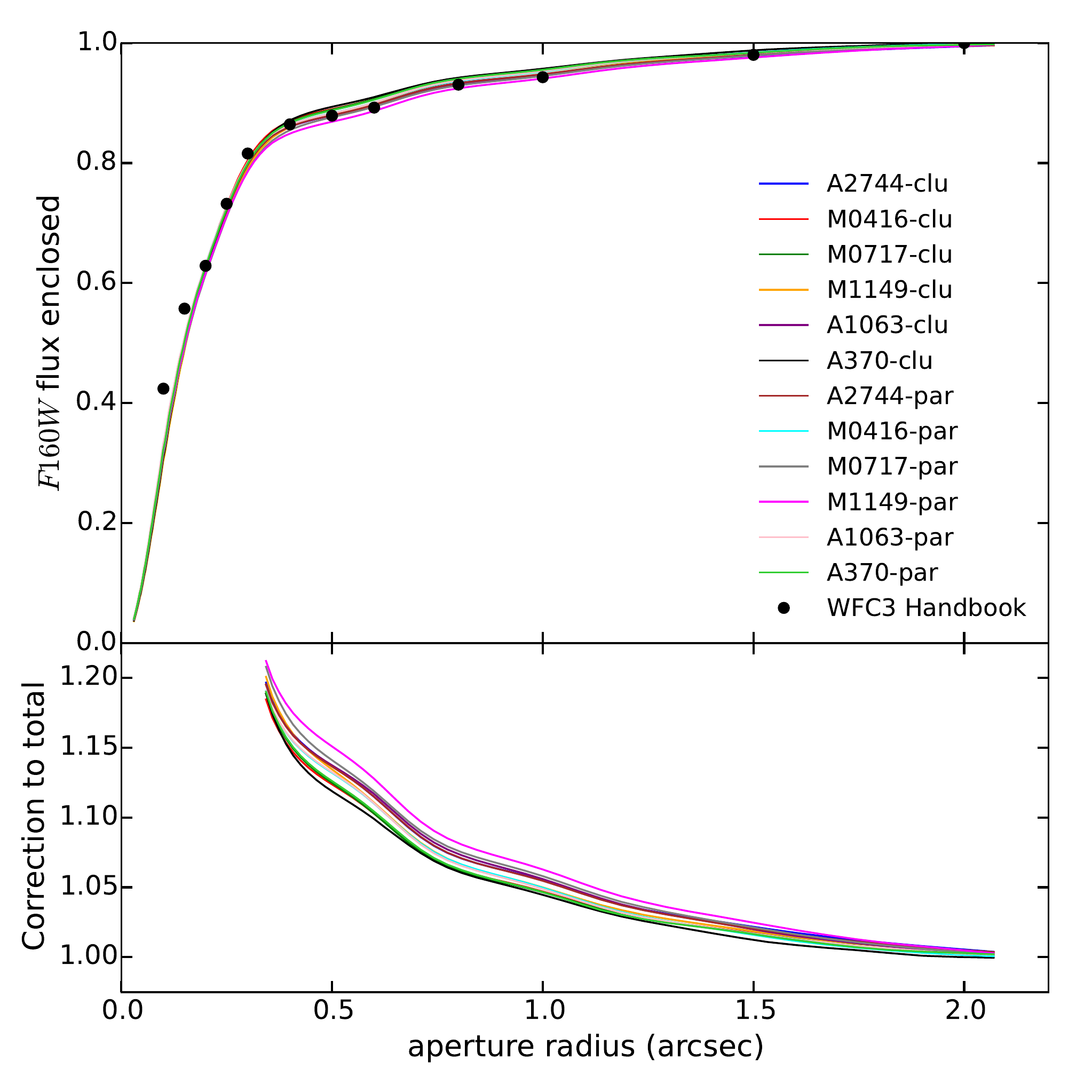}
\caption{$F160W$ growth curves.  Upper panel: the fraction of light enclosed as a function of radius relative to the total light within $2''$, $f(r)/f(2'')$, from the $F160W$ PSF stamp of each field.  The PSFs of the 12 fields are very consistent with each other. The black points show the encircled energy as a function of aperture size, also normalized to $2''$, from the WFC3 handbook. The empirical growth curves agree well with the theoretical expectation.  Lower panel: the correction to total flux for a point source with a circularized Kron radius equal to the aperture radius on the horizontal axis, derived as the inverse of the growth curves in the upper panel $(f(2'')/f(r))$. The minimum Kron radius is set to the aperture radius in which we measure photometry, $0\farcs35$, giving rise to a maximum correction of $\sim$1.19.}
\label{160 growth curves}
\end{figure}

In Figure \ref{psfs}, we demonstrate the PSF stamps at three different contrast levels for the ACS/$F814W$ and the WFC3/$F160W$ bands in the \mone\ cluster to expose the structure of the PSFs.  The structure of the PSFs shown are the core, the first Airy ring ($\sim 0.5$\%) and the diffraction spikes ($\sim 0.1$\%).  Furthermore, the growth curves (that is the fraction of light enclosed as a function of aperture size) for each of the fields are consistent with each other to $<$ 1\%, with almost identical curves at this scale (Figure \ref{160 growth curves}).  For context, we show the consistency of our growth curves with the encircled energy as a function of aperture provided by the WFC3 handbook (normalized to the radius of $2\farcs1 = 35$ pixels).

		%%%%%%%%%%%%%%%  PLOT  %%%%%%%%%%%%%%%%%

As demonstrated by \citet{Skelton2014}, we use a deconvolution code that fits a series of Gaussian-weighted Hermite polynomials to the Fourier transform of the stacked stars, to find the kernel that convolves each PSF to match the $F160W$ PSF (developed by I.\ Labb\'{e}).  In Figure \ref{growth curves}, we demonstrate the ratio of the growth curve in each band to that of the $F160W$ growth curve, before and after the convolution, for the \mone\ cluster.  The PSF-matching is excellent with an accuracy $< 1$~\% within a $0\farcs7$ diameter aperture for all the \hst\ bands and fields (see Appendix).

\subsection{HST Photometry}
\label{hst photometry}

We perform photometry for each \hst\ band with the same method described in \citet{Skelton2014}.  We summarize the steps below and alterations made to better suit the \hff\ data.  We run \sex\ in dual-mode for each \hst\ band, using the detection images described in Section~\ref{detection} and the PSF-matched \hst\ images described in Section~\ref{psf matching}, adopting an aperture diameter of $0\farcs7$ as the photometry aperture flux for all \hst\ bands.  We determine the total flux from the $F160W$ band, where the $F160W$ band has coverage, and the $F814W$ band otherwise (a few sources use the other detection bands depending on band coverage, i.e.\ ``bandtotal'' column in the catalogs).  We correct the \sex\ AUTO flux using the inverse of the fraction of light within a circular aperture that is equivalent to the Kron aperture determined from our growth curves.  For sources with AUTO flux radii smaller then the photometry aperture radius, we take the photometry aperture flux multiplied by the corresponding correction factor to be the total flux.

		%%%%%%%%%%%%%%%  PLOT  %%%%%%%%%%%%%%%%%

We estimate the uncertainty on the total flux using empty apertures of the background noise in increasing size within the noise-equalized images for each band.  For each aperture size, we measure the flux in more than 2000 apertures placed at random positions across the image excluding apertures that overlap with sources in the detection segmentation map.  We find including more apertures is not necessary for accurate error analysis and became difficult at larger radii for certain bands (e.g.\ the WFC3 bands).

\begin{figure*}[ht!]	%%%%%%%%%%%%%%%  PLOT  %%%%%%%%%%%%%%%%%
\epsscale{1.15}
\plottwo{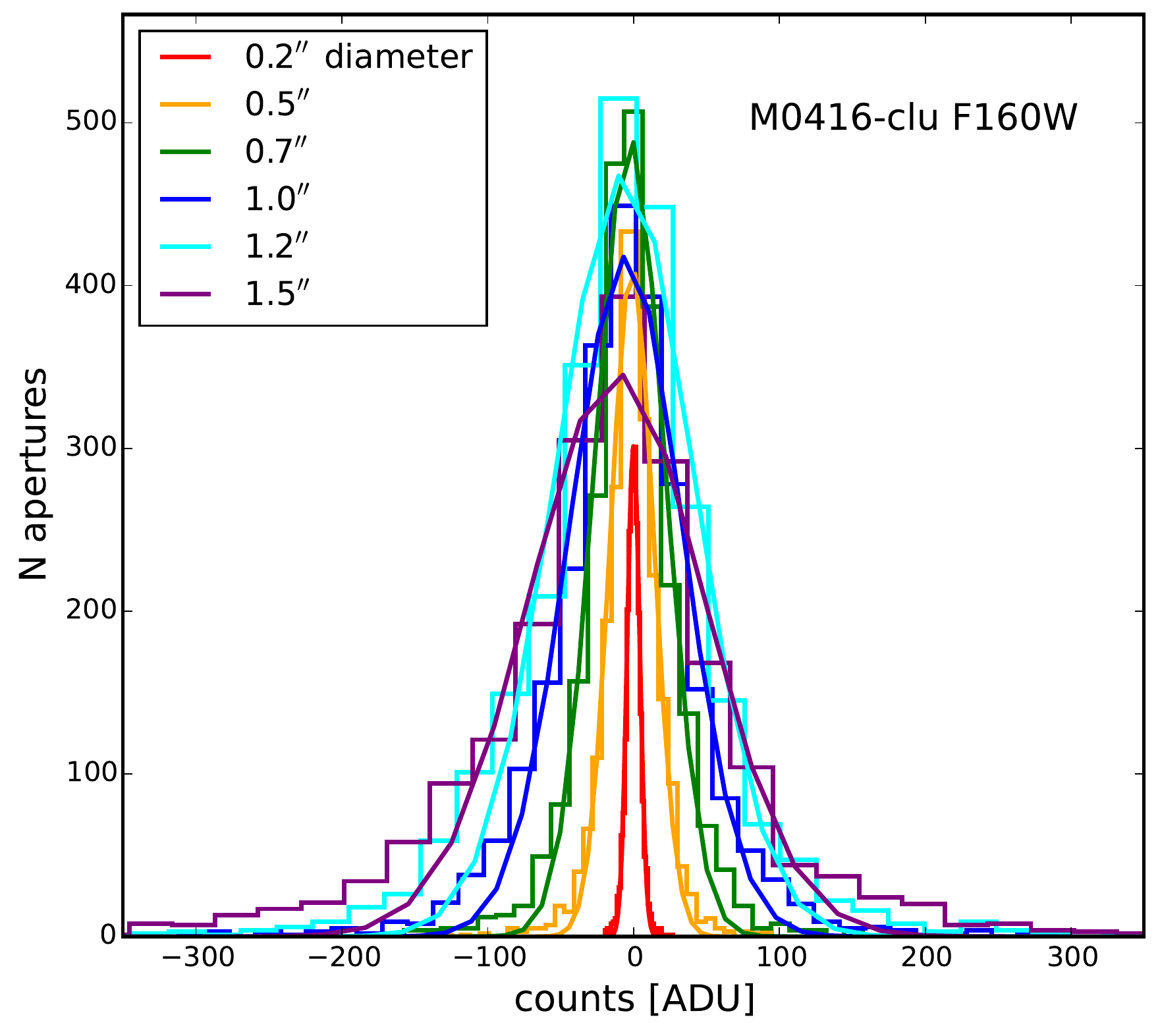}{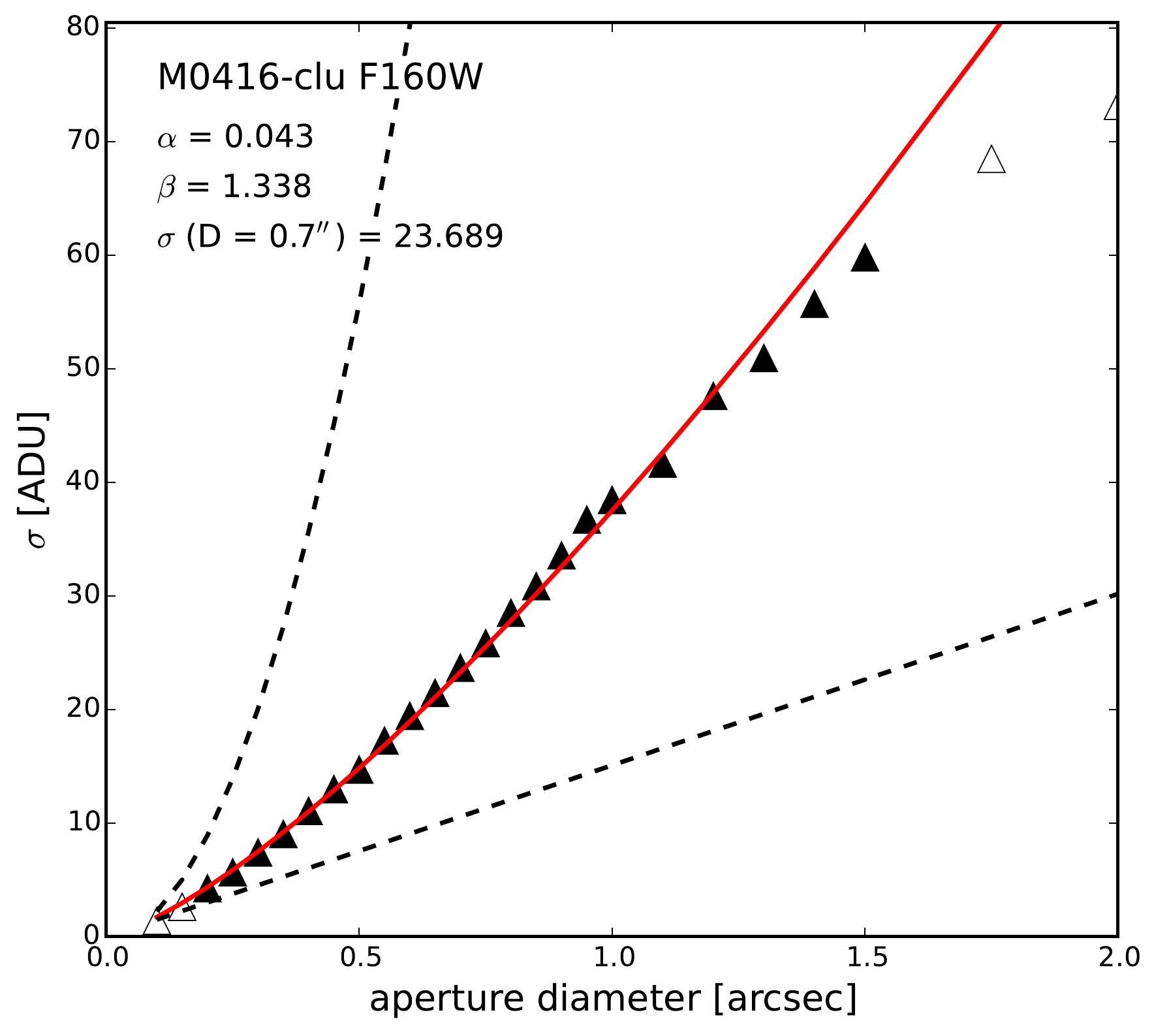}
\caption{Summed counts of different apertures sizes for empty regions sampling the image (left panel) and the scaling of the noise as a function of the aperture size (right panel) for the \mone\ $F160W$ cluster image (all other fields can be found in the Appendix).  The histograms are color-coded by aperture size and given in the figure.  The measured $\sigma$ are shown by the triangles. The solid line shows the power-law fit to the data, with the fit parameters given in the figure. The dashed lines show the linear ($\propto N$) and $N^ 2$ scalings, which correspond to no correlation and perfect correlation between the pixels, respectively.  The $\sigma$ listed is for the photometry aperture size (units of ADU).}
\label{empty aper}
\end{figure*}		%%%%%%%%%%%%%%%  PLOT  %%%%%%%%%%%%%%%%%

Figure \ref{empty aper} (left panel) demonstrates our results for the \mone\ cluster for each aperture size well-described by a Gaussian, with increasing width as aperture size increases.  The measured deviation is described as a function of aperture size in the \mone\ cluster noise-equalized $F160W$ image by fitting a power law to the trend.  We fit a power-law (solid line in Figure \ref{empty aper}, right panel) of the form
\begin{equation}
\sigma = \sigma (D=0\farcs7) \alpha N^{\beta},
\end{equation}
where $\sigma$(D$=0\farcs7$) is the standard deviation of the background pixels at the photometry aperture size (in ADU), $\alpha$ is the normalization and $1 < \beta < 2$ (dashed lines in the figure of $\beta = 1$ and $\beta = 2$ scalings).  The values for each field are given in Table \ref{noiseprop}.  We estimate the uncertainty from this analysis by dividing the median value from the square root of the weight at the position of the object within the circularized Kron radius \citep[see][for more details]{Skelton2014}.  This error term is added in quadrature to the Poisson error to calculate the final uncertainty of each source in the catalog.

\subsection{Low Resolution Photometry}
\label{low res phot}

The significant differences between the \hst\ data and, the ground-based $K_S$ and \spitzer/IRAC data image quality must be quantified, specifically the large differences in the PSF sizes of the \spitzer\ data.  This will allow for accurate information to be obtained without degrading the \hst\ images.  We use MOPHONGO, a code developed by one of us (I. Labb{\'e}), to perform photometry of these longer wavelength bands ($K_S$ and \spitzer/IRAC), as described in \citet{Labbe2006,Wuyts2007,Whitaker2011} following the steps of \citet{Skelton2014} (see their Section 3.5 for detailed description).

\begin{deluxetable}{lcc}		%%%%%%%%%%%%%  TABLE  %%%%%%%%%%%%%%%
\tablecaption{Power-law Parameters for Empty Aperture Errors \vspace{-6pt} \label{noiseprop}}
\tablecolumns{3}
\tabletypesize{\small}
\tablewidth{0pc}
\setlength{\tabcolsep}{18pt}
\tablehead{
\colhead{Field} & \colhead{$\alpha$} & \colhead{$\beta$}
}
\startdata
\aone -clu & 0.027 & 1.543 \\
\aone -par & 0.038 & 1.401 \\
\mone -clu & 0.043 & 1.338 \\
\mone -par & 0.038 & 1.395 \\
\mtwo -clu & 0.030 & 1.506 \\
\mtwo -par & 0.039 & 1.400 \\
\mthree -clu & 0.032 & 1.463 \\
\mthree -par & 0.042 & 1.354 \\
\atwo -clu & 0.028 & 1.526 \\
\atwo -par & 0.035 & 1.432 \\
\athree -clu & 0.036 & 1.416 \\
\athree -par & 0.033 & 1.457
 \enddata
\tablecomments{These parameters are for the $F160W$ band.}
\end{deluxetable}		%%%%%%%%%%%%%  TABLE  %%%%%%%%%%%%%%%

Briefly, the code uses a high-resolution image as a prior to estimate the contributions from neighboring blended sources in the lower resolution image.  We use the detection image as the high-resolution prior.  A map is created to cross-correlate the source positions in the two images.  Then the position-dependent convolution kernel that maps the higher resolution PSF to the lower resolution PSF is determined by fitting a number of point sources across each image.  The high resolution image is convolved with the local kernel to obtain a model of the low resolution image, with the flux normalization of individual sources as a free parameter.  We perform photometry on the original low-resolution image using an aperture appropriate for the size of the PSF (i.e., D$=0\farcs7$ and D=3'' for the $K_S$ and IRAC bands, respectively), with a correction applied for contamination from neighboring sources around each object as determined from the model.  Further flux corrections are applied to account for flux that falls outside of the aperture from the larger PSF.

\vspace{24pt}
\subsection{Flux Corrections}
\label{flux corr}

We correct for Galactic extinction using the values given by the NASA Extragalactic Database extinction law calculator\footnote{\url{http://ned.ipac.caltech.edu/help/extinction_law_calc.html}} at the center of each field, again using the same method presented in \citet{Skelton2014}.  However, we do not interpolate over the dataset for the filters in our catalogs but explicitly calculate the extinction for each field and filter.  The Galactic extinction values applied to our dataset are given in Table~\ref{hffds bands} for each field and filter.  We follow the rest of the flux corrections steps by \citet{Skelton2014} that are summarized briefly in the following paragraph.

\begin{deluxetable*}{rllllll}		%%%%%%%%%%%%%  TABLE  %%%%%%%%%%%%%%%
\tablecaption{Galactic Extinctions for the \textit{Hubble} Frontier Fields Filters \vspace{-6pt}
\label{hffds bands}}
\tablecolumns{7}
\tabletypesize{\footnotesize}
\tablewidth{0pc}
\setlength{\tabcolsep}{0pt}
\tablehead{
\colhead{Filter} & \colhead{\aone} & \colhead{\mone} & \colhead{\mtwo} & \colhead{\mthree} & \colhead{\atwo} & \colhead{\athree} \\
 & \colhead{clu / par} & \colhead{clu / par} & \colhead{clu / par} & \colhead{clu / par} & \colhead{clu / par} & \colhead{clu / par} }
\startdata
UVIS $F225W$ & \dots & 0.286 / \dots & 0.535 / \dots & 0.160 / \dots & 0.086 / \dots & \dots \\
$F275W$ & 0.072 / \dots & 0.225 / \dots & 0.420 / \dots & 0.126 / \dots & 0.067 / \dots & 0.178 / \dots \\
$F336W$ & 0.058 / \dots & 0.182 / \dots & 0.341 / \dots & 0.102 / \dots & 0.055 / \dots & 0.144 / \dots \\
$F390W$ & \dots & 0.160 / \dots & 0.298 / \dots & 0.089 / \dots & 0.048 / \dots & \dots \\
\noalign{\smallskip}
\hline
\noalign{\smallskip}
ACS $F435W$ & 0.047 / 0.044 & 0.148 / 0.152 & 0.276 / 0.275 & 0.083 / 0.086 & 0.044 / 0.044 & 0.117 / 0.110 \\
$F475W$ & \dots & 0.134 / \dots & 0.250 / \dots & 0.075 / \dots & 0.040 / \dots & 0.106 / \dots \\
$F555W$ & \dots & \dots & 0.214 / \dots & 0.064 / \dots & \dots & \dots \\
$F606W$ & 0.032 / 0.030 & 0.101 / 0.104 & 0.189 / 0.116 & 0.057 / 0.059 & 0.030 / 0.030 & 0.080 / 0.076 \\
$F625W$ & \dots & 0.091 / \dots & 0.170 / \dots & 0.051 / \dots & 0.027 / \dots & 0.072 / \dots \\
$F775W$ & \dots & 0.067 / 0.069 & 0.125 / \dots & 0.037 / \dots & 0.020 / \dots & \dots \\
$F814W$ & 0.020 / 0.019 & 0.062 / 0.064 & 0.117 / 0.116 & 0.035 / 0.036 & 0.019 / 0.019 & 0.049 / 0.047 \\
$F850LP$ & \dots & 0.051 / 0.052 & 0.095 / \dots & 0.028 / \dots & 0.015 / \dots & \dots \\
\noalign{\smallskip}
\hline
\noalign{\smallskip}
WFC3 $F105W$ & 0.013 / 0.012 & 0.040 / 0.041 & 0.074 / 0.074 & 0.022 / 0.023 & 0.012 / 0.012 & 0.031 / 0.030 \\
$F110W$ & \dots & 0.036 / \dots & 0.067 / \dots & 0.020 / \dots & 0.011 / \dots & 0.029 / \dots \\
$F125W$ & 0.010 / 0.009 & 0.030 / 0.031 & 0.056 / 0.055 & 0.017 / 0.017 & 0.009 / 0.009 & 0.024 / 0.022 \\
$F140W$ & 0.008 / 0.008 & 0.025 / 0.026 & 0.047 / 0.047 & 0.014 / 0.015 & 0.008 / 0.007 & 0.020 / 0.019 \\
$F160W$ & 0.007 / 0.006 & 0.021 / 0.022 & 0.039 / 0.039 & 0.012 / 0.012 & 0.006 / 0.006 & 0.017 / 0.016 \\
\noalign{\smallskip}
\hline
\noalign{\smallskip}
$K_S$ \; $2.2 \micron$ & 0.004 / 0.004 & 0.013 / 0.013 & 0.024 / 0.024 & 0.007 / 0.007 & 0.004 / 0.004 & 0.010 / 0.009 \\
\noalign{\smallskip}
\hline
\noalign{\smallskip}
IRAC $3.6 \micron$ & 0.002 / 0.002 & 0.007 / 0.008 & 0.014 / 0.014 & 0.004 / 0.004 & 0.002 / 0.002 & 0.006 / 0.005 \\
$4.5 \micron$ & 0.002 / 0.002 & 0.006 / 0.006 & 0.011 / 0.011 & 0.003 / 0.004 & 0.002 / 0.002 & 0.005 / 0.005 \\
$5.8 \micron$ & 0.002 / \dots & \dots & \dots & \dots & 0.002 / \dots & 0.004 / \dots \\
$8.0 \micron$ & 0.002 / \dots & \dots & \dots & \dots & 0.001 / \dots & 0.004 / \dots \\
\enddata
\vspace{-6pt}
\tablecomments{Galactic extinction values for the available filters for each field (see Section \ref{flux corr} for more details).  The cluster and parallel fields are designated by clu / par for the \hff.  We denote filters where no imaging data is available with ellipses (\dots).  All values are in AB magnitude.}
\end{deluxetable*}		%%%%%%%%%%%%%  TABLE  %%%%%%%%%%%%%%%

The fluxes provided in the catalogs are total fluxes. We correct the photometry aperture flux measured in each \hst\ band to a total flux by multiplying the ratio of the $F160W$ total flux to the $F160W$ flux measured in the 0\farcs{7} aperture.  The total flux for the $F160W$ reference band is calculated from the \sex 's AUTO flux and using the circularized Kron radius in combination with the $F160W$ growth curve (see Section \ref{hst photometry}).  However, the $F814W$ (in a few cases, $F105W$, $F125W$, or $F140W$) is used instead of the $F160W$, when the $F160W$ has no coverage.  We indicate this with ``bandtotal'' in the catalogs (see Table~\ref{cat cols} and Section \ref{cat format}).  The photometry aperture errors are converted to a total error by multiplying by the same correction as the fluxes.  We perform the same process for the $K_S$ and IRAC bands but for apertures of $0\farcs7$ and 3\arcsec, respectively.

\subsection{Point Source Classification}
\label{point source class}

\begin{figure*}[ht!]	%%%%%%%%%%%%%%%  PLOT  %%%%%%%%%%%%%%%%%
\epsscale{1.15}
\plottwo{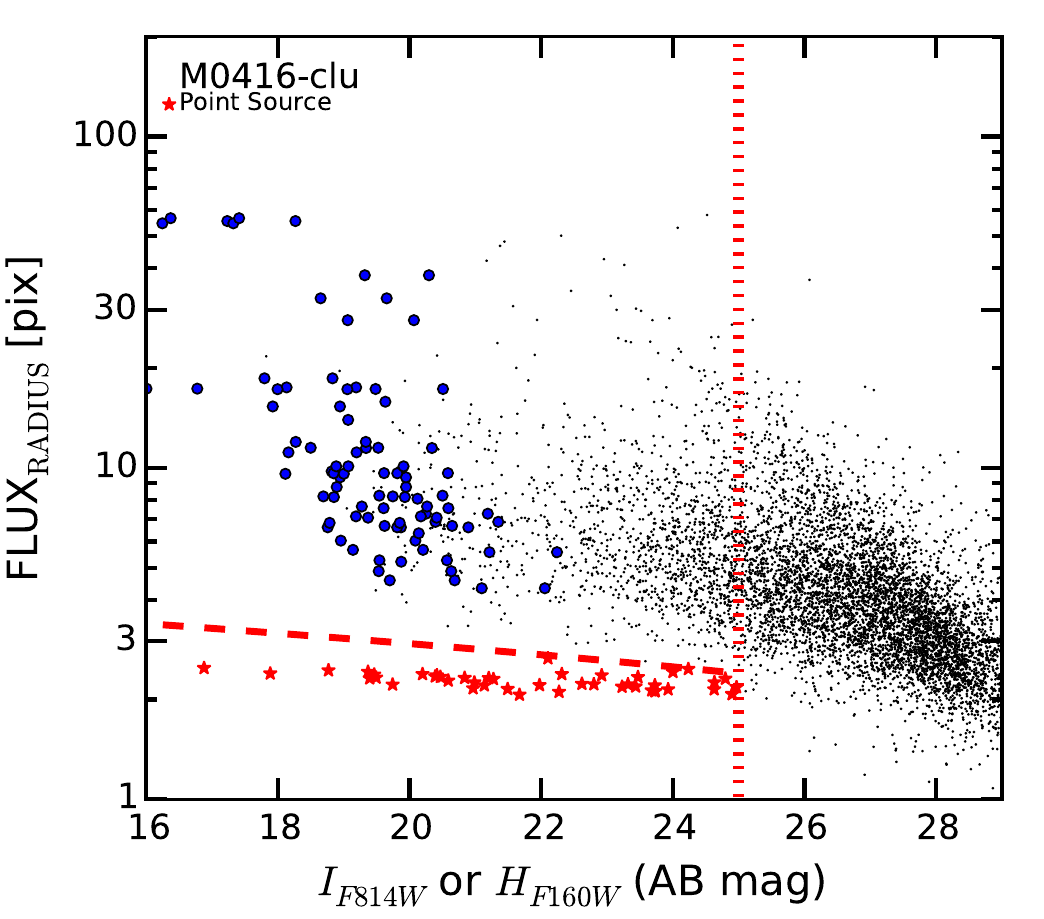}{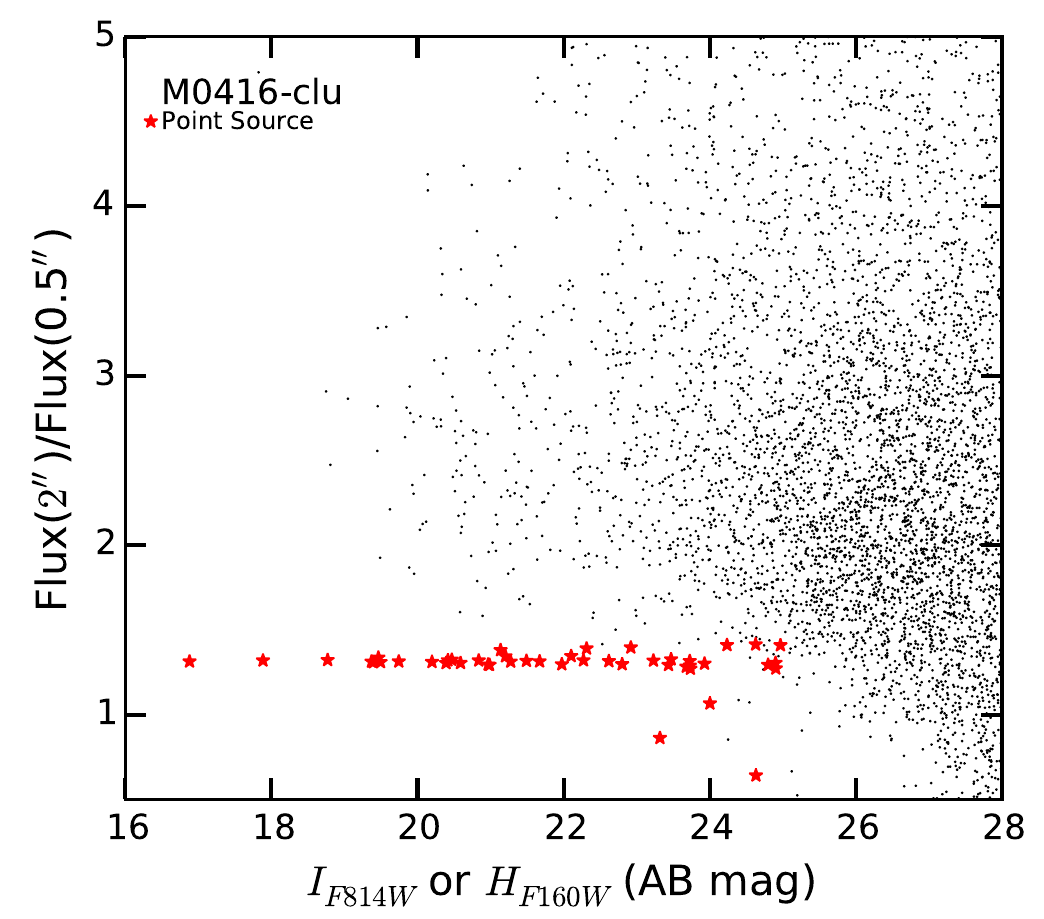}
\caption{The Left panel demonstrates \sex's FLUX$_{\mathrm{RADIUS}}$ against total $H_{F160W}$ magnitude when available or $I_{F814W}$ magnitude otherwise for the \mone\ cluster (all other fields can be found in the Appendix).  Objects classified as point sources in the catalog are shown with red stars, galaxies and uncertain classifications as black points, modeled out bCGs as blue circles.  The Right panel demonstrates an alternate method for selecting point sources using the ratio of fluxes in a large and small apertures.  The tightness of the stellar sequence in this ratio at brighter magnitudes ($H_{F160W} \lsim 24$~mag) allows for a more stringent classification, but the separation becomes less clear than the flux radius selection at fainter magnitudes.  In general, the two methods yield similar results for each field.}
\label{star flag fig}
\end{figure*}		%%%%%%%%%%%%%%%  PLOT  %%%%%%%%%%%%%%%%%

Compact or unresolved sources (i.e.\ point sources) have a tight correlation in size and magnitude, with fairly constant, small sizes as a function of magnitude.  We demonstrate this trend in Figure \ref{star flag fig} that shows the \sex\ FLUX\_RADIUS against total $H_{F160W}$ magnitude when available or the $I_{F814W}$ magnitude otherwise, for the \mone\ cluster (left panel).

\begin{deluxetable*}{ll}		%%%%%%%%%%%%%  TABLE  %%%%%%%%%%%%%%%
\tablecaption{Catalog Columns \vspace{-6pt}
\label{cat cols}}
\tablecolumns{2}
%\tabletypesize{12pt}
\tablehead{
\colhead{Column Name} & \colhead{Description}
}
\startdata
id & Unique identifier for HFF-DeepSpace \\
x  &  X centroid in image coordinates \\
y  & Y centroid in image coordinates \\
ra  & RA J2000 (degrees) \\
dec & Dec J2000 (degrees) \\
z\_spec  & Spectroscopic redshift, when available \\
flags\_{band}  & \sex\ extraction flags (\sex\ FLAGS parameter)\\
class\_star\_{band} & Stellarity index (\sex\ CLASS\_STAR parameter) \\
flux\_radius & Circular aperture radius enclosing half the total flux  (\sex\ FLUX\_RADIUS parameter, pixels) \\
star\_flag & Point source = 1, extended source = 0, uncertain source = 2 (source $\geq$~25 mag) \\
bandtotal & Either ``$F160W$'', ``$F140W$'', ``$F125W$'', ``$F105W$'', ``$F814W$'', ``bcg'' or ``none''; band used to derive total fluxes \\
f\_{band}  & Total flux for each band (zero point = 25)\\
e\_{band}  & $1\sigma$ error for each band (zero point = 25)\\
w\_{band}  & Weight relative to maximum exposure within image band (see Section \ref{cat format})\\
flag\_{band} & Identifies possibly problematic sources for each band (see Section \ref{flags})\\
use\_{band} & Identifies possibly problematic photometry for low resolution bands (see Section \ref{flags})\\
REFspecz & Literature reference for spectroscopic redshift\\
theta\_J2000 & Position angle of the major axis (counter-clockwise, measured from East)\\ 
kron\_radius  & \sex\ KRON\_RADIUS (pixels)\\
a\_image & Semi-major axis (\sex\ A\_IMAGE, pixels)\\
b\_image & Semi-minor axis (\sex\ B\_IMAGE, pixels)\\
use\_phot & Flag indicating source is likely to be a galaxy with reliable photometry (see Section \ref{use phot})\\
mwext\_band & Applied Milky Way extinction correction for each band (see Table \ref{hffds bands})\\
zpcorr\_band & Applied zero point correction for each band (see Table \ref{hff zps})
\enddata
\end{deluxetable*}		%%%%%%%%%%%%%  TABLE  %%%%%%%%%%%%%%%

Point sources can be separated cleanly from extended sources down to $H_{F160W}$ or $I_{F814W} \sim 25$~mag.  We provide a point source flag in the catalog based on the criteria here, as measured on the $F160W$ images when available and $F814W$ otherwise (a few sources utilize the other detection bands, i.e.\ the $F105W$, $F125W$ and $F140W$; this is based on their ``bandtotal'' band, refer to Table~\ref{cat cols}).  Objects are classified as point sources (star\_flag = 1) if they have $\mathrm{FLUX\_RADIUS} < -\, 0.11\, H_{F160W} + 5.15$, where $H_{F160W}$ is the total magnitude of the band used for total flux (i.e.\ ``bandtotal'').  We also perform visual inspection on the images to determine if any stars are missed or if any sources should be excluded from the above selection.  Due to the small effective areas of these fields with consequently low number of stars, this was a useful task to perform.  These sources are shown with red stars in Figure \ref{star flag fig}.  Sources fainter than 25~mag (dotted red line in figure) can not be identified accurately as point sources (unless by visual inspection) and are assigned star\_flag = 2.  All other objects are classified as extended, with star\_flag = 0.

Another method for classifying point sources is the ratio of fluxes in large ($2\arcsec$) and small ($0\farcs5$) apertures versus magnitude that provides a similar tight sequence for $I_{F814W}$ or $H_{F160W} \la 24$~mag (right panel of Figure~\ref{star flag fig}).  Both sequences prove to be useful diagnostics of the image quality, and demonstrate the dearth of stars in these small effective area fields.

\subsection{Flags}
\label{flags}

To better distinguish the quality of the photometry for the sources in the catalogs, we provide flags that allow straightforward selection of sources that have photometry of reasonably uniform quality.  For each photometric band, this flag\_band is set to 0 (i.e.\ ``OK'') if none of the following criteria are met (e.g.\ flag\_F160W = 0):

\begin{enumerate}
\item The photometry aperture overlaps with a masked region: flag set to flag\_band = 1.
\item The AUTO aperture from \sex\ overlaps with a masked region: flag\_band = 2.
\item Both the photometry and AUTO apertures overlap with a masked region: flag\_band = 3.  This occurs mostly for faint and extremely extended sources (e.g.\ gravitationally lensed arcs).
\item The source is a selected bCG for modeling (see Section \ref{selection of bcgs}) that could not be modeled out: flag\_band = 4.  This primarily applies to the UV bands as bCGs became to faint for modeling.
\item The weight value is $\leq$ 0 for any pixels associated with the source in the segmentation map: flag\_band = -1.
\end{enumerate}

We mask regions that are influenced significantly by any of the following:  bright stars that cause halos and large diffractions spikes, residual of a modeled out bCG, satellite trails, cosmic rays, and pixels that have weight values $\leq$ 0.\footnote{This weight value condition takes into account under-exposed regions of the science images flagging sources on the edges of the mosaics and in instrument chip gaps (e.g.\ CLASH and UV bands).}  For bad pixels not caught by cosmic ray detection and the weight images, the masking is done manually through visual inspection of each science image for the \hst\ photometric bands before and after the bCG modeling and sky subtraction steps (see Section \ref{bcg modeling}).

For the non-\hst\ bands, the $K_{S}$ and \spitzer/IRAC bands, we use a simplified `use' flag assignment due to the differences in the methods performed for photometry (i.e.\ MOPHONGO instead of \sex).  For each photometric band of the $K_{S}$ and \spitzer/IRAC, this `use\_band' flag is set to 1 (i.e.\ ``GOOD'') if none of the following criteria are met (e.g.\ use\_CH1 = 1).

\begin{enumerate}
\item If any of the flux, error or weight $\leq$ 0 and/or NaN/Inf values:  flag set to use\_band $= 0$ (i.e.\ ``BAD'').
\item The source is a selected bCG for modeling (see Section \ref{selection of bcgs}):  flag set to use\_band = 2.
\end{enumerate}

\begin{figure*}[htpb]	%%%%%%%%%%%%%%%  PLOT  %%%%%%%%%%%%%%%%%
\epsscale{1.1}
\plotone{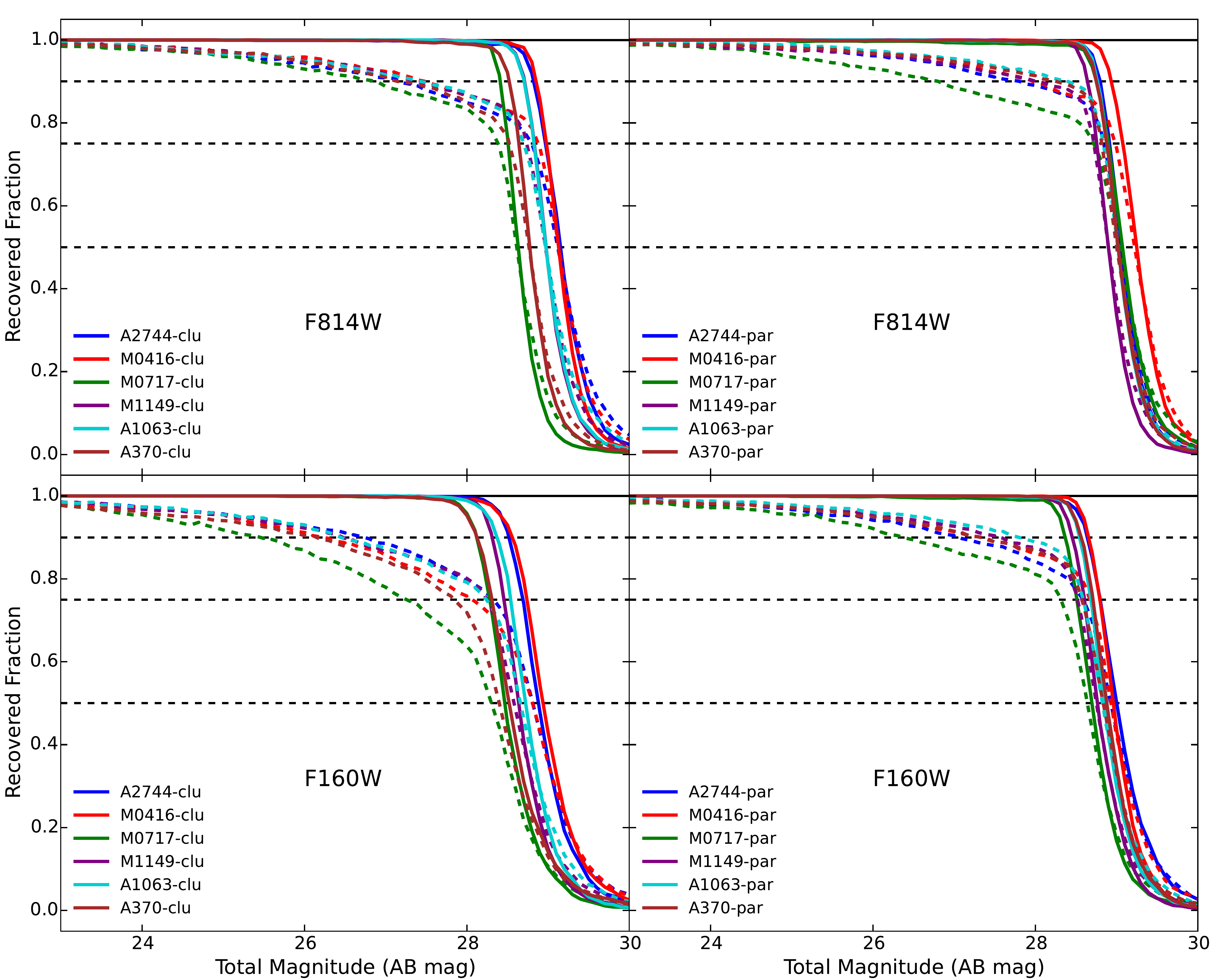}
\caption{Completeness fraction as a function of total magnitude for the deep region (reg1) in each field.  We show no-overlap criterion (solid lines) and ``allow'' overlap criterion (dashed lines) for the $I_{F814W}$ (top panels) and $H_{F160W}$ (bottom panels) bands (see Section~\ref{completeness}).  In both cases (no-overlap and ``allow'' overlap), the 75\% completeness fraction agrees within one magnitude for each field, respectively, in both bands.}
\label{deep complete overlap}
\end{figure*}		%%%%%%%%%%%%%%%  PLOT  %%%%%%%%%%%%%%%%%

\subsection{``use\_phot''}
\label{use phot}

We introduce use\_phot following \citet{Skelton2014} that selects ``OK'' sources in a consistent way.  By selecting sources with use\_phot = 1, this excludes stars (i.e.\ star\_flag = 0 or 2 are ``OK''), sources close to a bright star, S/N $\leq 3$ from the photometry aperture in the `bandtotal' band (see Section \ref{cat format}), ``non-catastrophic'' photometric redshift fit ($\chi_\mathrm{p} < 1000$, see Section \ref{photzs}) and ''non-catastrophic'' stellar population fit (log$(M) > 0$, see Section \ref{gal properties}).

The use\_phot flag selects approximately 80\% of all objects in the catalogs.  The flag is not very restrictive and is meant as a guide to inform the user of possibly problematic sources in the catalogs.  In most science cases, further cuts are required (particularly on magnitude, number of available photometric bands, and/or a stricter S/N ratio).  For studies of large samples, the `use\_phot' flag should be sufficiently reliable when combined with a magnitude criterion.  For an individual galaxy or small sample, we caution the reader to inspect the quality of the photometry for each source beyond the selection criteria.

\subsection{Catalog Format}
\label{cat format}

We provide a full photometric catalog for each of the six \hff\ clusters and associated parallels.  The catalogs contain total flux measurements and basic galaxy properties for 81315 objects in total - (9390, 6240), (7431, 7771), (6370, 5776), (6868, 5802), (7611, 5574) and (6795, 5687) for \aone, \mone, \mtwo, \mthree, \atwo\ and \athree, clusters and parallels, respectively.

A description of the columns in each photometric catalog is given in Table~\ref{cat cols}. All fluxes are normalized to an AB zero point of 25, such that
\begin{equation}
\mathrm{mag}_{AB} = -2.5\times\log_{10}(F)+25.
\end{equation}

The total fluxes and $1\sigma$ errors for every band listed in Table~\ref{hffds bands} are given in the photometric catalogs.  The structural parameters from \sex\ and the corrections to total fluxes are derived from the $F160W$ image, where there is $F160W$ coverage and the other detection bands otherwise.  The `bandtotal' column indicates which image was used to derive total fluxes.

We provide a weight column for each band to indicate the relative weight for each object compared to the maximum weight for that filter. In practice, the weight is calculated as the ratio of the weight at each object's position to the 95$^{th}$ percentile of the weight map.  We take the median weight value from a $7 \times 7$ grid of pixels around the central pixel of the source and divide by the 95$^{th}$ percentile pixel weight value of the image from the positive non-zero weights (i.e.\ no masked regions are used).  We use the 95$^{th}$ percentile weight, as opposed to the maximum, to avoid extreme values affecting the maximum weight.  Objects with weights greater than the 95$^{th}$ percentile weight have a value of 1 in the weight column.

\subsection{Completeness}
\label{completeness}

\begin{deluxetable*}{lllllll}	%%%%%%%%%%%%%  TABLE  %%%%%%%%%%%%%%%
\tablecaption{Completeness Fraction as a Function of $F814W$ Magnitude \vspace{-6pt}
\label{complete814}}
\tablecolumns{7}
\tabletypesize{\footnotesize}
\tablewidth{0pc}
\setlength{\tabcolsep}{0pt}
\tablehead{
 & & \colhead{No$-$Overlap} & & & \colhead{``Allow''$-$Overlap} & \\
\cline{3-3} \cline{6-6}
\colhead{Field} & \colhead{90\%} & \colhead{75\%} & \colhead{50\%} & \colhead{90\%} & \colhead{75\%} & \colhead{50\%} \\
 & reg1 (reg2) [reg3] & reg1 (reg2) [reg3] & reg1 (reg2) [reg3] & reg1 (reg2) [reg3] & reg1 (reg2) [reg3] & reg1 (reg2) [reg3]
}
\startdata
\aone -clu & 28.8 (28.5) [27.4] & 29.0 (28.7) [27.6] & 29.2 (28.9) [27.7] & 27.2 (27.8) [27.0] & 28.8 (28.6) [27.5] & 29.1 (28.9) [27.7] \\
\aone -par & 28.8 (28.5) [$\dots$] & 28.9 (28.6) [$\dots$] & 29.1 (28.8) [$\dots$] & 27.8 (27.9) [$\dots$] & 28.8 (28.5) [$\dots$] & 29.1 (28.7) [$\dots$] \\
\mone -clu & 28.9 (28.5) [26.3] & 29.0 (28.6) [26.5] & 29.1 (28.8) [26.7] & 27.5 (27.4) [26.3] & 28.9 (28.5) [26.4] & 29.1 (28.8) [26.7] \\
\mone -par & 28.9 (28.7) [$\dots$] & 29.1 (28.8) [$\dots$] & 29.2 (29.0) [$\dots$] & 28.0 (28.4) [$\dots$] & 29.0 (28.7) [$\dots$] & 29.2 (29.0) [$\dots$] \\
\mtwo -clu & 28.4 (28.1) [26.3] & 28.5 (28.2) [26.6] & 28.6 (28.3) [26.8] & 26.9 (27.2) [26.1] & 28.4 (28.1) [26.6] & 28.6 (28.3) [26.8] \\
\mtwo -par & 28.8 (28.4) [26.8] & 28.9 (28.5) [27.0] & 29.1 (28.7) [27.2] & 26.8 (27.4) [26.8] & 28.7 (28.4) [27.0] & 29.0 (28.6) [27.2] \\
\mthree -clu & 28.7 (28.4) [$\dots$] & 28.8 (28.5) [$\dots$] & 29.0 (28.7) [$\dots$] & 27.4 (27.7) [$\dots$] & 28.7 (28.4) [$\dots$] & 29.0 (28.6) [$\dots$] \\
\mthree -par & 28.6 (28.3) [26.1] & 28.8 (28.4) [26.3] & 28.9 (28.6) [26.5] & 28.0 (27.4) [26.1] & 28.7 (28.3) [26.3] & 28.9 (28.6) [26.5] \\
\mtwo -clu & 28.7 (28.4) [26.4] & 28.8 (28.5) [26.8] & 29.0 (28.7) [27.0] & 27.4 (27.4) [26.4] & 28.7 (28.5) [26.8] & 29.0 (28.7) [27.0] \\
\mtwo -par & 28.8 (28.4) [$\dots$] & 28.9 (28.6) [$\dots$] & 29.0 (28.7) [$\dots$] & 28.4 (28.1) [$\dots$] & 28.8 (28.5) [$\dots$] & 29.0 (28.7) [$\dots$] \\
\athree -clu & 28.5 (28.2) [26.7] & 28.6 (28.3) [27.2] & 28.8 (28.4) [27.5] & 27.1 (27.2) [26.6] & 28.5 (28.2) [27.1] & 28.8 (28.4) [27.5] \\
\athree -par & 28.8 (28.4) [$\dots$] & 28.9 (28.6) [$\dots$] & 29.0 (28.8) [$\dots$] & 28.3 (27.6) [$\dots$] & 28.8 (28.5) [$\dots$] & 29.0 (28.7) [$\dots$]
 \enddata
\tablecomments{Reg1 is the deepest region with Reg3 being the shallowest region for each field.  When Reg3 is too small for meaningful calculations of the completeness, no completeness value is given ($\dots$).  All values are in AB magnitude.}
\end{deluxetable*}		%%%%%%%%%%%%%  TABLE  %%%%%%%%%%%%%%%

\begin{figure*}[ht!]	%%%%%%%%%%%%%%%  PLOT  %%%%%%%%%%%%%%%%%
\epsscale{1.}
\plottwo{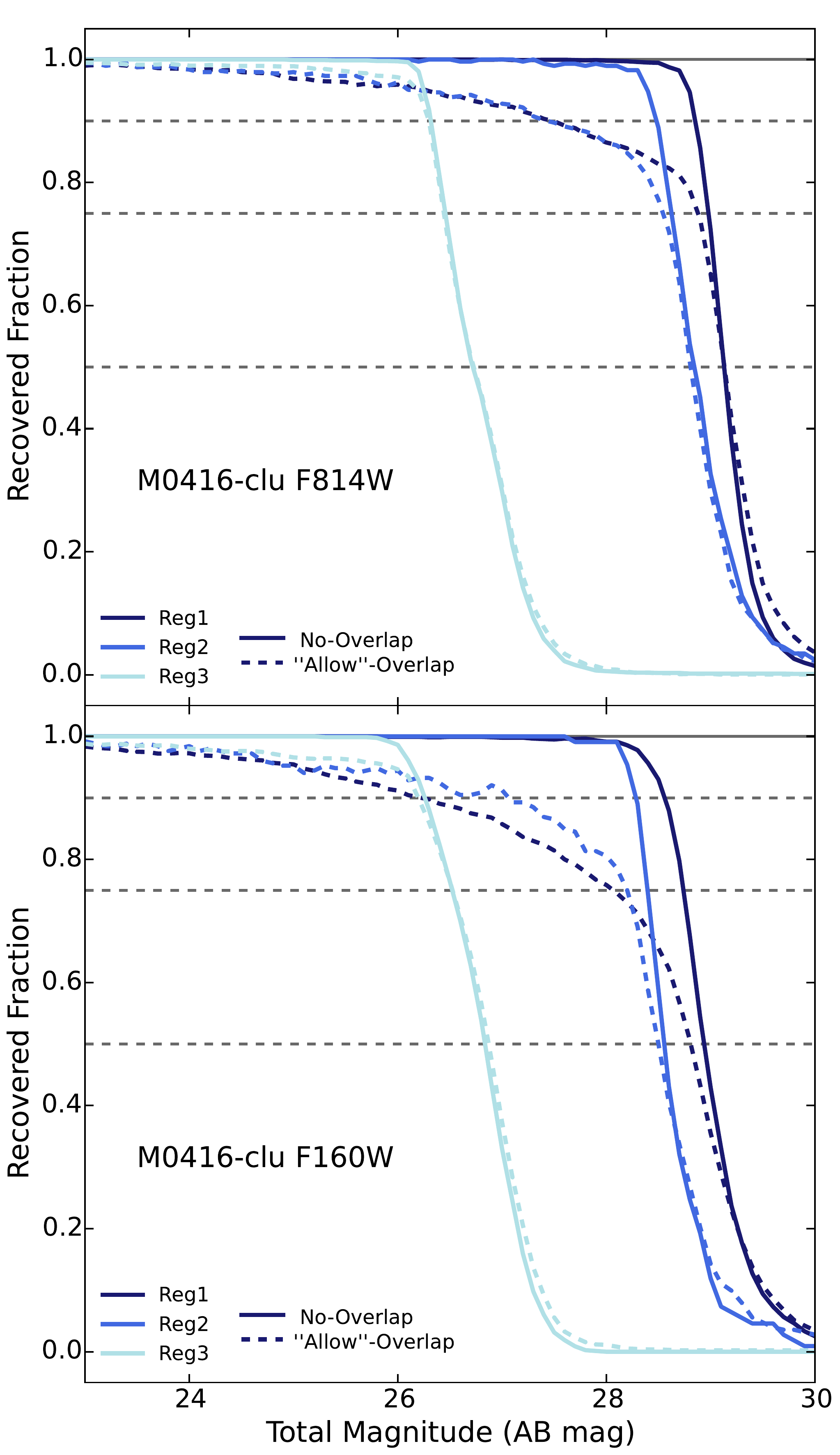}{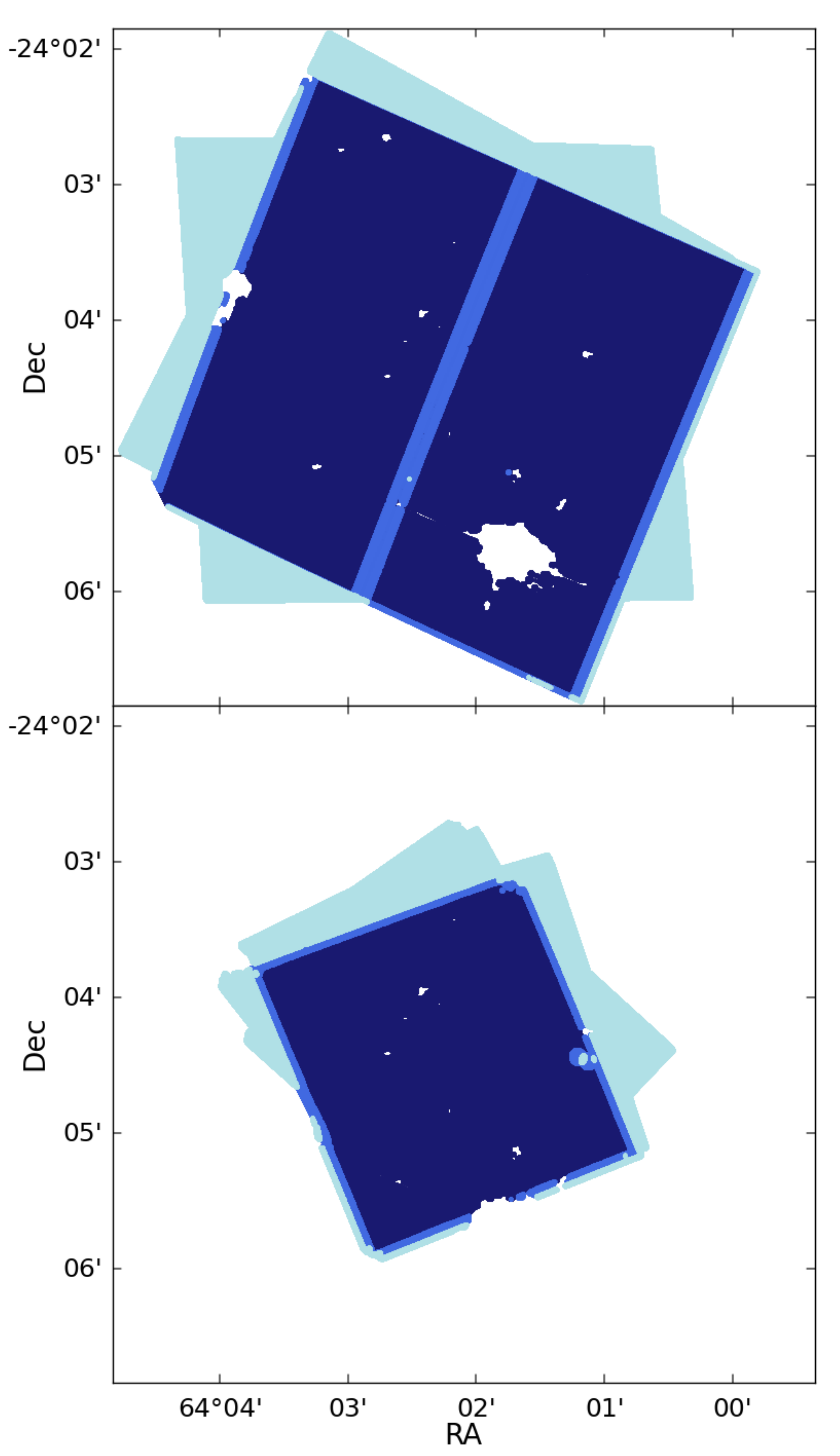}
\caption{We demonstrate the recovered completeness fraction of each region in the \mone\ cluster for both the no-overlap (solid lines) and ``allow'' overlap (dashed lines) criteria (all other fields can be found in the Appendix, see Section~\ref{completeness} for details).  The Top panels are the $F814W$ band and the Bottom panels of the $F160W$ band, where the areas for each region (Right panels) are shaded by region (reg1 is midnight blue; reg2 is royal blue; reg3 is powder blue) corresponding to the line colors (Left panels).}
\label{field complete all}
\end{figure*}		%%%%%%%%%%%%%%%  PLOT  %%%%%%%%%%%%%%%%%

\begin{deluxetable*}{lllllll}	%%%%%%%%%%%%%  TABLE  %%%%%%%%%%%%%%%
\tablecaption{Completeness Fraction as a Function of $F160W$ Magnitude \vspace{-6pt}
\label{complete160}}
\tablecolumns{7}
\tabletypesize{\footnotesize}
\tablewidth{0pc}
\setlength{\tabcolsep}{0pt}
\tablehead{
 & & \colhead{No$-$Overlap} & & & \colhead{``Allow''$-$Overlap} & \\
\cline{3-3} \cline{6-6}
\colhead{Field} & \colhead{90\%} & \colhead{75\%} & \colhead{50\%} & \colhead{90\%} & \colhead{75\%} & \colhead{50\%} \\
 & reg1 (reg2) [reg3] & reg1 (reg2) [reg3] & reg1 (reg2) [reg3] & reg1 (reg2) [reg3] & reg1 (reg2) [reg3] & reg1 (reg2) [reg3]
}
\startdata
\aone -clu & 28.5 (28.2) [26.5] & 28.7 (28.3) [26.6] & 28.9 (28.5) [26.8] & 26.7 (26.4) [26.3] & 28.3 (28.2) [26.5] & 28.8 (28.5) [26.7] \\
\aone -par & 28.6 (28.2) [$\dots$] & 28.8 (28.3) [$\dots$] & 29.0 (28.5) [$\dots$] & 27.1 (26.5) [$\dots$] & 28.6 (28.3) [$\dots$] & 28.9 (28.5) [$\dots$] \\
\mone -clu & 28.6 (28.3) [26.3] & 28.7 (28.4) [26.5] & 28.9 (28.6) [26.8] & 26.2 (27.1) [26.2] & 28.1 (28.2) [26.5] & 28.8 (28.5) [26.9] \\
\mone -par & 28.7 (28.3) [$\dots$] & 28.8 (28.4) [$\dots$] & 29.0 (28.6) [$\dots$] & 27.3 (27.2) [$\dots$] & 28.7 (28.4) [$\dots$] & 28.9 (28.6) [$\dots$] \\
\mtwo -clu & 28.1 (27.8) [26.3] & 28.3 (27.9) [26.5] & 28.5 (28.1) [26.7] & 25.5 (26.1) [26.2] & 27.3 (27.6) [26.4] & 28.3 (28.0) [26.7] \\
\mtwo -par & 28.4 (27.9) [26.7] & 28.5 (28.1) [26.8] & 28.7 (28.3) [27.0] & 26.4 (24.9) [26.6] & 28.3 (27.4) [26.8] & 28.6 (28.2) [27.0] \\
\mthree -clu & 28.3 (28.0) [26.1] & 28.5 (28.1) [26.4] & 28.6 (28.3) [26.9] & 26.5 (26.4) [25.9] & 28.2 (27.9) [26.4] & 28.6 (28.2) [26.9] \\
\mthree -par & 28.5 (28.0) [27.0] & 28.6 (28.2) [27.2] & 28.8 (28.3) [27.4] & 27.6 (27.2) [26.8] & 28.5 (28.1) [27.1] & 28.8 (28.4) [27.3] \\
\atwo -clu & 28.4 (28.0) [26.5] & 28.5 (28.1) [26.6] & 28.7 (28.3) [26.8] & 26.5 (26.5) [26.4] & 28.2 (27.9) [26.6] & 28.7 (28.3) [26.8] \\
\atwo -par & 28.5 (28.2) [26.8] & 28.7 (28.3) [27.0] & 28.8 (28.5) [27.1] & 27.8 (28.0) [26.7] & 28.6 (28.3) [26.9] & 28.8 (28.5) [27.1] \\
\athree -clu & 28.1 (27.8) [26.7] & 28.3 (27.9) [26.9] & 28.5 (28.1) [27.1] & 26.1 (26.4) [26.5] & 27.8 (27.7) [26.9] & 28.4 (28.1) [27.1] \\
\athree -par & 28.6 (28.1) [$\dots$] & 28.7 (28.3) [$\dots$] & 28.9 (28.5) [$\dots$] & 27.3 (28.0) [$\dots$] & 28.6 (28.3) [$\dots$] & 28.8 (28.5) [$\dots$]
\enddata
\tablecomments{Reg1 is the deepest region with Reg3 being the shallowest region for each field.  When Reg3 is too small for meaningful calculations of the completeness, no completeness value is given ($\dots$).  All values are in AB magnitude.}
\end{deluxetable*}		%%%%%%%%%%%%%  TABLE  %%%%%%%%%%%%%%%

The depth of the images varies from field to field and towards the edges of some fields (e.g.\ \aone\ cluster).  As a result, the completeness will depend on position, as well as different morphologies, magnitudes and sizes.  Here, we describe the completeness for point sources in the \hff\ cluster and parallel fields.  We measure the recovered fraction for the $H_{F160W}$ and $I_{F814W}$ bands of each field.  We do this by inserting fake stars, generated from the convolved PSF at random positions in the field, using the weight and segmentation maps to exclude pixels when determining random pixel locations.  First, we do not allow the fake stars to overlap with detected sources in the field.  Then, we allow the fake stars to overlap, to calculate the effect of blending in crowded fields.  We sample the recovered fraction of fake stars at magnitude intervals of 0.1 for about 2000 fake stars in each field and band.\footnote{At least 200 stars are inserted for the small effective areas of reg2 (medium depth region) but this does not impact our analysis, see Figure~\ref{field complete all}.}

Furthermore, we do this by dividing the images into deep and shallow regions (shallow regions are generally near the edges of the mosaics) for each field and band.  The following criteria are used to separate the deep region from the shallower regions:
\begin{equation} \label{complete region}
\begin{split}
reg1: & wht \geq 0.6 \times wht(95^{th}) \\
reg2: & 0.25 \times wht(95^{th}) \leq wht < 0.6 \times wht(95^{th}) \\
reg3: & wht < 0.25 \times wht(95^{th}) \\
& Area(reg3) > 0.05 \times Area(total)
\end{split}
\end{equation}
where $wht(95^{th})$ is the 95th percentile of the weight distribution, and Area(total) is the total area of all three regions.

We measure the completeness for the deeper regions (reg1 and reg2) in each field and band, but only measure the shallowest region (reg3) if the Area requirement of equation~\ref{complete region} is met.  We make a single-band detection image for each region in the same way as the detection image discussed previously (i.e.\ weighted\_mean / error image) and apply the same \sex\ parameters (see Section \ref{detection}).  However, we do lower the detection and analysis thresholds to account for the shallower depth of the single-band detection image (range of $2.5-4.0$ for $F814W$ and $1.1-1.4$ for $F160W$).  We run \sex\ in dual mode with the final residual image for each field and band as the measurement image ($F160W$ and $F814W$, see Figure \ref{modeling results}).

In Figure \ref{deep complete overlap}, we demonstrate the completeness fraction as a function of total magnitude for the deep region (reg1) in each field.  Tables~\ref{complete814} and \ref{complete160} list the 90\%, 75\% and 50\% completeness levels for each field of the no-overlap and ``allow'' overlap criteria for the $F814W$ and $F160W$ bands, respectively. The comparison between the no-overlap and ``allow'' overlap completeness levels shows that very deep fields like the \textit{Hubble} Frontier Fields, hence crowded fields, blending significantly affects the completeness.

Figure~\ref{field complete all} demonstrates the completeness fraction as a function of magnitude for each region in the \mone\ cluster.  The completeness fraction is measured for both the $F814W$ (top panels) and $F160W$ (bottom panels) bands, where the areas for each region (right panels) are shaded by region corresponding to the line colors (reg1 is midnight blue; reg2 is royal blue; reg3 is powder blue).  Furthermore, we show both the no-overlap (solid lines) and ``allow'' overlap (dashed lines) recovered fractions for each band (left panels).  In most cases, the deeper regions (reg1 and reg2) have similar recovered completeness fractions, where the shallowest region (reg3) differs significantly by about 2 magnitudes regardless of allowing overlap or no-overlap (see Tables \ref{complete814} and \ref{complete160} for specific values of each field and band).

\subsection{Number Counts}

\begin{figure*}[ht!]	%%%%%%%%%%%%%%%  PLOT  %%%%%%%%%%%%%%%%%
\epsscale{1.1}
\plotone{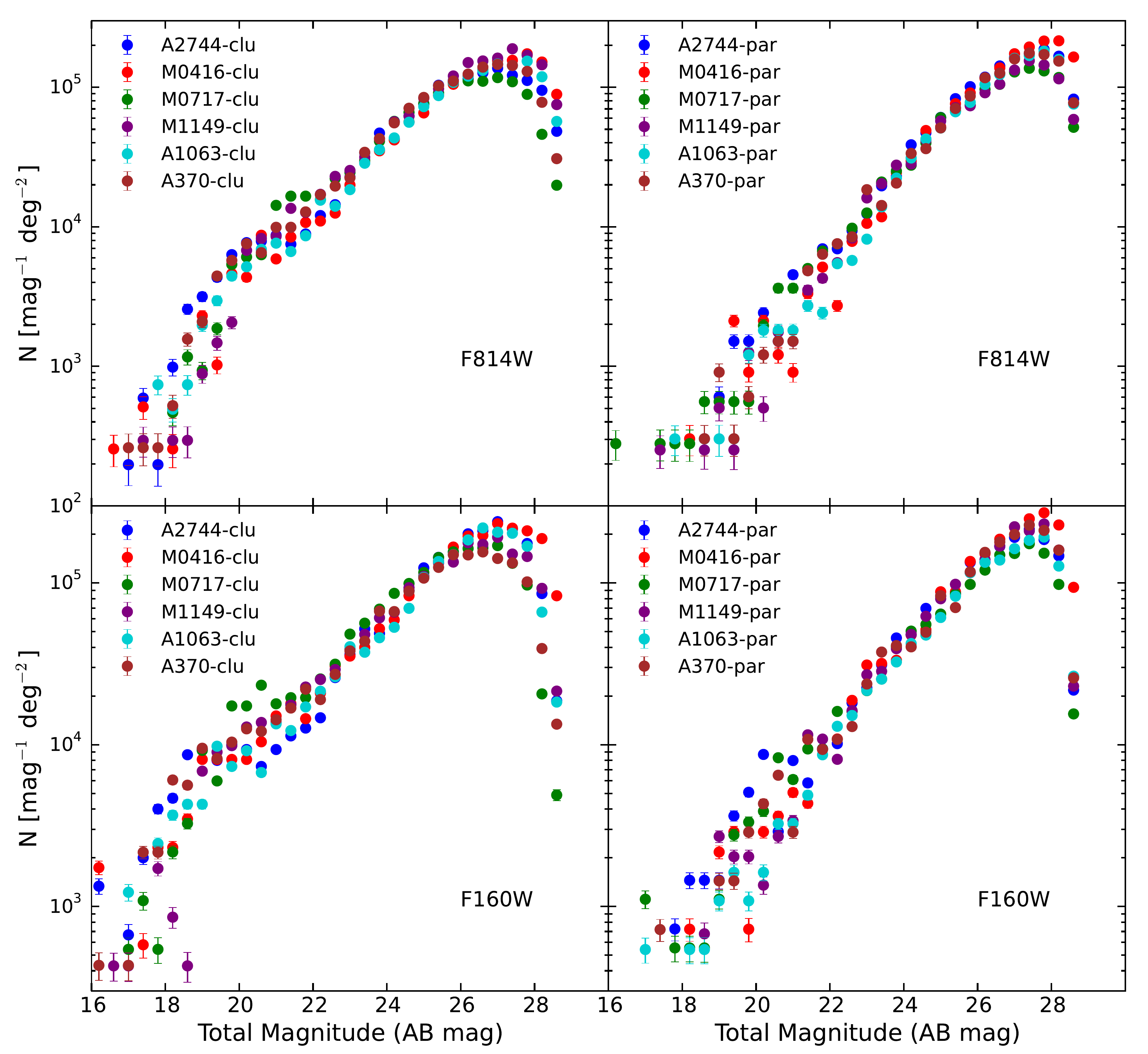}
\caption{Number counts per unit area.  The left panels show the number counts of galaxies (use\_phot = 1) per square degree in each of the six cluster fields as a function of $I_{F814W}$ or $H_{F160W}$ total magnitude (labeled in each panel), with no correction for incompleteness. In the right panels, we repeat this for the six parallel fields.  The error bars on all the data points represent Poisson errors.  Lensing magnification corrections have not been applied.}
\label{mag number counts}
\end{figure*}		%%%%%%%%%%%%%%%  PLOT  %%%%%%%%%%%%%%%%%

We determine the effective survey area of each of the six cluster and six parallel fields using the detection image of each field with the following steps.  For each of the detection band images, we create a map of the number of detection bands contributing to each pixel. We do not include regions masked out during photometry of each detection band as described in Section \ref{flags}.  Then, the science area for each field is calculated by adding up the number of unmasked pixels within the detection band area of our catalogs.  We follow this same procedure for single band effective areas, specifically the $I_{F814W}$ and $H_{F160W}$ bands.

The number density of galaxies (satisfying our ``use\_phot'' flag criteria in the \hff), as a function of the $H_{F160W}$ magnitude, is shown in Figure~\ref{mag number counts} (bottom panels). The bottom left panel shows the number counts for each of the six cluster fields, while the bottom right panel repeats this for the six parallel fields.  The error bars represent Poisson errors in both panels.  Considering the very small field of view of each pointing, the number counts are fairly consistent across the six cluster and six parallel fields.  Figure \ref{mag number counts} (top panels) show the number density of galaxies as a function of $I_{F814W}$ magnitude.  For both the $H_{F160W}$ and $I_{F814W}$ number density of galaxies figures, we use their respective effective areas given in Table~\ref{fields}.

\subsection{Photometry of Close Pairs}

We do extensive work to model out the light from the bCGs and ICL and ensure the quality of the final science images but this does not extend to remaining close pair sources.  To this end, we caution the reader that the photometry of sources may not account for systematic offsets from nearby sources in the formal uncertainties given in the catalogs.

The ground-based $K_S$ band and IRAC photometry is performed after subtracting a model for neighboring sources (see Section \ref{low res phot}), but the space-based photometry is performed directly on PSF-matched data without explicitly accounting for the flux of nearby sources.  \sex\ does attempt to mask and correct the aperture fluxes symmetrically for regions affected by overlapping sources (with the MASK\_TYPE parameter set to CORRECT).  As described in Section \ref{hst photometry}, the photometry aperture has a diameter of $0\farcs7$.  

We estimate the fraction of potentially affected sources in the catalogs by determining the number of sources with a distance smaller than the photometry aperture (i.e., with overlapping photometric apertures $< 0\farcs7$ and use\_phot = 1).  These fractions range from 11.3\% to 15.2\% in the six cluster fields, with an average of 13.2\%.  We repeat this for the six parallel fields and find similar amounts of close pairs with fractions ranging from 10.9\% to 13.6\% (average of 12.1\%).  If we assume that only the faintest overlapping source of the pair ($H_{F160W}$ > 25 mag) is affected, we determine that 5.0\% to 8.6\% (average of 6.4\%) of sources may have problematic \hst\ photometry due to contamination of a nearby source for both the clusters and parallel fields.

\section{Quality and Consistency Tests}
\label{photo check}

Here we assess the quality of our photometric catalogs.  We test whether the colors and uncertainties are reasonable and if there are offsets between the fields.

\subsection{Colors}
\label{colors}

\begin{figure*}[ht!]	%%%%%%%%%%%%%%%  PLOT  %%%%%%%%%%%%%%%%%
\epsscale{1.15}
\plotone{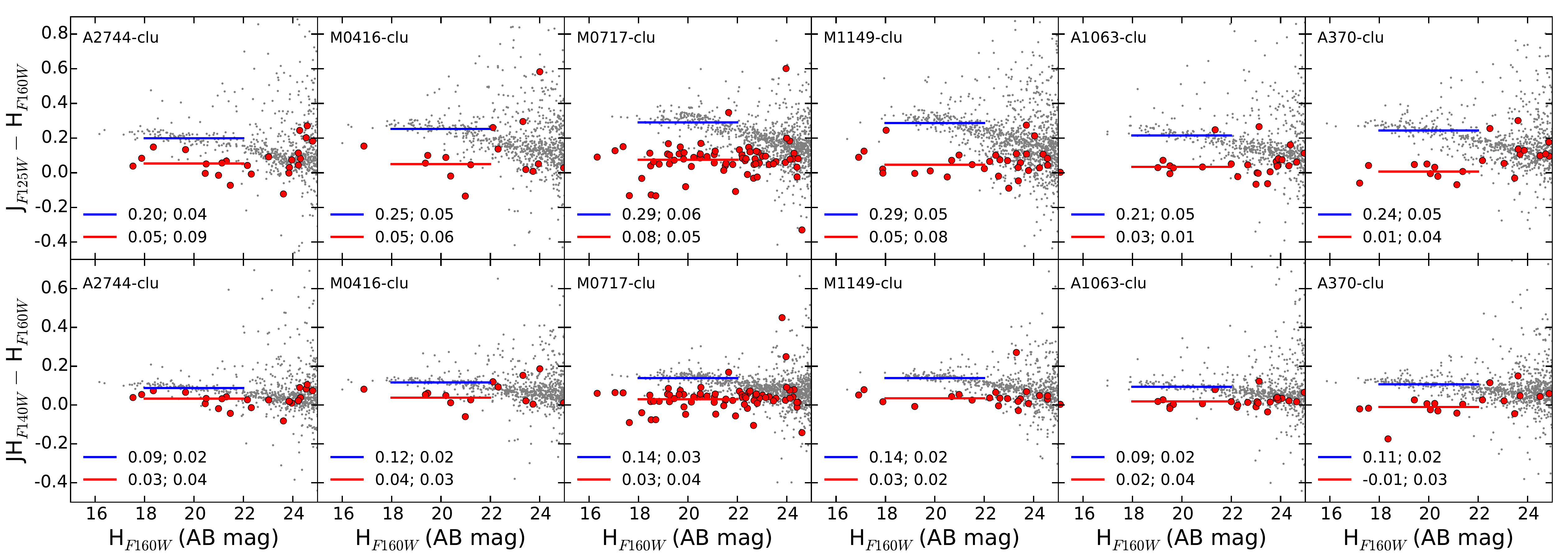}
\plotone{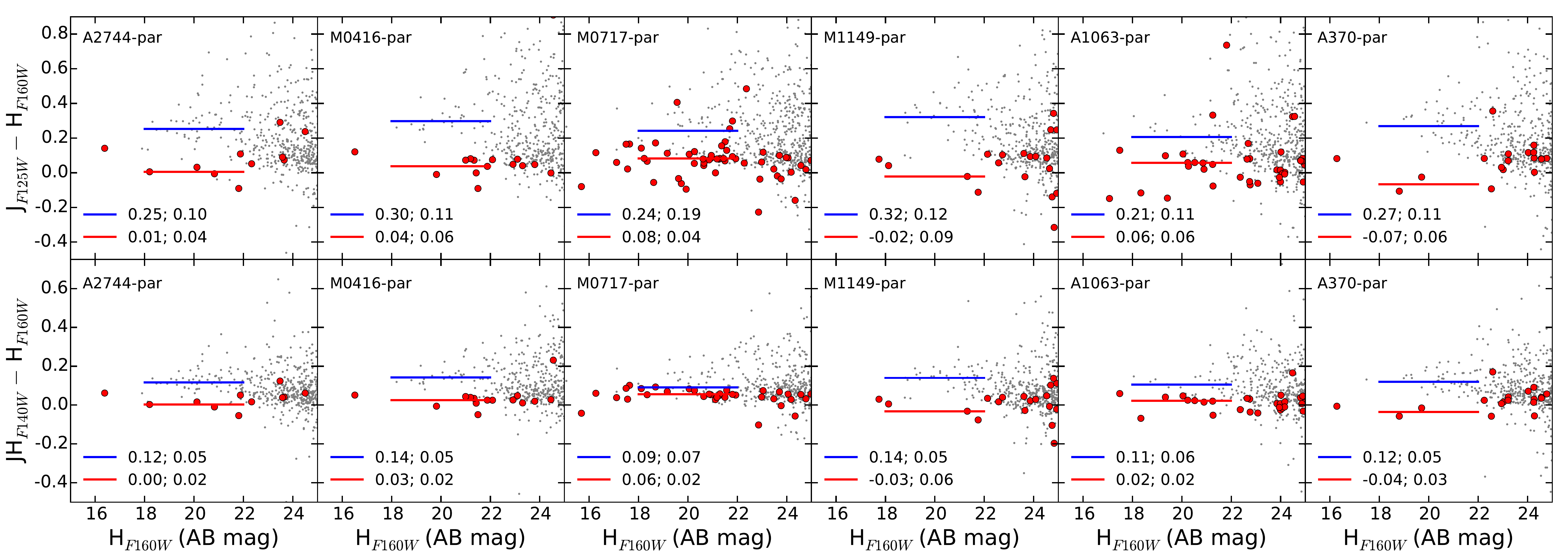}
\caption{$J_{F125W} -$ $H_{F160W}$ and $JH_{F140W} -$ $H_{F160W}$ colors versus $H_{F160W}$ magnitude for each of the twelve fields (clusters top half of figure and parallels bottom half of figure).  Point sources (star\_flag = 1) are shown in red and extended sources (use\_phot = 1) in gray.  The medians and their scatter ($\sigma_\mathrm{MAD}$) for point sources and extended sources in the range $18 < H_{F160W} < 22$ are labeled and shown by the red and blue lines, respectively for both clusters and parallels (see Section \ref{colors} for discussion).}
\label{colormag1}
\end{figure*}		%%%%%%%%%%%%%%%  PLOT  %%%%%%%%%%%%%%%%%

In order to determine if there are offsets between the fields, we look at $J-H$ colors in each field for stars (flag\_star = 1) and galaxies (use\_phot = 1), specifically the median observed colors of the two groups assuming they do not have a dependence on field.  Figure~\ref{colormag1} (top panels) demonstrates this comparison for the clusters.  We repeat this for the parallels (Figure~\ref{colormag1}, bottom panels).  For each panel, the red and blue lines show the median color in the magnitude range $18<H_{F160W}<22$ for stars and galaxies, respectively, with the median and scatter from the median absolute deviation \citep[MAD,][$\sigma_\mathrm{MAD}$]{Beers1990} values listed in each panel, respectively.  The top row, for each set of panels, shows the relation between $J_{F125W} - H_{F160W}$ color and $H_{F160W}$ magnitude in each cluster and parallel field.  The bottom row, for each set of panels, shows the relation between $JH_{F140W}-H_{F160W}$ color and magnitude.  The scatter seen for the stars and galaxies is expected due to the fact that not all stars and galaxies have similar colors.

We find that the median WFC3 colors show very little field dependence for the clusters and parallels, when considering the expected scatter of the data and in the case of the stars having low statistics for some of the fields (e.g.\ \mthree\ and \athree\ parallels).  Although, the median colors of the galaxies for the clusters do vary slightly due to the different redshifts.  The median galaxy colors are reddest in \mtwo\ and \mthree\ clusters, which are the highest redshift.  The field-to-field variation ($\sigma_\mathrm{MAD}$) in the median $J_{F125W} - H_{F160W}$ color is $\sim0.037$ mag for stars and $\sim0.059$ mag for galaxies.  The $JH_{F140W}- H_{F160W}$ colors show even less variation between fields for the stars and galaxies than the $J_{F125W}-H_{F160W}$ colors: the field-to-field variation ($\sigma_\mathrm{MAD}$) in the median color of stars (galaxies) is $\sim0.015$ ($\sim0.030$)~mag.

\subsection{Total Fluxes}
\label{total flux}

\begin{figure*}[ht!]	%%%%%%%%%%%%%%%  PLOT  %%%%%%%%%%%%%%%%%
\epsscale{1.15}
\plotone{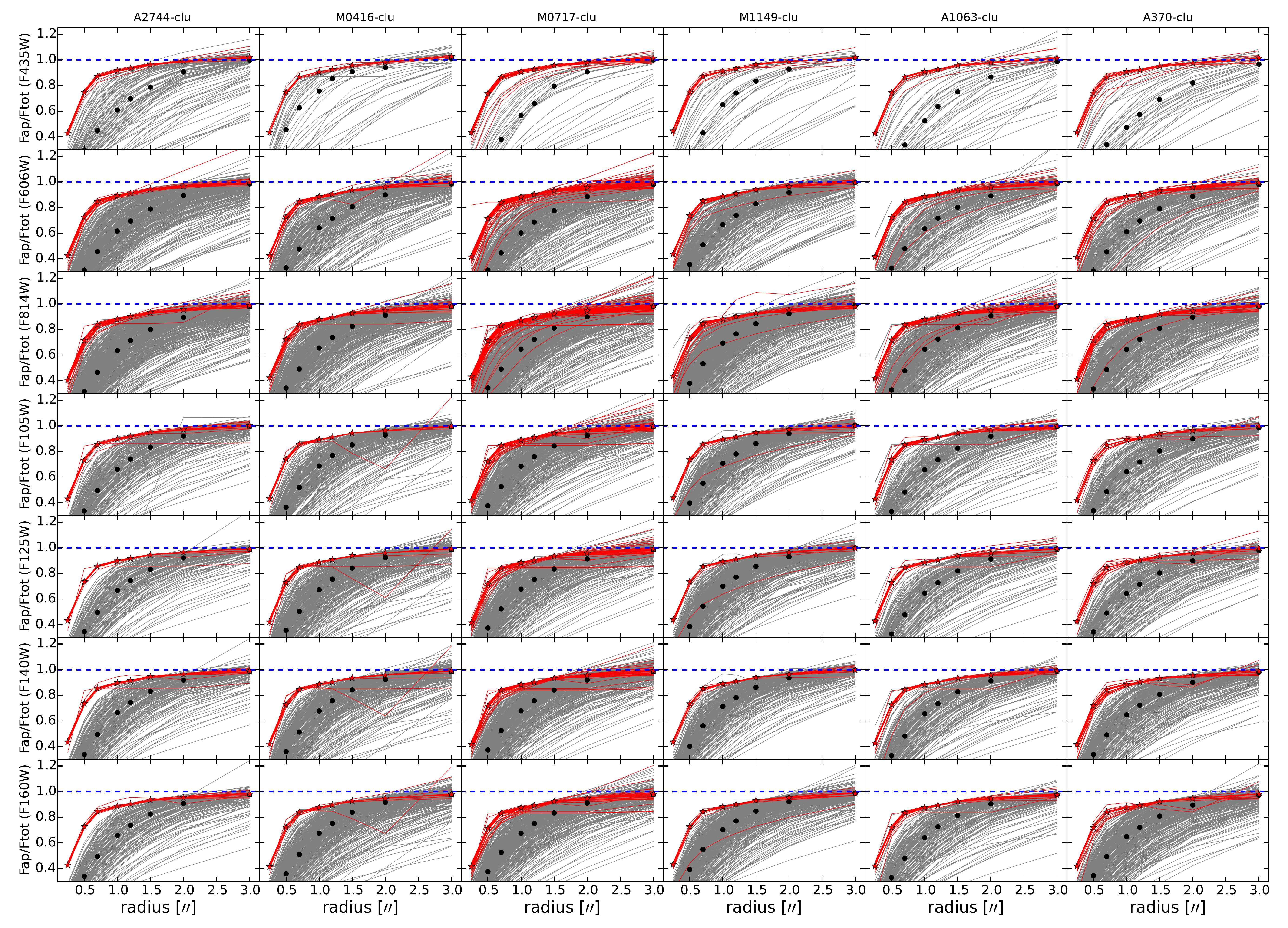}
\caption{Ratio of aperture flux to total flux given in the catalogs as a function of aperture radius in each \hff\ cluster and \hff\ band.  We select sources with high signal-to-noise ratios (S/N > 50) for better comparison.  In each case a few hundred extended sources were chosen randomly from the catalog (except for the $F435W$ band) with the requirement that each source satisfy use\_phot = 1 in addition to the S/N cut.  Point sources (star\_flag = 1) are shown in red and extended sources in gray.  The median values for point sources and extended sources are shown by the large stars and filled circles, respectively.  We find good agreement between the derived total fluxes and the direct measurements of flux in $3^{\prime\prime}$ apertures, to within a percent for point sources.  Furthermore, the measurements are consistent across all clusters with few spurious sources.}
\label{flux aper ratio clu}
\end{figure*}		%%%%%%%%%%%%%%%  PLOT  %%%%%%%%%%%%%%%%%

\begin{figure*}[ht!]	%%%%%%%%%%%%%%%  PLOT  %%%%%%%%%%%%%%%%%
\epsscale{1.15}
\plotone{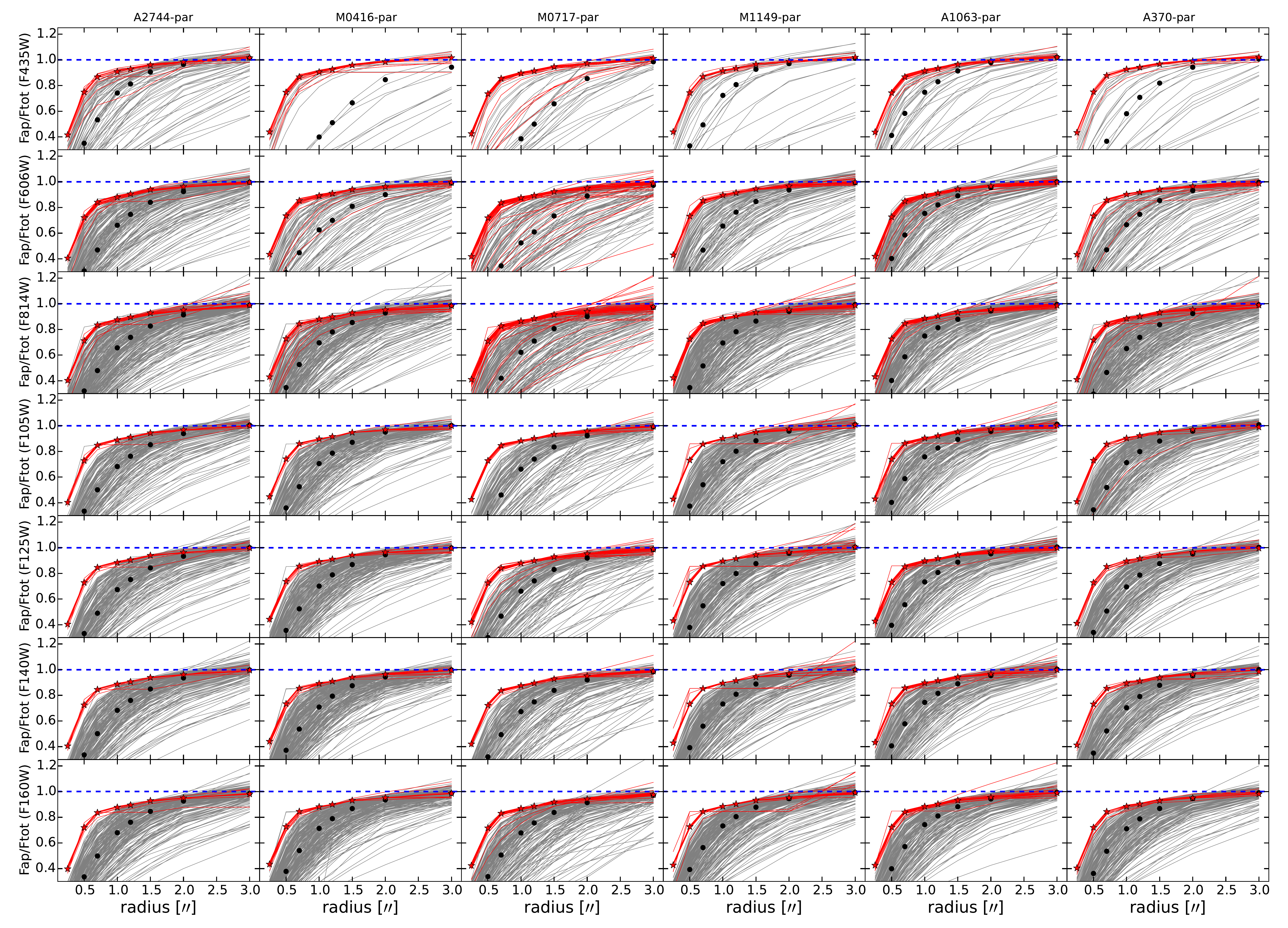}
\caption{Same as Figure \ref{flux aper ratio clu} but for the parallel fields.  Point sources (star\_flag = 1) are shown in red and extended sources in gray.  The median values for point sources and extended sources are shown by the large stars and filled circles, respectively.  We find good agreement between the derived total fluxes and the direct measurements of flux in $3^{\prime\prime}$ apertures, to within a percent for point sources.  Furthermore, the measurements are consistent across all parallels with few spurious sources.}
\label{flux aper ratio par}
\end{figure*}		%%%%%%%%%%%%%%%  PLOT  %%%%%%%%%%%%%%%%%

The total fluxes in the catalogs are based on measurements from \sex\ AUTO aperture, corrected on an object-by-object basis for flux falling outside of this aperture (see Section~\ref{hst photometry}).  In the previous section, we assess the quality of the catalogs using colors.  This is an important first step as colors of objects are determined with higher accuracy than their total fluxes.  This is usually the case, as colors and more generally the shapes of the SEDs (for derived quantities) are measured using carefully matched aperture photometry.  As described in Section~\ref{flux corr}, total fluxes are determined empirically for the $F160W$ band, where there is coverage, and $F814W$ elsewhere (except for a few cases for which, the $F105W$, $F125W$ or $F140W$ are used; i.e.\ ``bandtotal'').  All other bands are corrected to a total flux using the ratio of total flux to the photometry aperture flux from the ``bandtotal'' band for each source.  As a result, the {\em shapes} of the SEDs in our catalog are based on psf-matched photometry using a reference aperture of $0\farcs 7$, and their {\em normalizations} are based on the total ``bandtotal'' band flux \citep[same procedure as][for $3D-HST$]{Skelton2014}.

We test the accuracy of the total flux measurements by measuring the flux of sources in varying sized apertures.  We measure fluxes in aperture sizes of 0.25, 0.5, 0.7, 1, 1.2, 1.5, 2 and 3$\arcsec$ for all the available bands for each field primarily focusing on the \hff\ bands for our analysis here.  Figures \ref{flux aper ratio clu} and \ref{flux aper ratio par} show $F_\mathrm{ap}/F_\mathrm{tot}$, the ratio of these aperture magnitudes to the total flux given in the catalogs as a function of aperture size for sources with S/N $> 50$.  Stars are shown as red lines, with median values indicated by open star symbols.  The growth curves show little scatter, and reach values that are within 1\% of unity for an aperture radius of $3\arcsec$.  As our correction to total fluxes for the \hst\ photometry is based on psf-matched mosaics for each band to the $F160W$ from the growth curves of stars, this is not surprising.  The grey curves and black points show growth curves of extended sources (i.e.\ galaxies).  There is a large variation in the curves, reflecting the large variation in the apparent sizes and shapes of galaxies.  However, the median growth curves again reach unity (within 3\%) at the 3$\arcsec$ aperture size and behave similarly for the growth from small to larger radii in all \hff\ filters and all twelve fields.  This implies that our correction to total fluxes (and the PSF-correction for extended sources) is correct to within a few percent for the median flux ratio at large radii.  We do note this is for high S/N sources and the uncertainties likely will become larger for lower S/N cuts.

\subsection{Error Estimates}
\label{errors}

\begin{figure*}[ht!]	%%%%%%%%%%%%%%%  PLOT  %%%%%%%%%%%%%%%%%
\epsscale{1.15}
\plotone{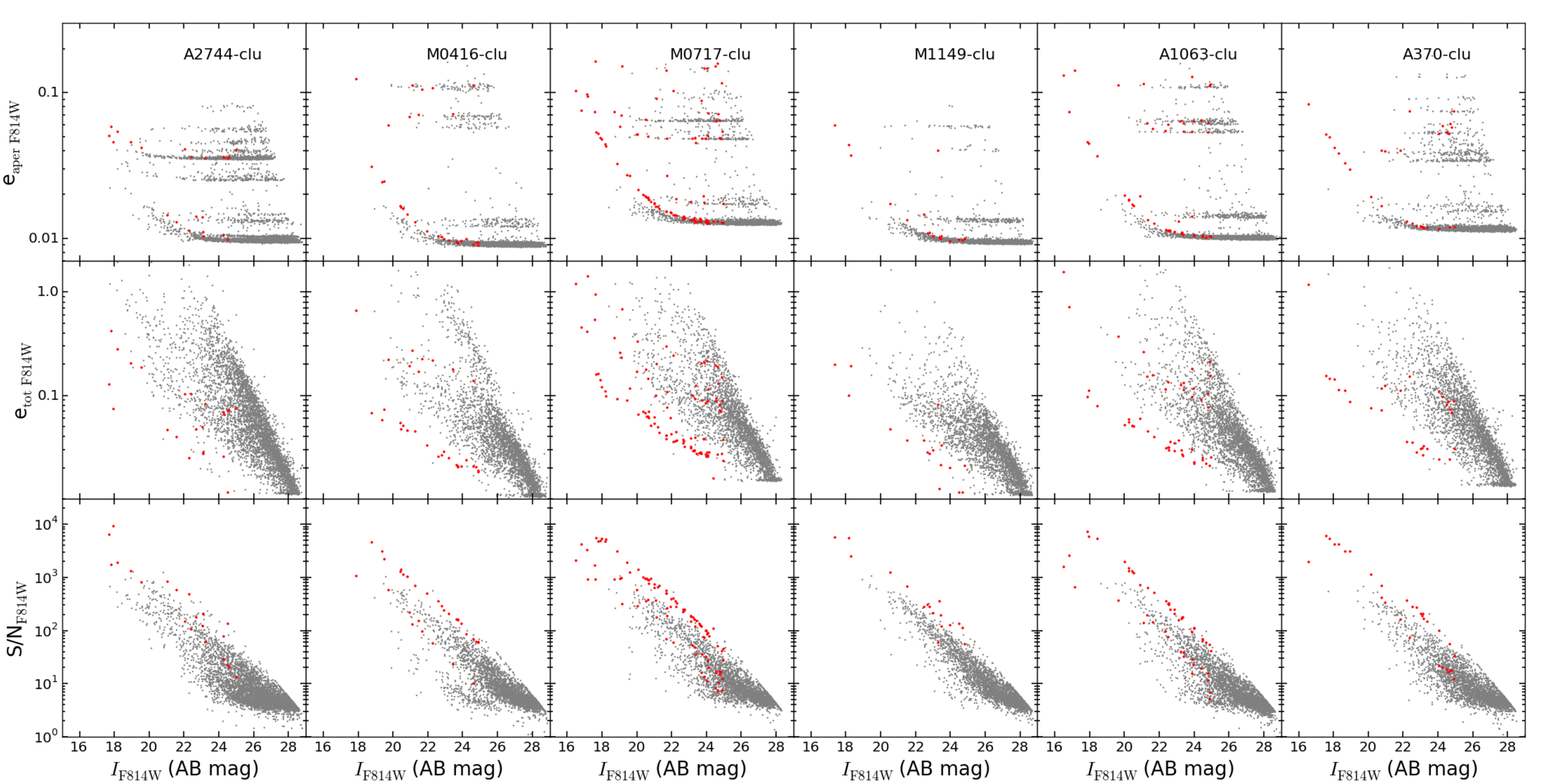}
\plotone{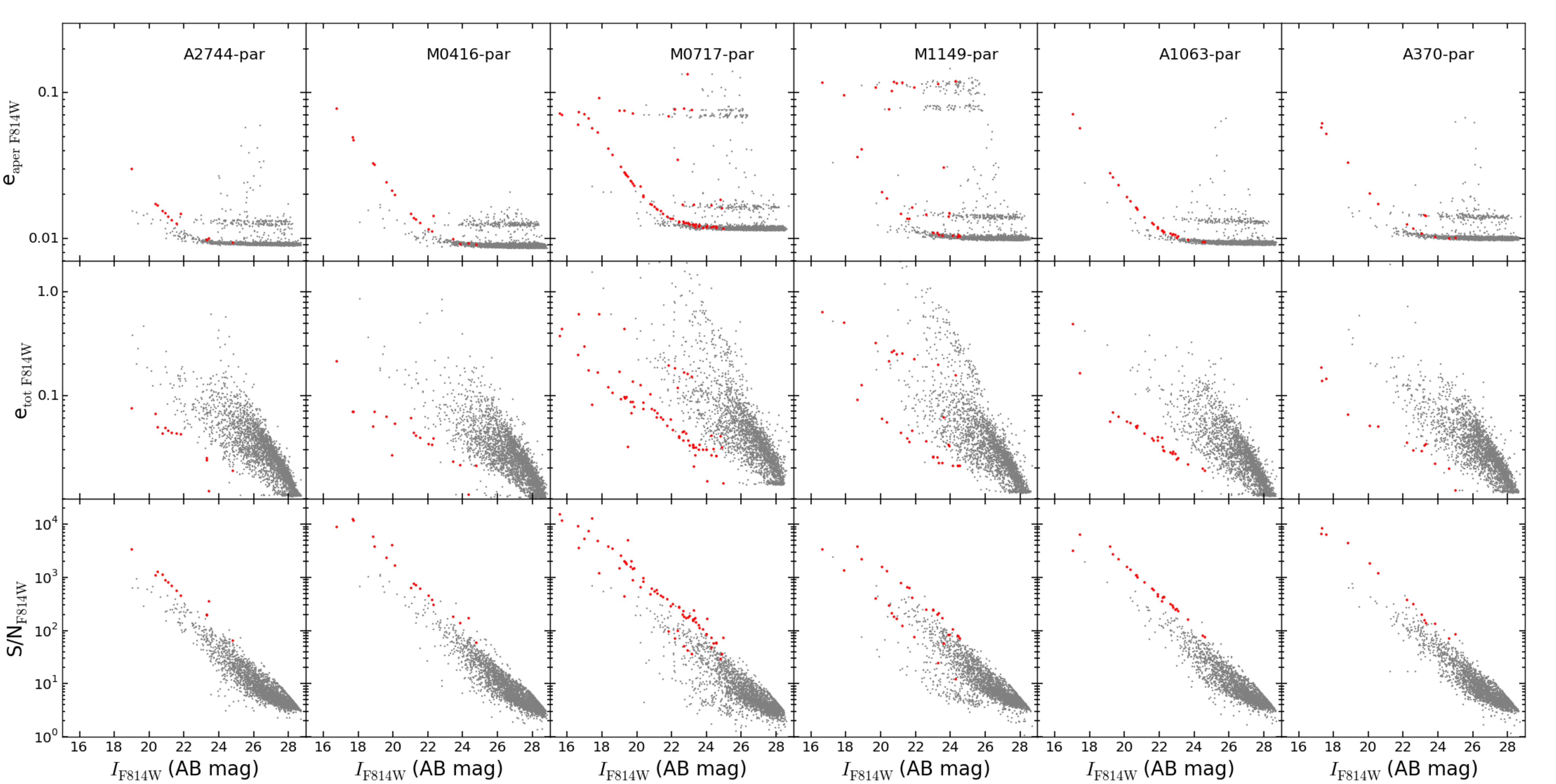}
\caption{$F814W$ error distributions in each of the twelve fields.  The units are mag$_\mathrm{AB} = -$log(value) + 25.  Galaxies defined with use\_phot = 1 are shown in gray and point sources in red (star\_flag = 1).  Top panels:  $F814W$ errors within an aperture of $0\farcs7$ versus magnitude.  The variable depths across each mosaic give rise to the discrete levels.  Most of the sources fall within the deepest part of the mosaics with photometry aperture errors reaching the lowest values.  Middle panels:  Total $F814W$ error versus magnitude from the catalogs (see Section \ref{hst photometry}).  Bottom panels:  Total $F814W$ S/N versus magnitude.  In general, point sources have the highest S/N at a given magnitude, while extended sources form the lower envelope of the distribution.}
\label{errs814}
\end{figure*}

\begin{figure*}[ht!]	%%%%%%%%%%%%%%%  PLOT  %%%%%%%%%%%%%%%%%
\epsscale{1.15}
\plotone{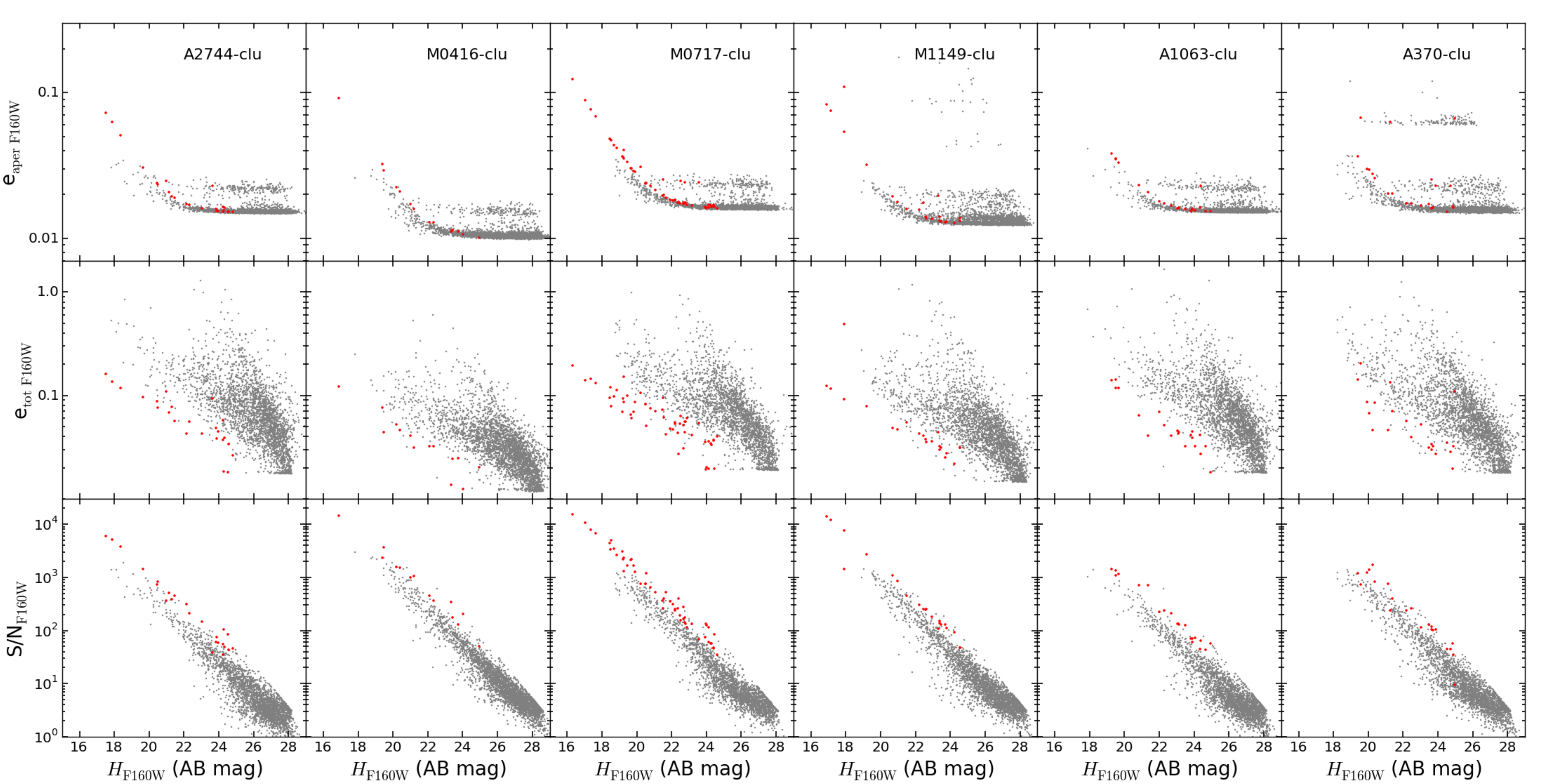}
\plotone{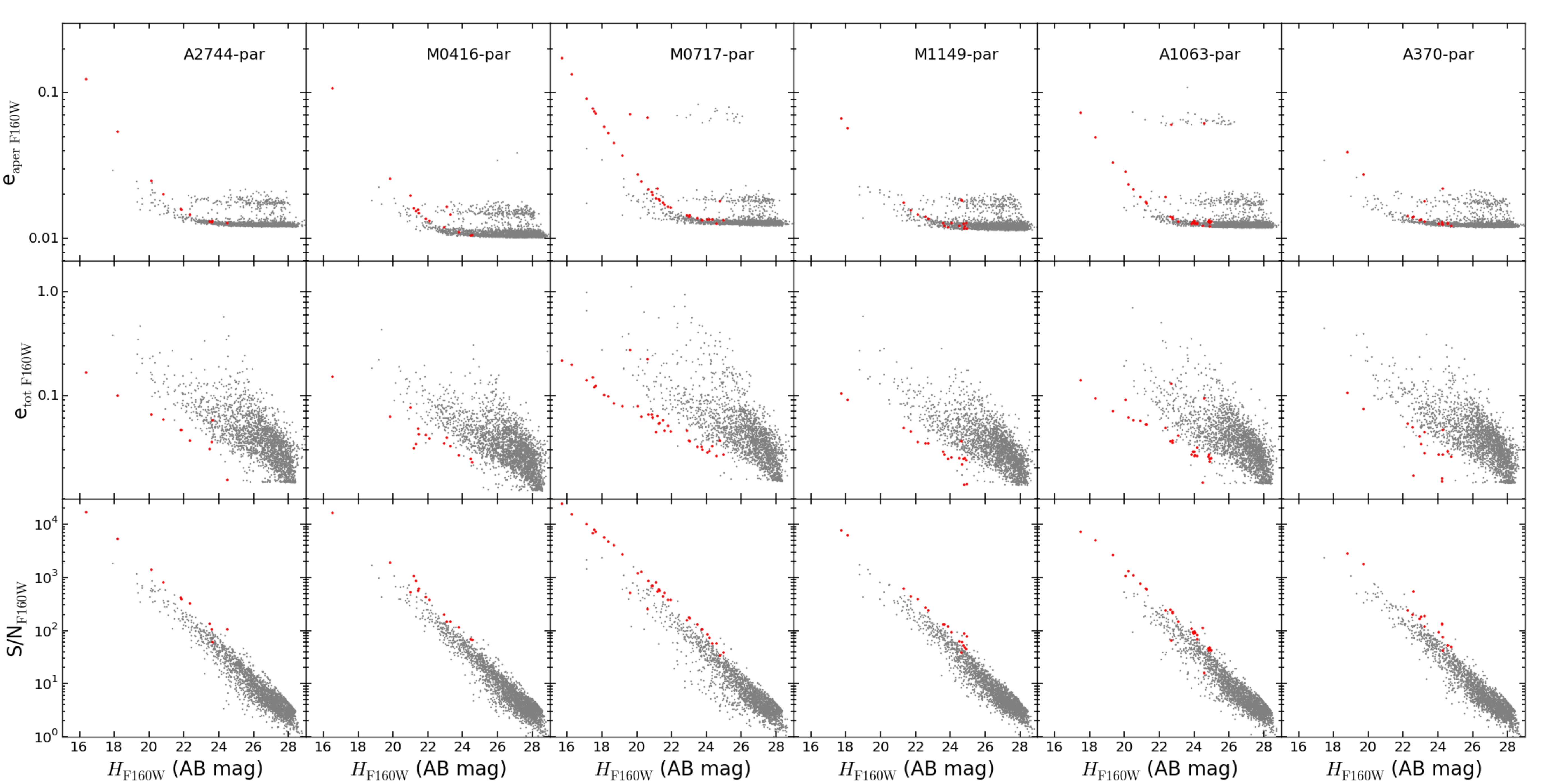}
\caption{Same as Figure \ref{errs814} but for the $F160W$ error distributions in each of the twelve fields with consistent results.  The units are mag$_\mathrm{AB} = -$log(value) + 25.}
\label{errs160}
\end{figure*}

As described in Section~\ref{hst photometry} following the procedure in \citet{Skelton2014}, we ensure that the error for each source is adjusted to take into account the photometric weight at its position.  Furthermore, the total error, in part, is determined by placing ``empty apertures'' in each of the mosaics to determine the width of the distribution in various sized apertures for flux measurements.  In Figures \ref{errs814} and \ref{errs160}, we show the errors as a function of $I_{F814W}$ and $H_{F160W}$ magnitude, designated by the ``bandtotal'' column of the catalogs in each field.

		%%%%%%%%%%%%%%%  PLOT  %%%%%%%%%%%%%%%%%

		%%%%%%%%%%%%%%%  PLOT  %%%%%%%%%%%%%%%%%

The top panels show the errors in our standard photometric aperture of $0\farcs 7$.  The scatter in the error at fixed magnitude is caused by the variation in the depth of the $I_{F814W}$ and the $H_{F160W}$ mosaics.  The stripes reflect the weights, and hence the errors, that largely show the depth of a particular position in the mosaic, based on the number of exposures for each source's position (e.g., see Figure \ref{deep complete overlap}).  These discrete levels are more prominent in the $I_{F814W}$ mosaics due to more orientations during the observations.  Stars (red points) fall in the same bands as galaxies (gray points), as their aperture fluxes are measured in the same $0\farcs 7$ aperture.  The distributions differ from field to field, as the depths are not identical.  At the bright end, each discrete level turns up from the error being dominated by the Poisson error.  In particular, this effect is most obvious for the lowest level of each field.

The middle panels show the ``total'' errors from the photometry. These errors are determined from the empty aperture errors using the power-law fit at the number of pixels in the circularized Kron aperture (see right panel of Figure~\ref{empty aper} and Section~\ref{hst photometry}) added in quadrature to the Poisson error for each source, and then scaled to total using the AUTO-to-total flux correction (see Section~\ref{flux corr}).  The stripes are blurred in these panels, as the scatter in the error is now dominated by the variation in the Kron aperture size at fixed magnitude and the Poisson error contribution is mostly smoothed out at brighter magnitudes.  The range in the Kron aperture sizes reflects the sizes of galaxies at fixed magnitude.  Stars (red points) are now offset from galaxies, as the total flux of stars is measured in a smaller aperture than the total flux of extended sources.  The estimated errors are smaller for their total fluxes.  A few point sources fall within the extended sources envelope at discrete levels in each field (e.g.\ \mtwo\ and \atwo\ clusters in Figure~\ref{errs814}) due to the empty aperture error being dependent on the depth (see Section~\ref{hst photometry}) that varies across the mosaics as discussed previously.  These point sources are found in the shallower areas of the mosaics.

In the bottom panels, the S/N of the sources is given as a function of magnitude.  The S/N is calculated by dividing the total $I_{F814W}$ and $H_{F160W}$ flux by their respective estimated total error.  The relation of the S/N of stars (red points) with magnitude shows very little scatter, demonstrating the small scatter in the errors of stars in the middle panels.  Furthermore, the discrete levels persist, from the middle panels, but form tighter sequences similar to the top panels.  The errors in the total magnitudes of galaxies are typically much larger and thus have a larger scatter.  This should be taken into account when assessing the depth of the $I_{F814W}$ and $H_{F160W}$ mosaics (see Section~\ref{completeness}).

\vspace{24pt}
\section{Redshifts, Rest-frame Colors, and Stellar Population Parameters}
\label{derived properties}

We use the photometric catalogs to derive photometric redshifts, rest-frame colors, and stellar population parameters of the galaxies for all 12 fields.  We note these derived parameters depend significantly on the assumptions and methodology used to derive them \citep[see, e.g.,][]{Brammer2008, Kriek2009}.  We choose to use the photometric fitting codes EAZY \citep{Brammer2008} and FAST \citep{Kriek2009} to accomplish these tasks.  We describe a ``default'' set of parameters, described below, that we provide with the release of the photometric catalogs.  We stress that the released catalogs of the stellar population properties and the rest-frame luminosities do not include the corrections for the lensing magnifications.

\begin{figure*}[ht!]	%%%%%%%%%%%%%%%  PLOT  %%%%%%%%%%%%%%%%%
\epsscale{1.}
\plotone{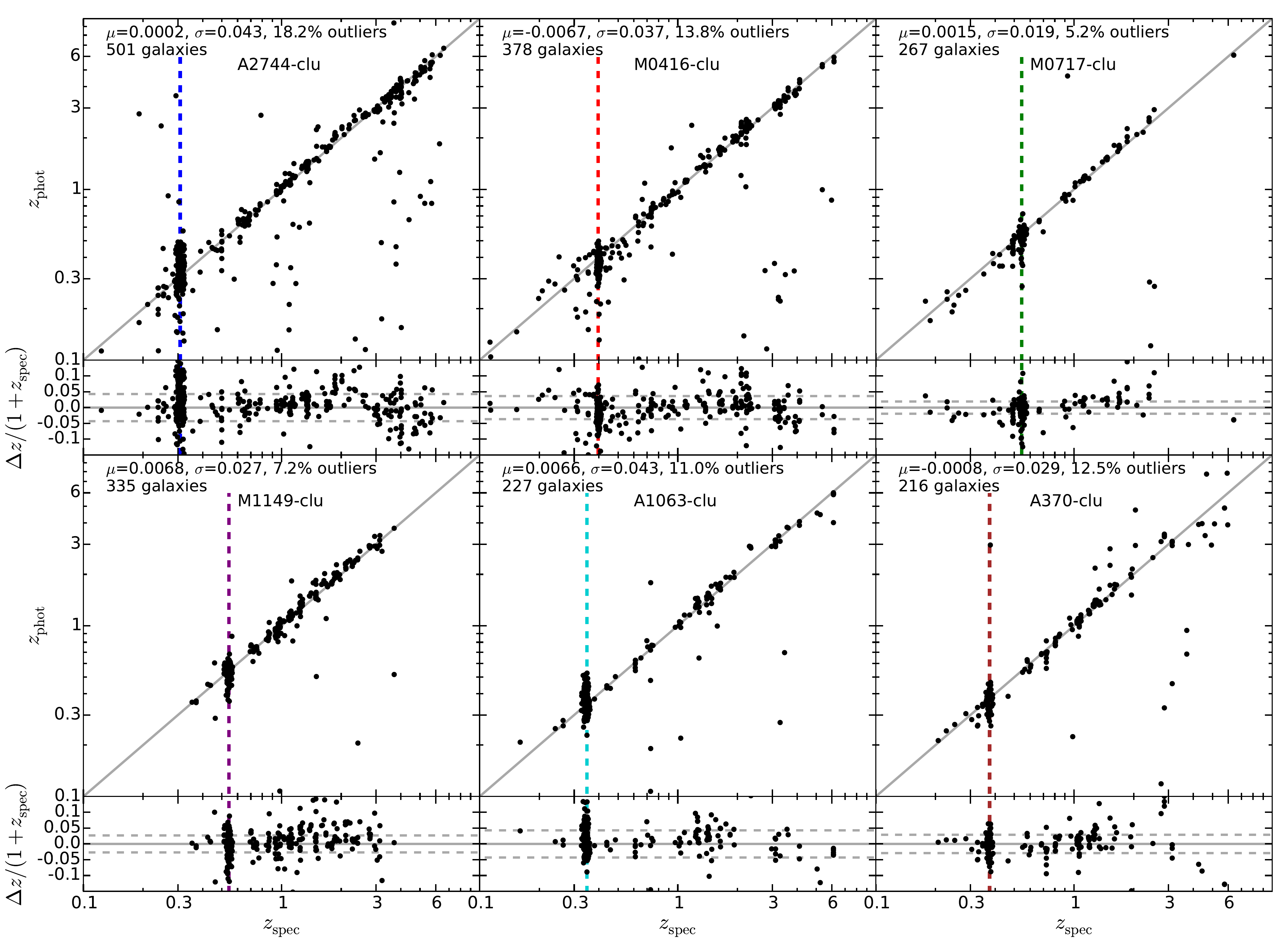}
\plotone{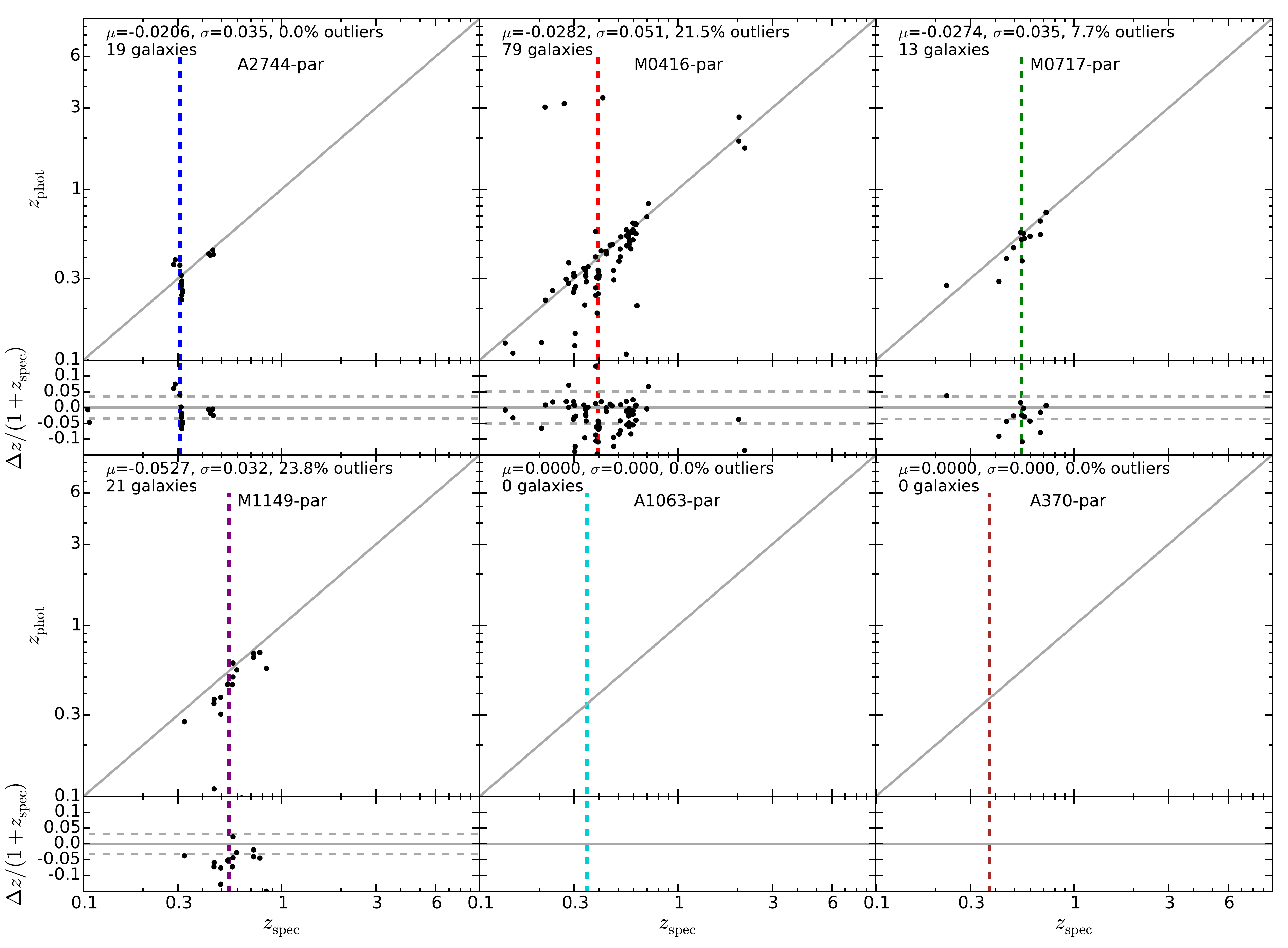}
\caption{Comparison of estimated photometric redshifts from our analysis and confirmed published spectroscopic redshifts from the literature for all twelve fields.  The bi-weight mean $\mu = (z_{\mathrm{phot}} - z_{\mathrm{spec}})/(1+z_{\mathrm{spec}})$, NMAD scatter $\sigma$, \% of objects with $|z_{\mathrm{phot}} - z_{\mathrm{spec}}|/(1+z_{\mathrm{spec}}) > 0.1$ and the number of galaxies in each comparison are shown in the upper left of the panel for each field. The lower panels of each field show the difference between the photometric and spectroscopic redshifts over $1+z_{\mathrm{spec}}$. The gray dashed lines indicate $\pm \sigma_{\mathrm{NMAD}}$ in each case.  The vertical dashed line in each panel indicates the cluster redshift.  The gray solid lines in each panel indicate the unity relation between $z_{\mathrm{phot}}$ and $z_{\mathrm{spec}}$.}
\label{zphot vs zspec}
\end{figure*}		%%%%%%%%%%%%%%%  PLOT  %%%%%%%%%%%%%%%%%

\subsection{Spectroscopic Redshifts}
\label{speczs}

\begin{deluxetable}{lrr}		%%%%%%%%%%%%%  TABLE  %%%%%%%%%%%%%%%
\tablecaption{Spectroscopic Redshift Matches \label{spec-z matches} \vspace{-6pt} }
\tablecolumns{3}
%\tablewidth{0pc}
%\setlength{\tabcolsep}{0pt}
\tablehead{
\colhead{Field} & \colhead{Matches} & \colhead{Source = ``OK'' Matches} \\
& \colhead{(\# galaxies)} & \colhead{(\# galaxies)} }
\startdata
\aone -clu & 546 & 501 \\
\aone -par & 22 & 19 \\
\mone -clu & 389 & 378 \\
\mone -par & 80 & 79 \\
\mtwo -clu & 294 & 267 \\
\mtwo -par & 17 & 13 \\
\mthree -clu & 344 & 335 \\
\mthree -par & 22 & 21 \\
\atwo -clu & 237 & 227 \\
\atwo -par & 0 & 0 \\
\athree -clu & 221 & 216 \\
\athree -par & 0 & 0
\enddata
\tablecomments{Spectroscopic redshift matches from the literature (see Section \ref{speczs}). The last column (Source = ``OK'') is designating sources with use\_phot = 1.}
\end{deluxetable}		%%%%%%%%%%%%%  TABLE  %%%%%%%%%%%%%%%

As the \hff\ cluster and parallel fields are very small areas on the sky, we search the literature to find spectroscopic redshifts of sources that targeted the \hff\ clusters.  These redshifts are used to assess the quality of photometric redshifts in Section \ref{photzs} and used in place of the estimated photometric redshifts, when available, to derive rest-frame colors and stellar population parameters (see Sections \ref{rfcols} and \ref{gal properties}).  The spectroscopic redshifts in our catalogs are obtained by cross-matching the positions of sources within $0\farcs5$ to a number of publicly available catalogs.  We select only secure (i.e.\ reliable according to the reference) spectroscopic redshifts to ensure only quality redshifts are used.  However, we have the possibility of sources having multiple references with a spectroscopic redshift.  In these cases, we give more weight to references with more robust measurements of the spectroscopic redshift (i.e.\ a better reported $\delta z$, usually non-grism data).  When multiple references are comparable, for the same source, we select the first spectroscopic redshift in our list of references (described below) and cite that reference in the catalog (column REFspecz).  This is done for simplicity and does not affect our analysis of derived parameters in the following sections.

The total number of sources with spectroscopic redshifts for each cluster and parallel field and how many of those sources have use\_phot = 1 are given in Table \ref{spec-z matches}.  We use spectroscopic redshifts from the following literature catalogs:  \aone\ has five catalogs from GLASS\footnote{Redshift catalogs for GLASS were downloaded from \url{https://archive.stsci.edu/prepds/glass/}.}, Brammer et al.\ (in prep), \citet{Mahler2017}, \citet{Owers2011} and \citet{Richard2014}; \mone\ has seven catalogs from \citet{Balestra2016}, Brammer et al.\ (in prep), \citet{Caminha2016}, \citet{Ebeling2014}, GLASS, \citet{Grillo2015} and \citet{Jauzac2014}; \mtwo\ has four catalogs from Brammer et al.\ (in prep), \citet{Ebeling2014}, GLASS and \citet{Limousin2016}; \mthree\ has five catalogs from Brammer et al.\ (in prep), \citet{Ebeling2014}, GLASS, \citet{Grillo2016} and \citet{Smith2009}; \atwo\ has five catalogs from Brammer et al.\ (in prep), \citet{Diego2016}, GLASS, \citet{Karman2016} and \citet{Richard2014}; and \athree\ has four catalogs from Brammer et al.\ (in prep), GLASS, \citet{Lagattuta2016} and \citet{Richard2014}.  Some of these catalogs utilize previous references that we keep track of and give the original reference, when possible, in our photometric catalogs for the spectroscopic redshift column~(REFspecz).

\subsection{Photometric Redshifts and Zero Point Corrections}
\label{photzs}

We use the EAZY code \citep{Brammer2008}\footnote{\url{https://github.com/gbrammer/eazy-photoz/}} to estimate photometric redshifts by fitting the SED of each source with a linear combination of 12 galaxy templates.  These templates are derived with the method used for the original EAZY templates \citep[after][]{BR2007}, but now using Flexible Stellar Population Synthesis (FSPS) models \citep{Conroy2009,Conroy2010} and trained on the UltraVISTA photometric catalogs \citep{Muzzin2013c}.  We use the default template error function scaled by a factor of 0.2, which helps to account for systematic wavelength-dependent uncertainties in the templates, and a redshift prior based on the $F160W$ apparent magnitudes.\footnote{For more detailed information refer to the documentation on the current version of the EAZY code.}

\begin{figure*}[ht!]	%%%%%%%%%%%%%%%  PLOT  %%%%%%%%%%%%%%%%%
\epsscale{1.15}
\plotone{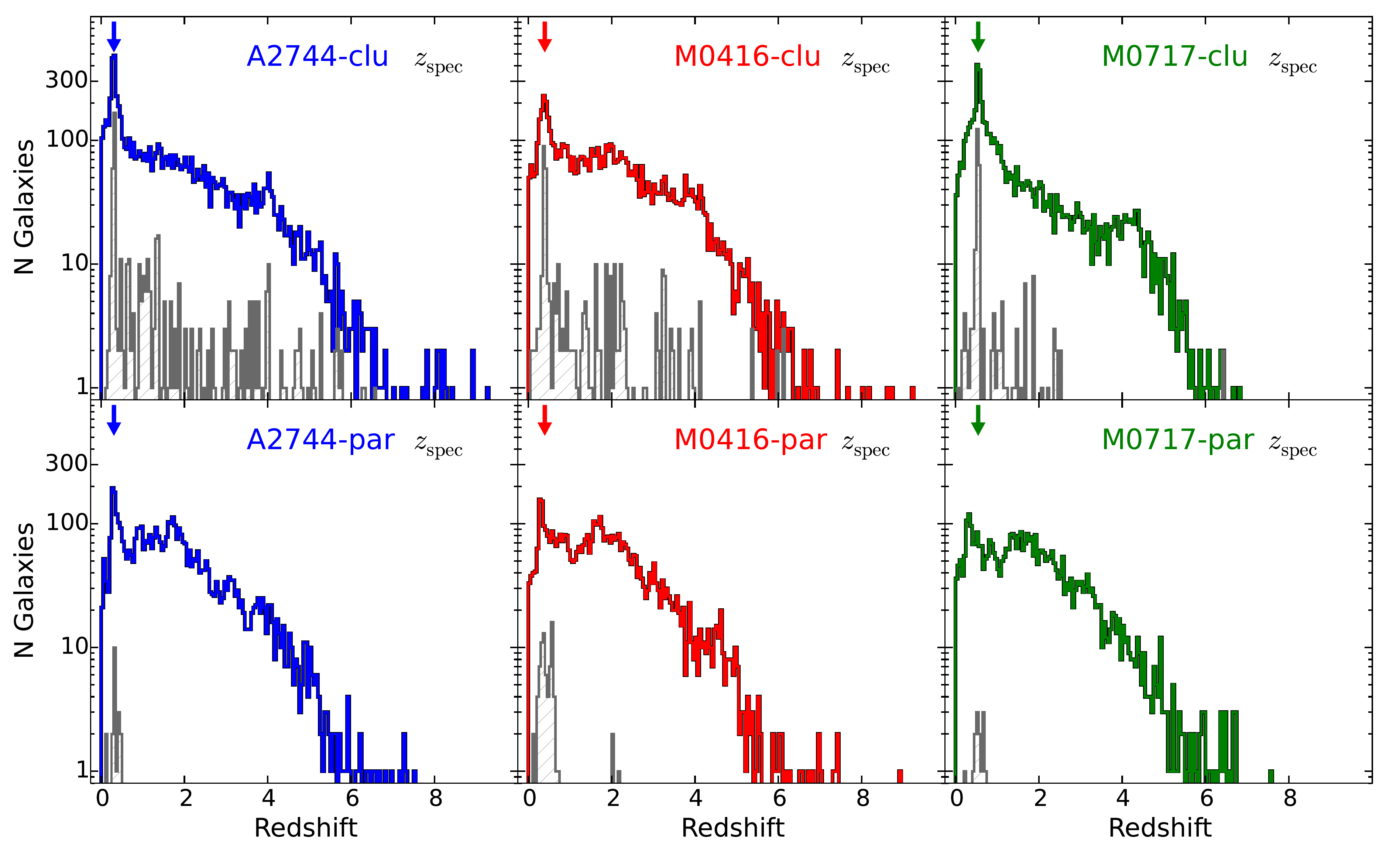}
\plotone{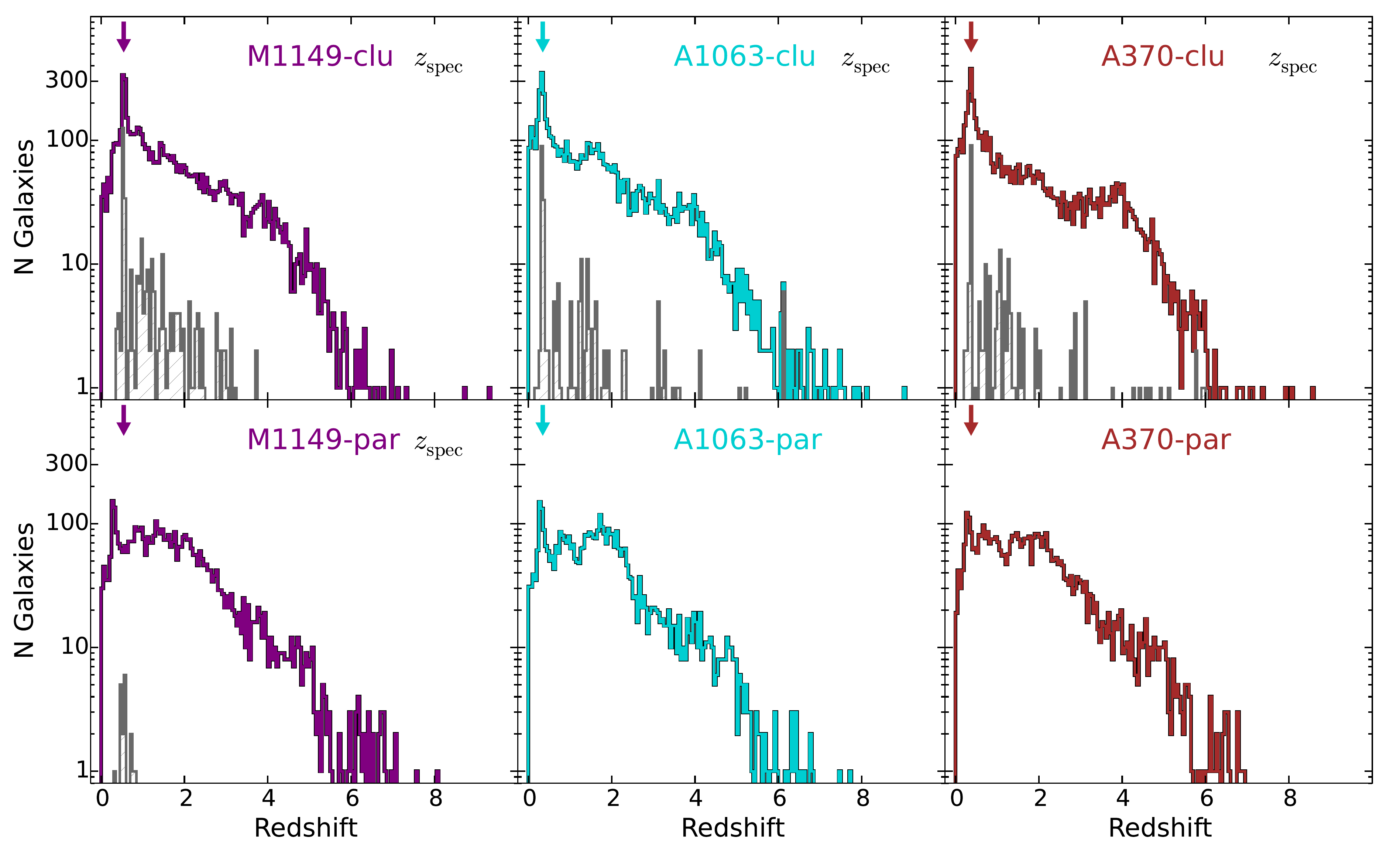}
\caption{Redshift distributions of the estimated photometric redshifts (empty histograms) and spectroscopic redshifts (hatched histograms) for the clusters (top) and parallels (bottom) for all \hff.  A magnitude cut corresponding to the 90\% completeness limits has been applied (utilizing Tables \ref{complete814} and \ref{complete160}).}
\label{z histograms}
\end{figure*}		%%%%%%%%%%%%%%%  PLOT  %%%%%%%%%%%%%%%%%

Furthermore, we correct the photometry for empirically determined zero point correction factors as in \citet{Skelton2014}.  Zero point corrections are determined for each band of all fields (see Table \ref{hff zps}).  We settle on a single set of zero point corrections for all fields due to the small area and limited number of spectroscopic redshifts in each field to determine the zero point corrections from SED fitting.  The listed zero point corrections are applied to the catalogs, and the corrected photometry is used for the redshift estimates and stellar population parameters, presented in the following sections.  We use the spectroscopic redshift ($z_\mathrm{spec}$), if available, or the peak of the photometric redshift distribution (EAZY's \texttt{z\_peak}) as the galaxy redshift, unless otherwise noted.

We find good agreement between the spectroscopic and photometric redshifts by estimating the scatter in each field, the average $\sigma_\mathrm{nmad} = 0.034$ across all fields with spectroscopic redshifts (\atwo\ and \athree\ parallels do not have any spectroscopic redshifts).  Figure~\ref{zphot vs zspec} demonstrates the comparison of the photometric redshifts to spectroscopic redshifts for each field.  Among sources with spectroscopic redshifts, there are few outliers (i.e.\ failures): (18.2, 0.0)\%, (13.8, 21.5)\%, (5.2, 7.7)\%, (7.2, 23.8)\%, (11.0, N/A)\%, (12.5, N/A)\% for \aone, \mone, \mtwo, \mthree, \atwo\ and \athree, respectively, for the clusters and parallels (cluster, parallel), where we define an outlier as $|z_{\mathrm{phot}} - z_{\mathrm{spec}}|/(1+z_{\mathrm{spec}}) >$~0.1.  We note that, whereas the mean and sigma in $\Delta z/(1+z_{\mathrm{spec}})$ of the \aone\ cluster is quantitatively similar to the other cluster fields, its fraction of outliers is significantly larger.  We investigate the SEDs of the outliers in the \aone\ cluster, but find no obvious problem with the photometry.  No systematic trend is found when $\Delta z/(1+z_{\mathrm{spec}})$ is plotted as a function of the $F160W/F814W$ magnitude of the source or as a function of the brightness of the subtracted model (that includes the background) in either the $F160W$ or the $F814W$ bands.  This provides confidence that our bCGs/ICL subtraction does not introduce systematic effects in the derivation of the photometric redshifts.

\begin{figure*}[ht!]	%%%%%%%%%%%%%%%  PLOT  %%%%%%%%%%%%%%%%%
\epsscale{1.15}
\plotone{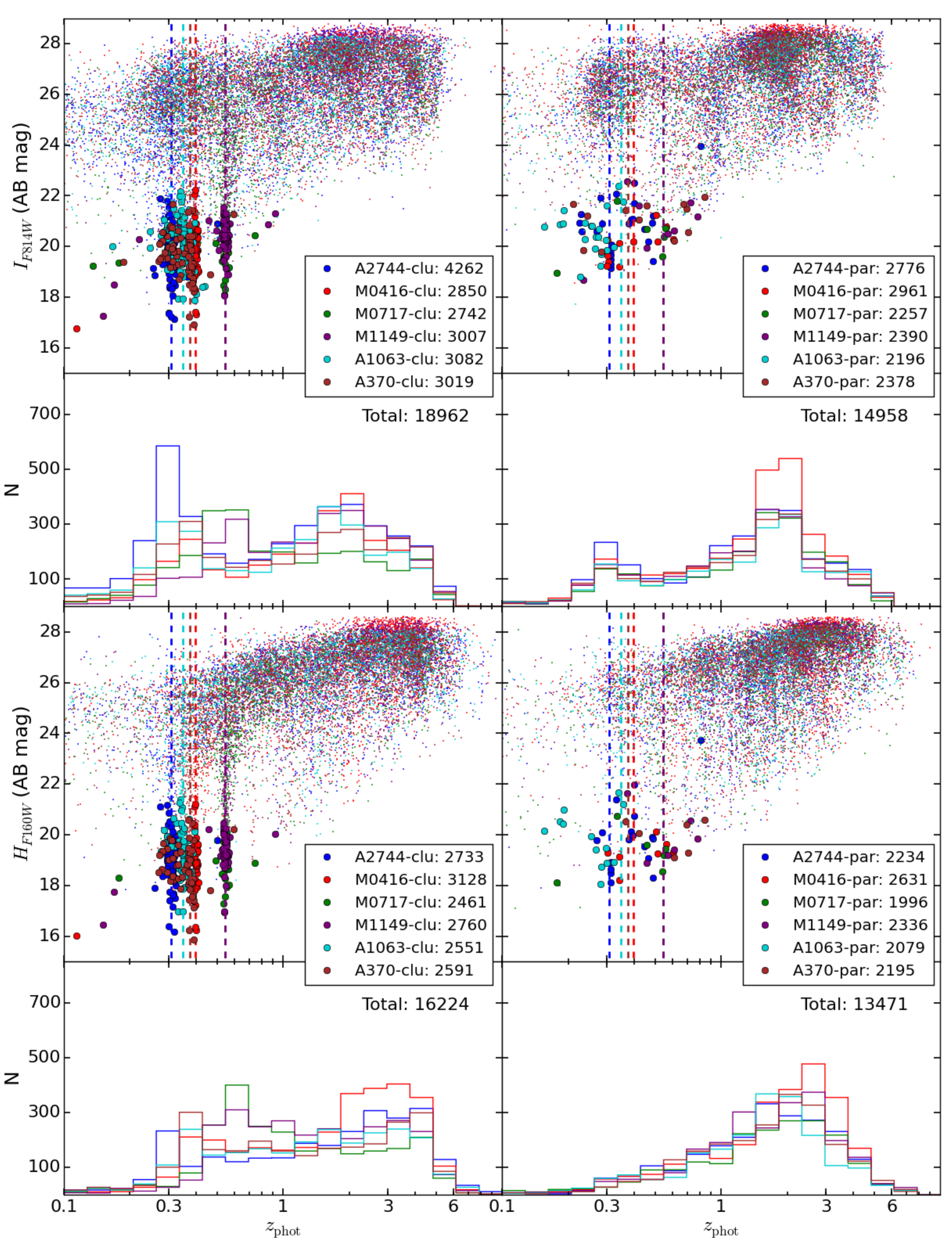}
\caption{The distribution of $I_{F814W}$ (top panels) and $H_{F160W}$ (bottom panels) apparent magnitudes with photometric redshift ($z_\mathrm{peak}$ from EAZY, $z_\mathrm{phot}$ in figure), spectroscopic redshift when available, color-coded by field for both the clusters (left panels) and parallels (right panels).  The modeled out bCGs are shown as circles (top panels), color-coded by field the same as the points (see legend in each panel).  The lower half of the panels show the number of galaxies as a function of $z_\mathrm{phot}$ broken down into the contribution from each field (for sources with use\_phot = 1 and ``bandtotal'' = $F814W$, top panels, and use\_phot = 1 with ``bandtotal'' = $F160W$, bottom panels), again color-coded the same.  A magnitude cut corresponding to the 90\% completeness limits has been applied (utilizing Tables \ref{complete814} and \ref{complete160}).}
\label{mag vs redshift}
\end{figure*}		%%%%%%%%%%%%%%%  PLOT  %%%%%%%%%%%%%%%%%

In Figure~\ref{z histograms}, we show the photometric redshift distributions of sources (selected with use\_phot = 1) in each of the fields, using $z_\mathrm{peak}$ (open histograms).  The spectroscopic redshift distributions are shown by the hashed histograms.  For the clusters, the over-densities correspond to the redshift of the cluster as indicated by the peaks in both the $z_\mathrm{phot}$ and $z_\mathrm{spec}$ distributions.  Though clearly less pronounced than in the cluster fields themselves, an excess of galaxies at the cluster redshifts can be seen in the parallel fields $\sim$6 arcmin (1.5$-$2~Mpc) away from the cluster core.  These observations are even more evident as the cluster over-densities are illustrated by the distribution of apparent magnitudes with $z_\mathrm{peak}$ (see Figure~\ref{mag vs redshift}, the circles correspond to modeled bCGs) and in the mass distributions (see Figures~\ref{massz160}, \ref{massz814} and Section \ref{gal properties}).  In the lower panel of Fig.~\ref{mag vs redshift}, we show the number of galaxies as a function of $z_\mathrm{peak}$ and give the number of galaxies for each field for $I_{F814W}$ (with use\_phot = 1 and bandtotal = $F814W$) and $H_{F160W}$ (with use\_phot = 1 and bandtotal = $F160W$).

In Figure \ref{z histograms}, we noticed small peaks at a redshift of $z\sim4$ in a few of the cluster fields (e.g., \aone, \mtwo\ and \athree).  We investigated this further to determine if there was a problem with the SED fitting resulting in mis-identified sources preferentially placed at $z\sim4$.  We first considered the effect of lensing that may be a likely cause of these $z\sim4$ peaks as they are not present in the parallel fields and not present at the same redshift in each cluster.  We looked at the distribution of galaxies, particularly in the range of $3.5 < z < 4.5$ for all fields.  The galaxies at $z\sim4$ tend to cluster around the massive galaxies in the cluster as expected from the available lensing maps (see Section \ref{lensing factors}) for the clusters but are mostly distributed evenly in the parallel fields.  This is further supported by the fact that we also modeled out bCGs in the parallel fields; suggesting that the apparent enhancement of $z\sim4$ galaxies in some cluster fields is not due to modeling or SED fitting, but, at least partly, to lensing magnification by the clusters.  However, we also explored the possibility that a few of these sources in the clusters could be mis-identified globular clusters (GCs) that become visible after modeling out the bCGs.  We selected sources around the very bright massive galaxies in a few of the clusters.  The most notable are in the \mone, \mtwo\ and \athree\ clusters.  We selected many of the sources near the modeled out galaxies that have $z_\mathrm{peak}\sim4$ and could potentially be GCs from their spatial distribution, and checked their SEDs and SED fits.  Among the objects with reliable modeling of the observed SEDs, only a small fraction of these sources are likely mis-identified GCs with secondary peaks in the EAZY redshift probability functions consistent with the redshifts of the modeled out galaxies.  Although we can not rule out some level of contamination from low-redshift GCs to the $z\sim4$ galaxy population, the mis-identified GCs do not appear to represent a significant contribution to the peaks at $z\sim4$ noticed in Figure~\ref{z histograms}, with the \athree\ cluster having the most possible GC candidates ($< 10$ in all).  This further supports our conclusion that lensing magnification is likely the main origin of the $z\sim4$ peaks.

Lastly, we compared our photometric redshifts against the available ASTRODEEP \citep{Merlin2016,Castellano2016,DC2017} catalogs, which have been publicly released for the \aone, \mone, \mtwo\ and \mthree\ clusters and parallels.  To make accurate comparisons as meaningful as possible, we have selected sources based on similar selection criteria (see Figures \ref{astromatch} and \ref{astrohist} in the Appendix for selection criteria).  Comparing the redshift histograms in Figure \ref{astrohist}, we find qualitatively very similar redshift distributions.  In regards to the $z\sim4$ redshift sources, we find both ours and ASTRODEEP's distributions show small comparable peaks for the \aone, \mone, \mtwo\ and \mthree\ clusters.  This is further supported by comparing the matched sources' photometric redshifts, albeit, with some scatter (see Figure \ref{astromatch}).  The above discussion provides evidence that the peaks at $z\sim4$ are likely real and due to lensing magnification of the clusters, while mis-identified sources (e.g., GC contamination) are most likely small or negligible contributions to our photometric redshift catalogs.

\begin{deluxetable}{rc}		%%%%%%%%%%%%%  TABLE  %%%%%%%%%%%%%%%
\tablecaption{Zero Point Corrections for the Hubble Frontier Fields Filters \vspace{-6pt}
\label{hff zps}}
\tablecolumns{2}
\tabletypesize{\footnotesize}
\tablewidth{0pc}
\setlength{\tabcolsep}{0pt}
\tablehead{
\colhead{Filter} & \colhead{Zero Point Correction} }
\startdata
UVIS $F225W$ & 1.1652 \\
$F275W$ & 0.9855 \\
$F336W$ & 0.9936 \\
$F390W$ & 1.0094 \\
\noalign{\smallskip}
\hline
\noalign{\smallskip}
ACS $F435W$ & 1.0352 \\
$F475W$ & 1.0067 \\
$F555W$ & 1.0287 \\
$F606W$ & 1.0105 \\
$F625W$ & 0.9931 \\
$F775W$ & 0.9923 \\
$F814W$ & 0.9990 \\
$F850LP$ & 0.9999 \\
\noalign{\smallskip}
\hline
\noalign{\smallskip}
WFC3 $F105W$ & 1.0173 \\
$F110W$ & 1.0178 \\
$F125W$ & 1.0117 \\
$F140W$ & 1.0105 \\
$F160W$ & $\equiv 1.0$ \\
\noalign{\smallskip}
\hline
\noalign{\smallskip}
$K_S$ \;\; HAWK-I & 0.9505 \\
MOSFIRE & 1.0221 \\
\noalign{\smallskip}
\hline
\noalign{\smallskip}
IRAC $3.6 \micron$ & 1.0464 \\
$4.5 \micron$ & 1.0193 \\
$5.8 \micron$ & 1.0072 \\
$8.0 \micron$ & 0.9970
\enddata
%\tablecomments{}
\end{deluxetable}		%%%%%%%%%%%%%  TABLE  %%%%%%%%%%%%%%%

\subsection{Rest-frame Colors}
\label{rfcols}

From the photometric catalogs, it is easy to calculate observed colors, but to compare galaxies at different redshifts rest-frame colors need to be used.  These can be determined robustly from EAZY as we have a large set of observed-frame photometry in each of the fields.  The EAZY templates and best-fitting redshift, spectroscopic redshift if available, are used for each galaxy to determine its rest-frame luminosity in a series of filters.  The rest-frame luminosities are calculated individually rather than multiple filters; for more information on how the rest-frame colors are calculated, see \citet{Brammer2011}.  Along with the photometric catalogs, we provide a catalog that contains the rest-frame luminosities in a variety of commonly used filters (Johnson-Cousins B and R; Johnson-Morgan B; Johnson U, B and V; SDSS ugriz; 2MASS J, H and K; UV 1600 and 2800; Tophat 1400, 1700, 2200, 2700 and 2800).

\begin{figure*}[ht!]	%%%%%%%%%%%%%%%  PLOT  %%%%%%%%%%%%%%%%%
\epsscale{1.15}
\plotone{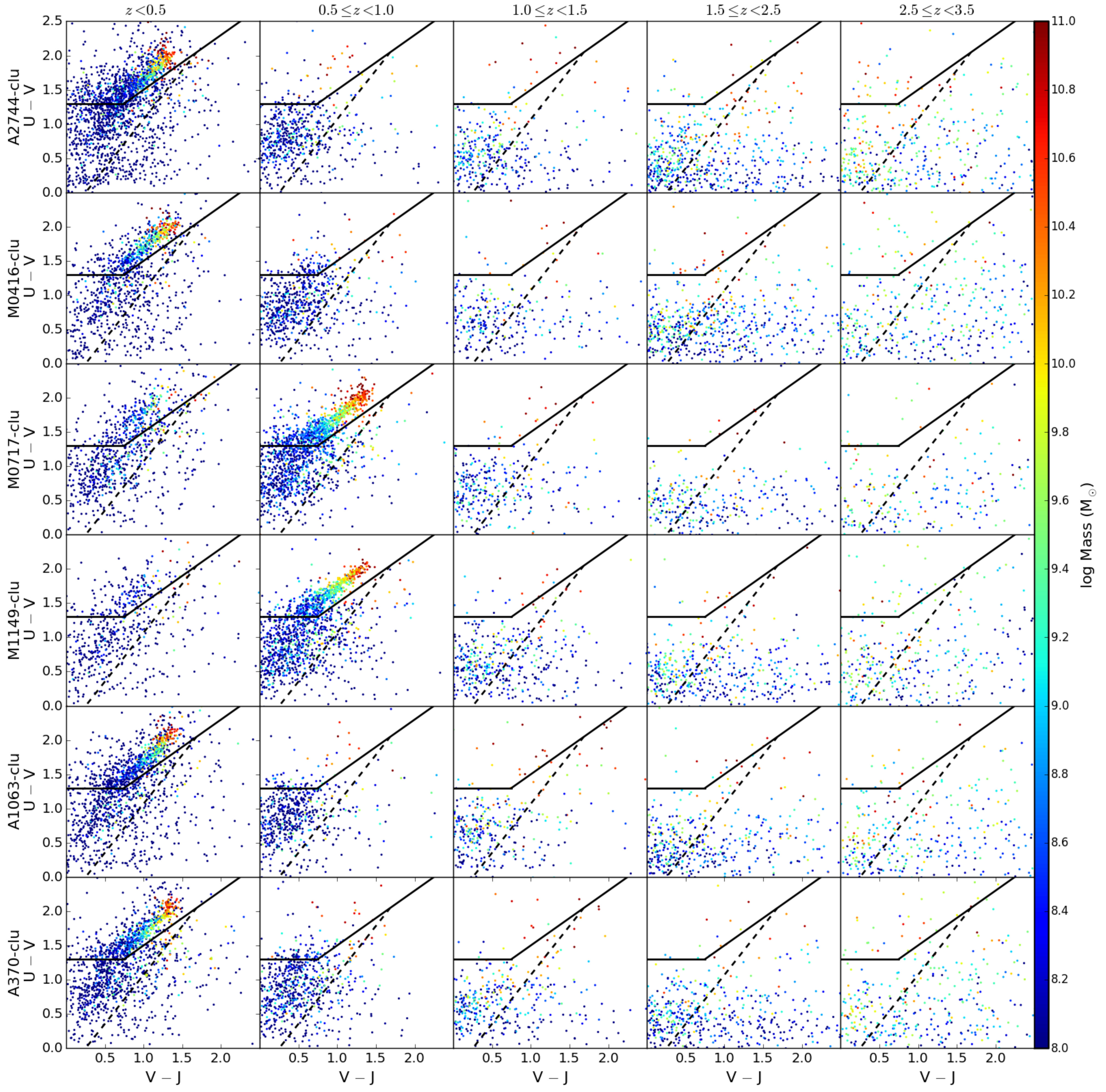}
\caption{Color-color selection of the \hff\ cluster fields.  Each row is a different cluster (field name given in vertical axis label), with redshift increasing from left to right (designated at the top of each column).  Quiescent and star-forming galaxies are separated by the color-color selection given in Section~\ref{rfcols} (solid black lines) and the additional selection of dusty star-forming galaxies (dashed black line).  The sources (use\_phot = 1) are color-coded by the estimated stellar mass for each cluster (color bar on the right side, see Section~\ref{gal properties}).  Lensing magnification corrections have not been applied.  A magnitude cut corresponding to the 90\% completeness limits has been applied (utilizing Tables \ref{complete814} and \ref{complete160}).}
\label{uvj diagrams clusters}
\end{figure*}		%%%%%%%%%%%%%%%  PLOT  %%%%%%%%%%%%%%%%%

\begin{figure*}[ht!]	%%%%%%%%%%%%%%%  PLOT  %%%%%%%%%%%%%%%%%
\epsscale{1.15}
\plotone{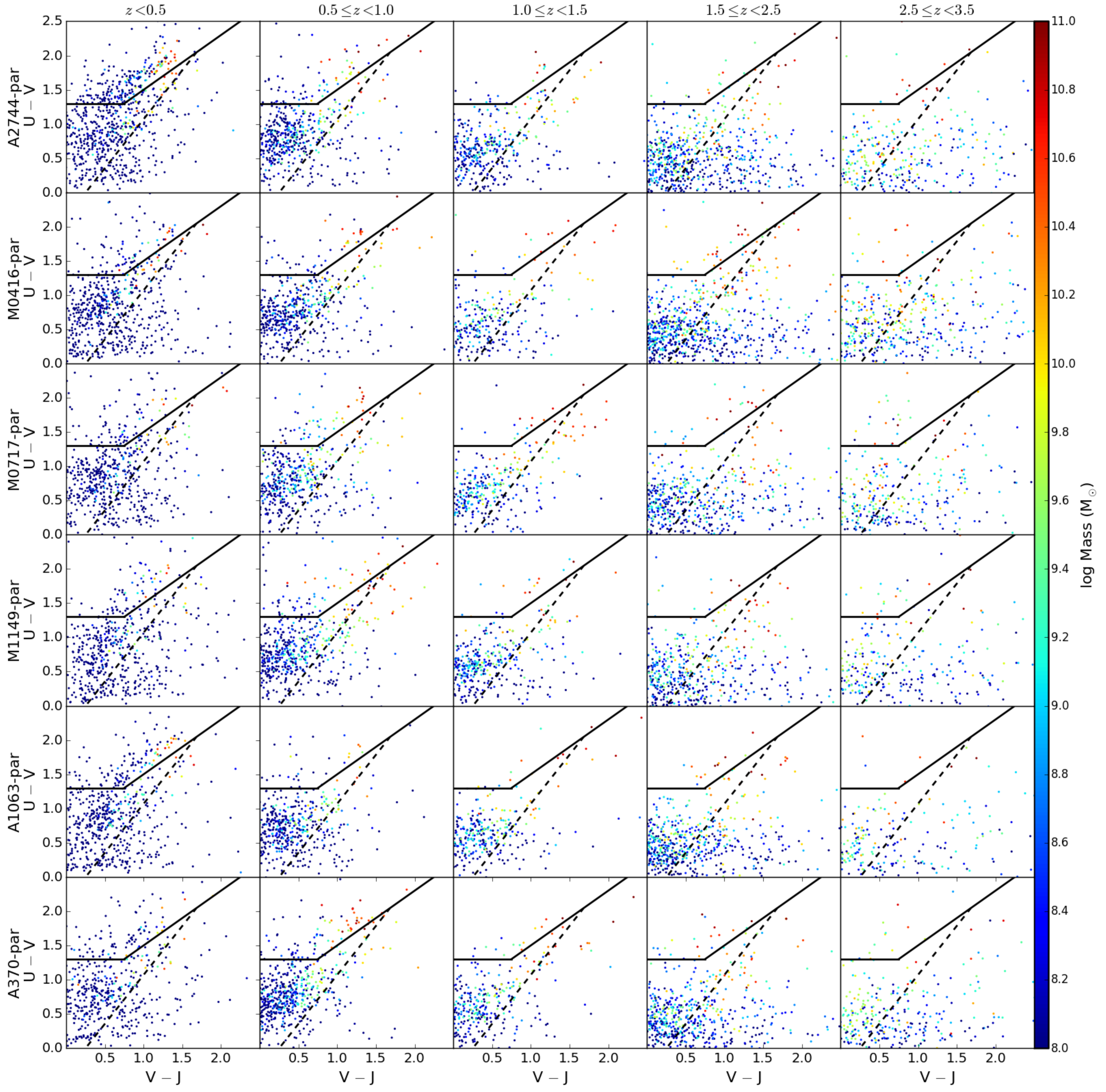}
\caption{Same as Figure~\ref{uvj diagrams clusters} but for the \hff\ parallel fields.  Lensing magnification corrections have not been applied.  A magnitude cut corresponding to the 90\% completeness limits has been applied (utilizing Tables \ref{complete814} and \ref{complete160}).}
\label{uvj diagrams parallels}
\end{figure*}		%%%%%%%%%%%%%%%  PLOT  %%%%%%%%%%%%%%%%%

We can further assess the galaxy populations of each field using a color-color analysis of the rest-frame photometry.  For this, we use ``UVJ'' diagrams \citep{Labbe2005,Wuyts2007,Williams2009} to separate the galaxy population into quiescent and star-forming galaxies for each field.  This diagram shows the rest-frame $U-V$ color versus the rest-frame $V-J$ color.  In Figures~\ref{uvj diagrams clusters} and \ref{uvj diagrams parallels}, we use $U - V < 1.3$ for $V - J < 0.75$ and $U - V < 0.8(V - J) + 0.7$ for $(V - J) \geq 0.75$ (solid black lines), also for dusty star-forming galaxies $(U - V) < 1.43(V - J) - 0.36$ \citep[dashed black line, see][for justification of criteria]{Martis2016}.  Each row represents one of the fields, with redshift increasing from left to right, as shown at the top of each column.  Quiescent galaxies with low levels of star formation that are red in $U-V$ (upper left region) are separated from similarly red (in $U-V$), dusty star-forming galaxies, with the star-forming galaxies having bluer $U-V$ and $V-J$ colors.  Galaxies (selected with use\_phot = 1) are color-coded by mass, with the most massive galaxies in red and least massive galaxies in blue (the color bar in the figure at the right gives the stellar mass breakdown).  The clusters are identified clearly by a strong quiescent galaxy sequence in their respective redshift bin.  The majority of low mass galaxies lie in the star-forming ``blue cloud'' at all redshifts.  In the highest redshift bins ($1.5 < z \leq 3.5$) many massive galaxies lie within the star-forming regions and appear to be red due to higher levels of dust rather than older stellar populations.

\subsection{Stellar Population Parameters}
\label{gal properties}

\begin{figure*}[ht!]	%%%%%%%%%%%%%%%  PLOT  %%%%%%%%%%%%%%%%%
\epsscale{1.15}
\plotone{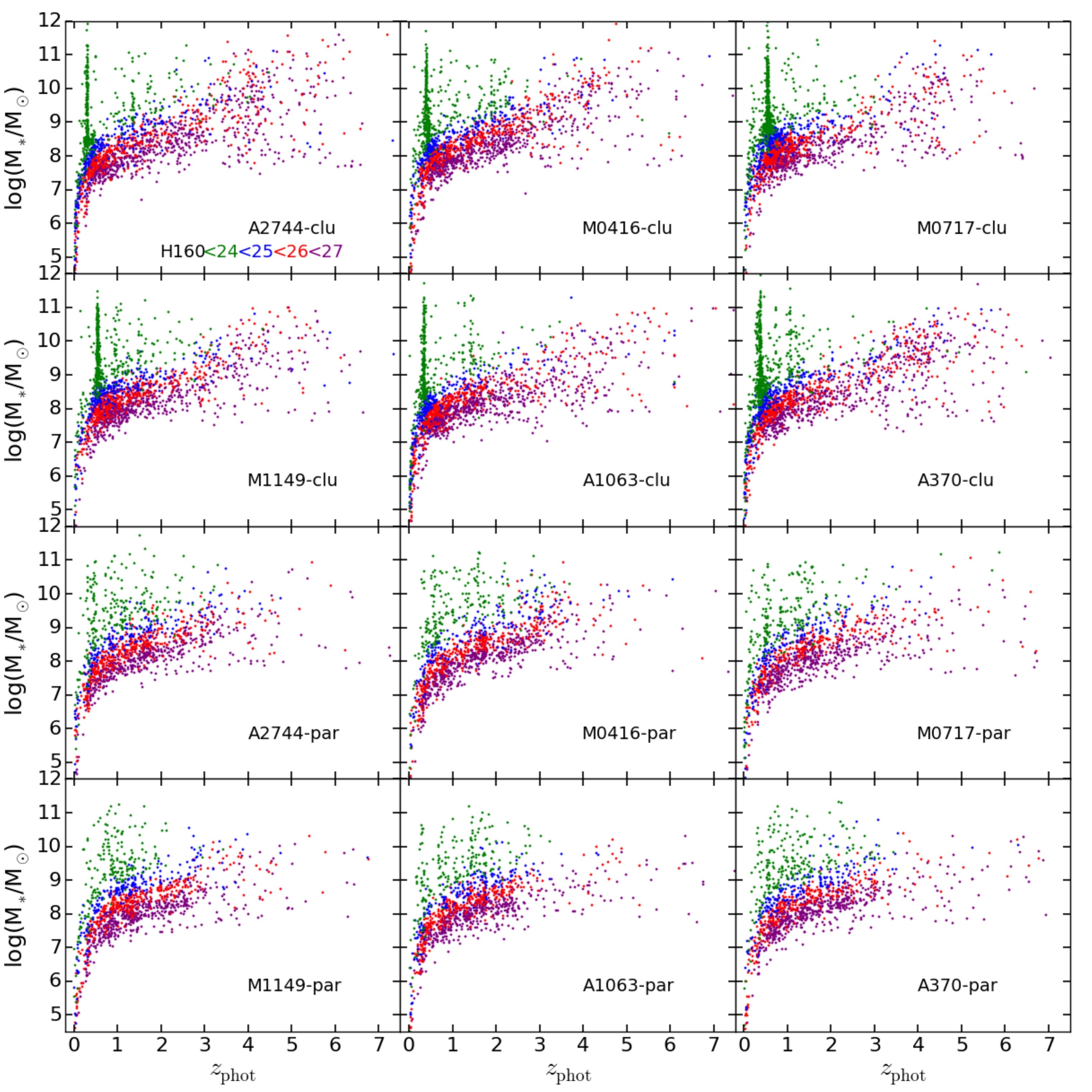}
\caption{Photometric redshift versus stellar mass from EAZY and FAST for the $H_{F160W}$ band.  The points are color-coded by magnitude, galaxies (use\_phot = 1 and ``bandtotal'' = $F160W$) with $H_{F160W} < 24$ are green, $24 \leq H_{F160W} < 25$ are blue, $25 \leq H_{F160W} < 26$ are red and $26 \leq H_{F160W} < 27$ are purple.  The over densities seen as peaks (top panels) in photometric redshift correspond to the redshift of the cluster for each of those fields.  Lensing magnification corrections have not been applied.  A magnitude cut corresponding to the 90\% completeness limits has been applied (utilizing Tables \ref{complete814} and \ref{complete160}).}
\label{massz160}
\end{figure*}		%%%%%%%%%%%%%%%  PLOT  %%%%%%%%%%%%%%%%%

\begin{figure*}[ht!]	%%%%%%%%%%%%%%%  PLOT  %%%%%%%%%%%%%%%%%
\epsscale{1.15}
\plotone{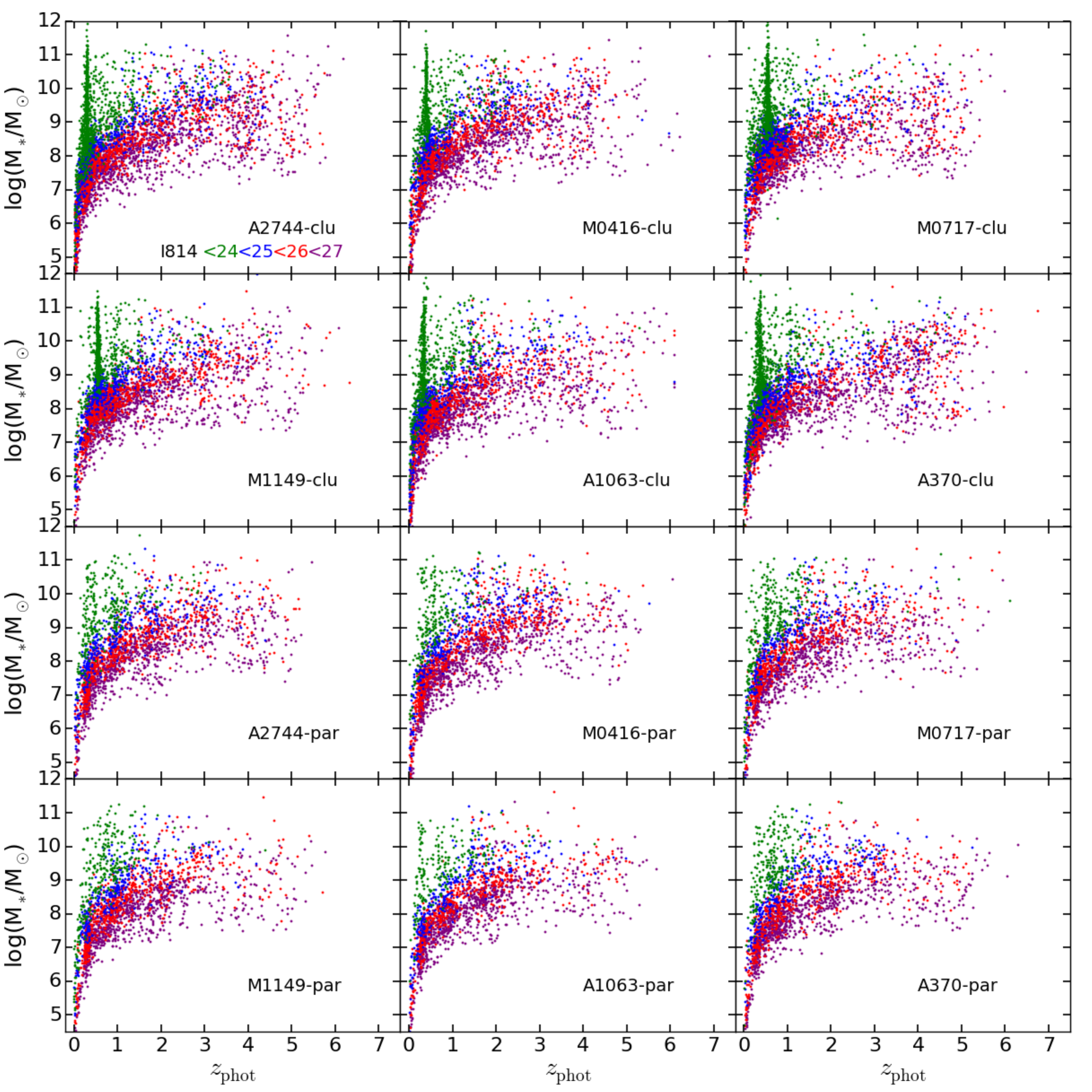}
\caption{Same as Figure~\ref{massz160} but for the $I_{F814W}$ band.  The points are color-coded by magnitude, galaxies (use\_phot = 1) with $I_{F814W} < 24$ are green, $24 \leq I_{F814W} < 25$ are blue, $25 \leq I_{F814W} < 26$ are red and $26 \leq I_{F814W} < 27$ are purple.  The over densities seen as peaks (top panels) in photometric redshift correspond to the redshift of the cluster for each of those fields.  Lensing magnification corrections have not been applied.  A magnitude cut corresponding to the 90\% completeness limits has been applied (utilizing Tables \ref{complete814} and \ref{complete160}).}
\label{massz814}
\end{figure*}		%%%%%%%%%%%%%%%  PLOT  %%%%%%%%%%%%%%%%%

We use the FAST code from \citet{Kriek2009}\footnote{\url{http://w.astro.berkeley.edu/~mariska/FAST.html}} to estimate stellar masses, star formation rates, ages and dust extinctions, given the photometric redshift from EAZY ($z\_\mathrm{peak}$, see Section \ref{photzs}) and the spectroscopic redshift, when available.  We use similar input parameters as \citet{Skelton2014}.  The input parameters are the \citet{BC2003} stellar population synthesis model library with a \citet{Chabrier2003} IMF, solar metallicity, exponentially declining star formation histories with a minimum e-folding time of $\log_{10}(\tau/yr) = 7$, a minimum age of 10 Myr, $0 < A_V < 6$~mag and the \citet{Calzetti2000} dust attenuation law.  Although, we derive star formation rates, dust absorption, and star formation histories for many of the galaxies; we note these quantities are uncertain when derived primarily from optical and near-IR photometry \citep[e.g., see][]{Wuyts2012}.  The stellar population parameters are provided in separate catalogs for each field.  The stellar masses and $M/L$ ratios are relatively well-constrained as they mostly depend on the rest-frame optical colors of the galaxies, and these are well-covered by our photometry (observed wavelengths of $0.2 - 8$~\micron).

In Figure~\ref{massz160}, we show the distributions of the galaxy stellar masses as a function of the photometric redshift ($z_\mathrm{peak}$), for sources that fall within the $F160W$ area.  The points are color-coded according to the galaxy's $H_{F160W}$ magnitude, with the brightest galaxies in green ($H_{F160W} < 24$), galaxies with $24 \leq$ $H_{F160W} < 25$ in blue, galaxies with $25 \leq$ $H_{F160W} < 26$ in red and galaxies with $26 \leq$ $H_{F160W} < 27$ in purple.  In Figure~\ref{massz814}, we show these distributions for all sources with $I_{F814W}$ magnitude (color-coded the same).  Again for the cluster fields, we see clearly the over-densities corresponding to the redshift of the clusters with a large population at higher redshifts of relatively bright sources (likely due to lensing of sources by the clusters).  For the parallel fields, the distribution is mostly uniform with few very massive galaxies that populate the clusters themselves.

\subsection{Lensing Magnifications}
\label{lensing factors}

\begin{figure*}[ht!]	%%%%%%%%%%%%%%%  PLOT  %%%%%%%%%%%%%%%%%
\epsscale{1.2}
\plotone{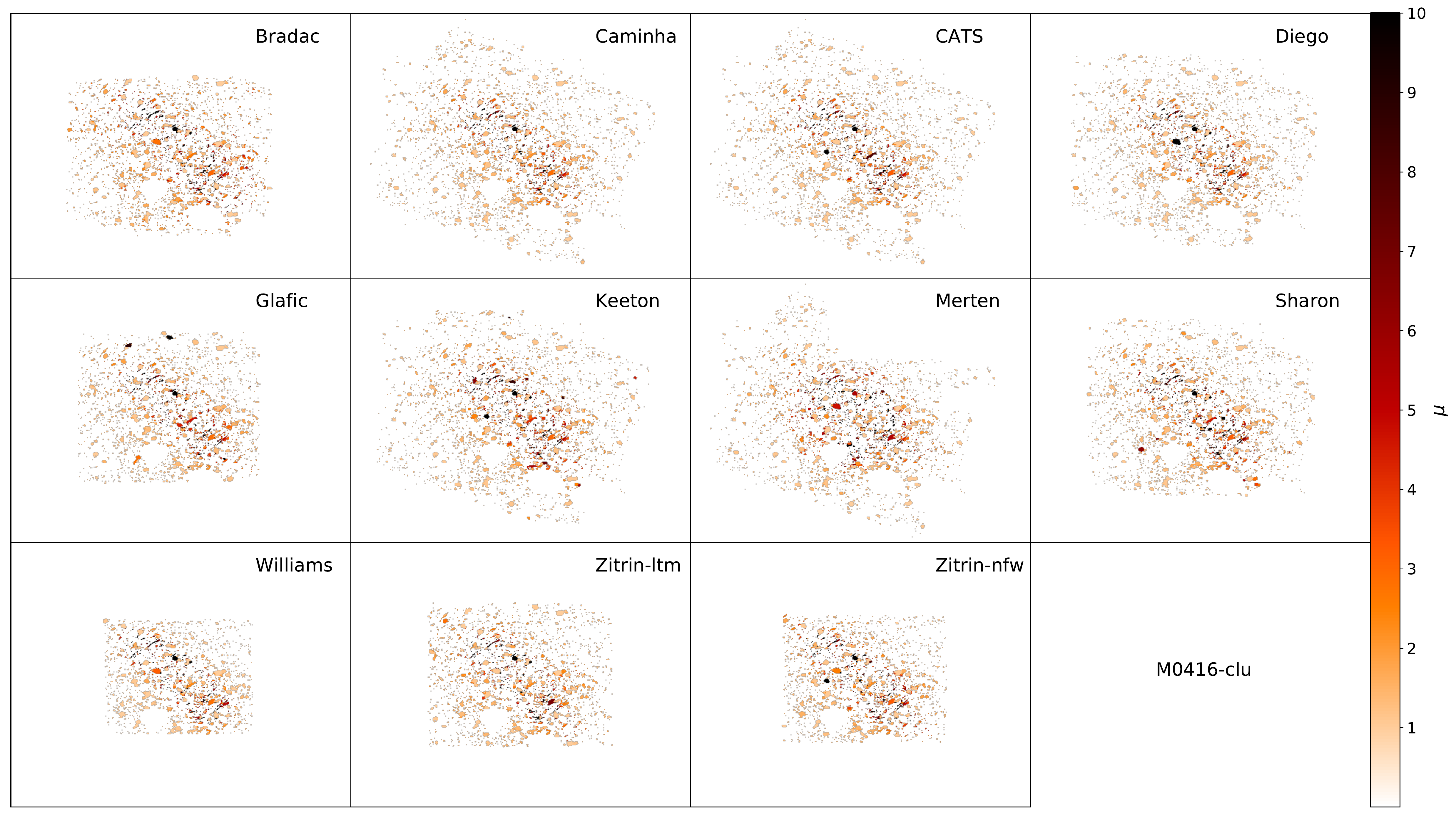}
\caption{The magnification factors from the lensing models for each group of the \mone\ cluster (see Section~\ref{lensing factors}).  The produced segmentation map (see Section~\ref{detection}) is populated with the derived lensing magnifications specific to each source for the various group's lensing model (labeled in each panel).  The darkest sources are the most heavily magnified and the lightest are the least (the color bar at the right gives the lensing magnification breakdown).  It is also worth noting that the coverage areas vary for each group's lensing model (all other clusters can be found in the Appendix).}
\label{lensing maps field}
\end{figure*}		%%%%%%%%%%%%%%%  PLOT  %%%%%%%%%%%%%%%%%

We use available lensing models for all the \hff\ fields to derive magnification values of the sources in our catalogs.  Many independent groups have contributed reliable models of the lensing maps for the \hff\ clusters based on a common set of input data before the \hff\ observing campaign to help facilitate data analysis.  Most of these groups have continued to update the lensing maps to improve and include data from the \hff\ observing campaign.  Several groups assume that the cluster galaxies trace the cluster mass substructure to derive models:  the CATS \citep[P.I. Ebeling, e.g.][]{Jauzac2014}, Sharon \citep[e.g.][]{Johnson2014}, \citep{Caminha2017} use Lenstool, similarly P.I.\ Keeton using Lensmodel, the GLAFIC model \citep{Oguri2010,Ishigaki2015}, and the two different parameterizations (LTM and NFW) provided by the Zitrin team \citep[e.g.][]{Zitrin2013}.  Other models that are provided, by P.I.'s Williams \citep[e.g.][]{Grillo2015}, Brada\v{c} \citep[e.g.][]{Bradac2009} and Merten \citep[e.g.][]{Merten2011}, do not assume that cluster mass is traced by its member galaxies but are constrained only by lensing observables.  The remaining models, P.I.'s Diego and Bernstein using WSLAP+ \citep{Diego2005,Diego2007,Diego2016b}, assume the mass distribution is built as a superposition of Gaussian functions and a compact component that traces the light of the cluster members.

\begin{figure*}[ht!]	%%%%%%%%%%%%%%%  PLOT  %%%%%%%%%%%%%%%%%
\epsscale{1.2}
\plotone{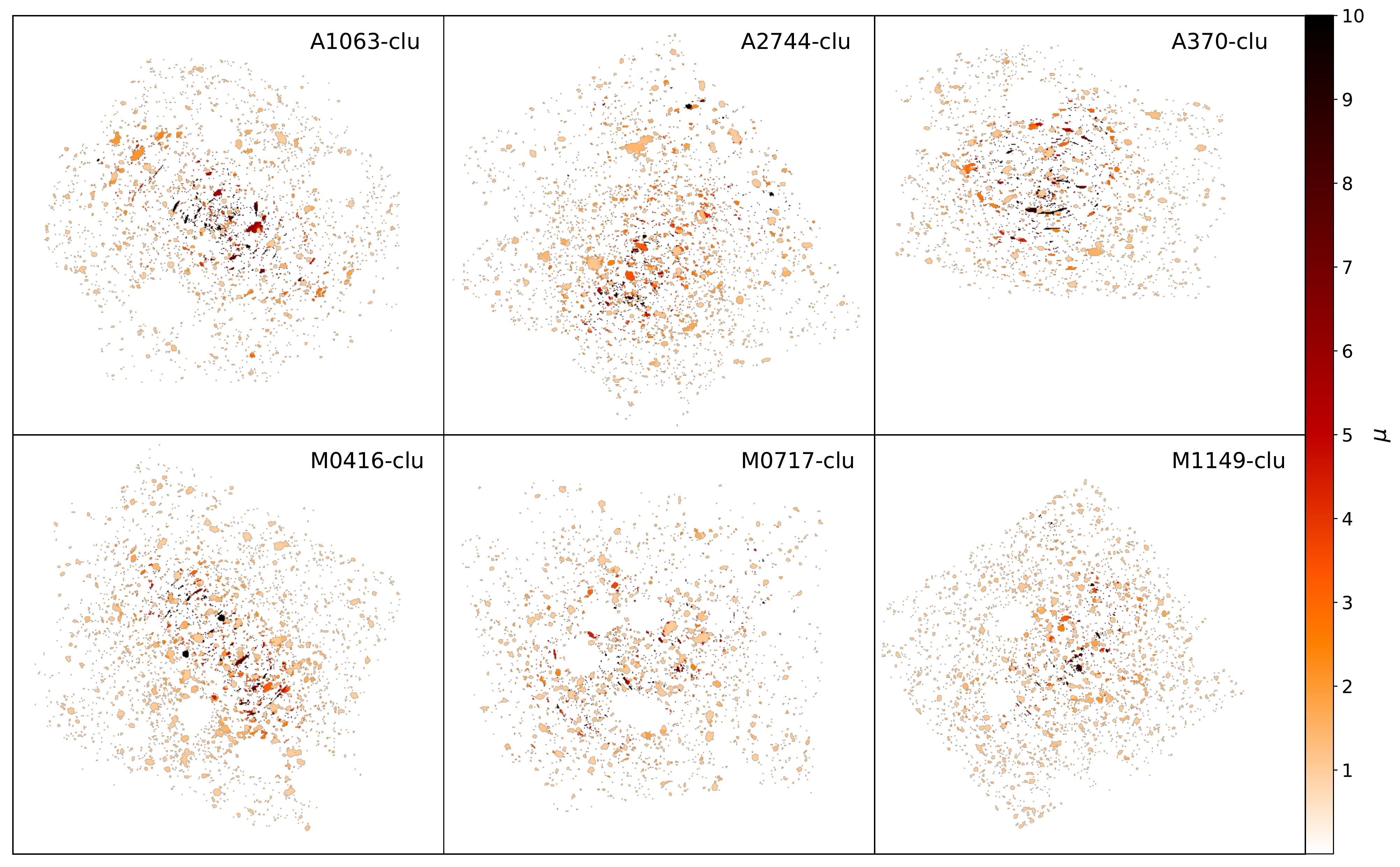}
\caption{The magnification factors from the lensing model CATS (arbitrarily chosen) for all six clusters (see Section~\ref{lensing factors}).  The produced segmentation map (see Section~\ref{detection}) for each cluster (labeled in each panel) is populated with the derived lensing magnifications specific to each source.  The darkest sources are the most heavily magnified and the lightest are the least (the color bar at the right gives the lensing magnification breakdown).}
\label{lensing maps all}
\end{figure*}		%%%%%%%%%%%%%%%  PLOT  %%%%%%%%%%%%%%%%%

Each team has provided publicly available shear and mass surface density maps.  A detailed description of different models can be found on the \hff\ lensing website\footnote{\url{https://archive.stsci.edu/prepds/frontier/lensmodels/}} and references therein. Among the available maps only the Merten models partially cover the parallel pointings for some of the fields.  For each source, we assign shear ($\gamma$) and mass surface-density ($\kappa$) values by matching the right ascension (RA) and declination (Dec) from our catalogs to the corresponding pixel in the shear and mass maps.  Then, we derive the magnification as
\begin{equation} \label{lens mag}
\mu =  \frac{1}{(1 - \kappa \times \frac{D_{LS}}{D_S})^2 - (\gamma \times \frac{D_{LS}}{D_S} )^2}
\end{equation}
where, $D_{LS} = D_A(z_L, z_S)$ and $D_S = D_A(0, z_S)$ with $D_A(0, z_S)$ being the angular diameter distance to the redshift of the source, and $z_L$ being the redshift of the lensing cluster.

In each cluster, we derive lensing magnifications for sources in our catalogs that have an estimated photometric redshift (see Section \ref{photzs}), spectroscopic redshift when available, for each group's most recent lensing model.  From this analysis, we include two lensing magnification catalogs for each field in our data release.  In both catalogs, each column is a different group's derived lensing magnification of the sources; one catalog lists errors of the magnification factors derived from the model uncertainties, whereas the other catalog lists the errors of the magnification factors caused by the photometric redshift uncertainties.  If a source does not have an estimated photometric redshift, the source is flagged with a $-99$ value.  If a source does not fall within the area of the lensing map for each group, the source is flagged with a $-50$ value.  For sources with use\_phot = 0, these are flagged with a $-1$ value.  These flag values are consistent between both lensing magnification catalogs for each field.  We leave it up to the user to determine the desired approach for estimating the best magnification of the sources in our photometric catalogs and the various groups' derived lensing magnifications.  We note that the stellar population models derived from FAST and released are not corrected for lensing magnification.

In Figure~\ref{lensing maps field}, we show the differences between the lensing models for each group of the \mone\ cluster.  The produced segmentation map is populated with the derived lensing magnifications specific to each source for the various group's lensing model (labeled in the figure).  The darkest sources are the most heavily magnified and the lightest are the least (the color bar in the figure at the right gives the lensing magnification breakdown).  It is also worth noting that the coverage areas vary for each group's lensing model.  Figure~\ref{lensing maps all} shows the derived lensing magnifications for the CATS lensing model of all six clusters (lensing model arbitrarily chosen).

\section{Summary}
\label{summary}

We present the data products and multi-wavelength photometric catalogs produced by the \hff -DeepSpace project for the \hff\ observing campaign.  The survey covers $\sim$165~arcmin$^2$ in the six clusters \aone, \mone, \mtwo, \mthree, \atwo, \athree\ and accompanying parallel fields with \hst/ACS and \hst/WFC3 imaging.  The details of the data reduction are given in Section \ref{redux}.  In addition to the \hff\ \hst\ data, we include \hst/UVIS, ultra-deep $K_S$, \spitzer/IRAC and any other available bands from the \hst/ACS and WFC3 instruments (see Section~\ref{data} and Table~\ref{image sources}).  We make all the images that have been generated available on our website\footnote{\url{http://cosmos.phy.tufts.edu/~danilo/HFF/Download.html}} with the catalogs.  Each of the images is on the same astrometric system as the \hff/WFC3 $F160W$ mosaics.

We apply consistent methodology to produce multi-wavelength photometric catalogs and data products for all twelve of the fields.  The \sex\ software \citep{Bertin1996} is used to detect sources on a noise-equalized combination of the $F814W$, $F105W$, $F125W$, $F140W$ and $F160W$ images.  Using the four \hff\ WFC3 bands and the \hff\ ACS/$F814W$, we exploit the maximum survey area without sacrificing the depth of the \hff, specifically for the WFC3 bands.  As described in Section \ref{bcg modeling}, we model out many of the bright cluster members and occasionally other bright sources in the fields.  We take great care to achieve accurate cluster models of the bCGs and ICL modeled out ($< 1$\% uncertainty of the total flux for the bCGs).  We carefully measure the flux and errors of the objects in each field and band, taking into account the differences in image resolution between the \hst\ and lower resolution $K_S$ and IRAC photometry (see Section~\ref{hst photometry} and \ref{low res phot}).

Furthermore, we test that the results are consistent for all twelve fields and the total magnitudes and errors agree well with the expected behavior that each source includes only light associated with it (see Section~\ref{photo check}).  The resulting photometric catalogs span a broad wavelength range from UV to near-IR ($0.2-8$~\micron) and are of excellent quality, as demonstrated by the analysis throughout this work.  We use EAZY \citep{Brammer2008} to derive photometric redshifts and achieve an average scatter ($\sigma_\mathrm{NMAD} \sim 0.034$) between the photometric and spectroscopic redshifts for all fields with an average significant outlier fraction of $\sim12$\% in all fields (i.e.\ 10/12 fields, \atwo\ and \athree\ parallels do not have any spectroscopic redshift matches).  We provide rest-frame colors based on the best-fitting EAZY templates, as well as stellar masses and stellar population parameters for all the galaxies based on fits to their observed photometry (see Section~\ref{derived properties}) and gravitational lensing magnification factors (see Section~\ref{lensing factors}).  Furthermore, different methodologies are useful to understand possible systematic uncertainties between various groups' catalogs of the \hff, as ours are not the only available catalogs \citep[e.g\ the catalogs of the ASTRODEEP collaboration][]{Merlin2016,Castellano2016,DC2017}.

This work, by our \hff-DeepSpace team, concludes the first phase of even more ambitious projects, as outlined in the Introduction.  Future work will describe the grism spectroscopy that accompanies these data sets, and help improve the measurements of the redshifts, stellar masses and other stellar properties.  Furthermore, these photometric catalogs will be an important aide in designing future surveys as well as planning follow-up programs with current and future observatories (i.e.\ \jwst, \textit{GMT}, \textit{TMT} and others) to answer key questions remaining about first light, reionization, the assembly of galaxies and many more topics, most notably, by gaining access to high-redshift sources that are otherwise inaccessible without the strong lensing clusters and power of the \hst.

\acknowledgements

We thank the anonymous referee for valuable comments that improved the quality of this work.  DM, HS, DLV, NM, EKF acknowledge the very generous support of the National Science Foundation under Grant Number 1513473 and by HST-AR-14302, provided by NASA through a grant from the Space Telescope Science Institute, which is operated by the Association of Universities for Research in Astronomy, Incorporated, under NASA contract NAS5-26555. KN acknowledges support by HST-AR-14553, provided by NASA through a grant from the Space Telescope Science Institute, which is operated by the Association of Universities for Research in Astronomy, Incorporated, under NASA contract NAS5-26555. DM and EKF acknowledges support from the Tufts University Faculty Research Fund, Grants-In-Aid, and Graduate/Undergraduate Summer Scholar programs. This work was supported by NASA Keck PI Data Awards, administered by the NASA Exoplanet Science Institute. We gratefully acknowledge funding support from the STScI Director's Discretionary Research Fund.  We acknowledge funding by NWO grant 614.001.302.  Some of the data presented in this study were obtained at the W.M. Keck Observatory from telescope time allocated to NASA through the agency's scientific partnership with the California Institute of Technology and the University of California. The Observatory was made possible by the generous financial support of the W. M. Keck Foundation. The authors wish to recognize and acknowledge the very significant cultural role and reverence that the summit of Mauna Kea has always had within the indigenous Hawaiian community. We are most fortunate to have the opportunity to conduct observations from this mountain. This research made use of Astropy, a community-developed core Python package for Astronomy (Astropy Collaboration et al. 2013), open-source Python modules Numpy, Scipy, Matplotlib, and scikit-image, and NASA's Astrophysics Data System (ADS).  We would like to thank the \hff\ data team (A.\ M.\ Koekemoer, J.\ Mack, J.\ Lotz, J.\ Anderson, R.\ Avila, E.\ Barker, D.\ Borncamp, H.\ Gunning, B.\ Hilbert, H.\ Khandrika, R.\ Lucas, C.\ Martlin, S.\ Ogaz, B.\ Porterfield, M.\ Robberto, B.\ Sunnquist) for generously providing public releases of the reduced and mosaicked data.  This work utilizes gravitational lensing models produced by PIs Brada\v{c}, Natarajan \& Kneib (CATS), Merten \& Zitrin, Sharon, Williams, Keeton, Bernstein and Diego, and the GLAFIC group. This lens modeling was partially funded by the HST Frontier Fields program conducted by STScI. STScI is operated by the Association of Universities for Research in Astronomy, Inc. under NASA contract NAS 5-26555. The lens models were obtained from the Mikulski Archive for Space Telescopes (MAST).

\clearpage

\appendix
\label{app1}

\subsection{Redshift Distribution Comparisons to Other Catalogs}

\vspace{-12pt}
\begin{figure*}[ht!]	%%%%%%%%%%%%%%%  PLOT  %%%%%%%%%%%%%%%%%
\epsscale{1.1}
\plotone{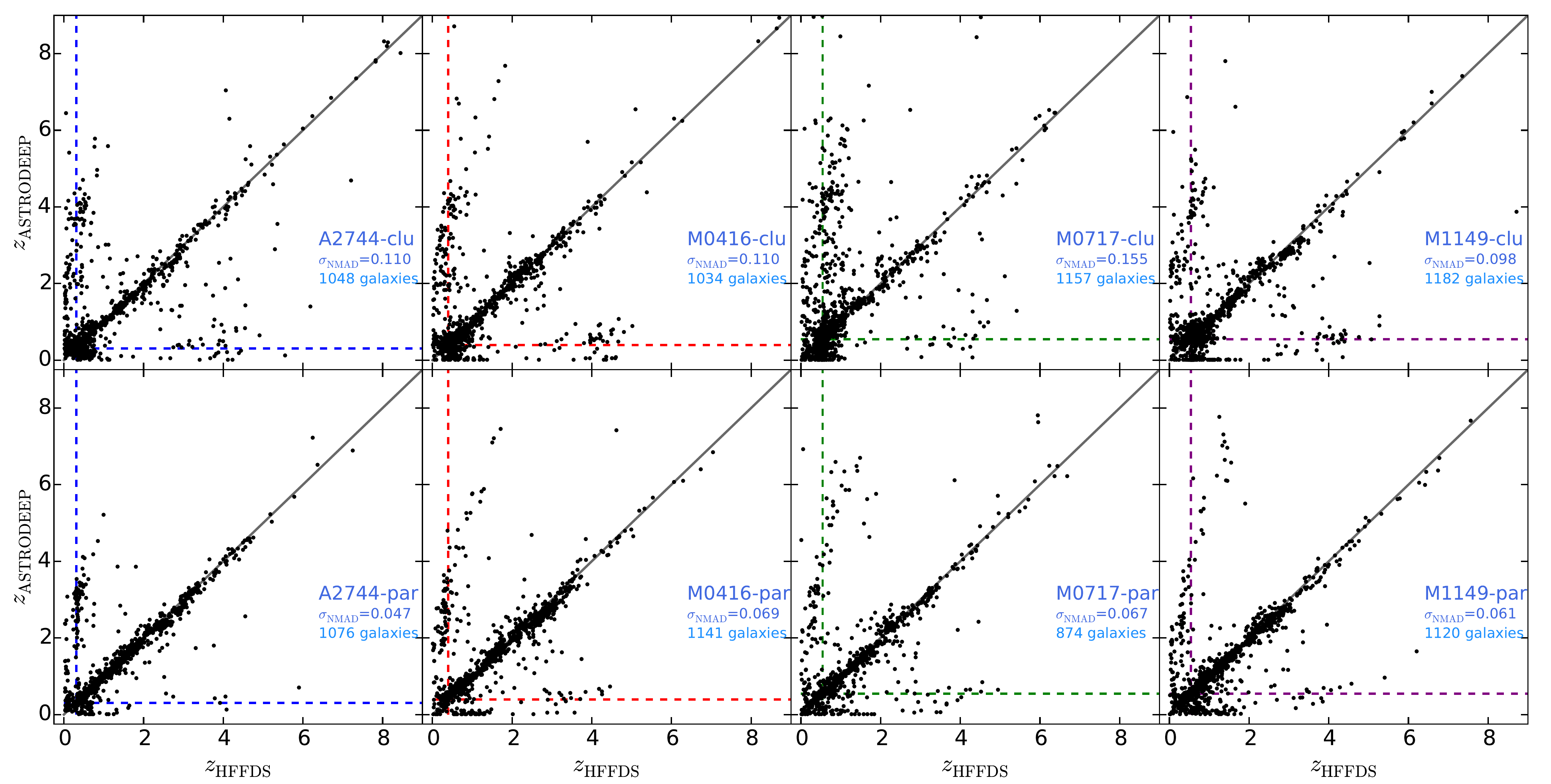}
\caption{Comparison of the \hff-DeepSpace photometric redshifts ($z_\mathrm{HFFDS}$) to ASTRODEEP \citep[$z_\mathrm{ASTRODEEP,}$][]{Merlin2016,Castellano2016,DC2017}.  Galaxies were selected to be brighter than AB magnitude of 27 in the $F160W$ band (corresponding roughly to the 90\% completeness limit for both catalogs) and matching sources within D$=0.5$~\arcsec.  The scatter ($\sigma_\mathrm{NMAD}$) and number of matched galaxies are given for each field in their respective panel.  The unity relation (solid line) and redshift of the cluster (dashed lines) are marked in each panel.}
\label{astromatch}
\end{figure*}		%%%%%%%%%%%%%%%  PLOT  %%%%%%%%%%%%%%%%%

\begin{figure*}[ht!]	%%%%%%%%%%%%%%%  PLOT  %%%%%%%%%%%%%%%%%
\epsscale{1.1}
\plotone{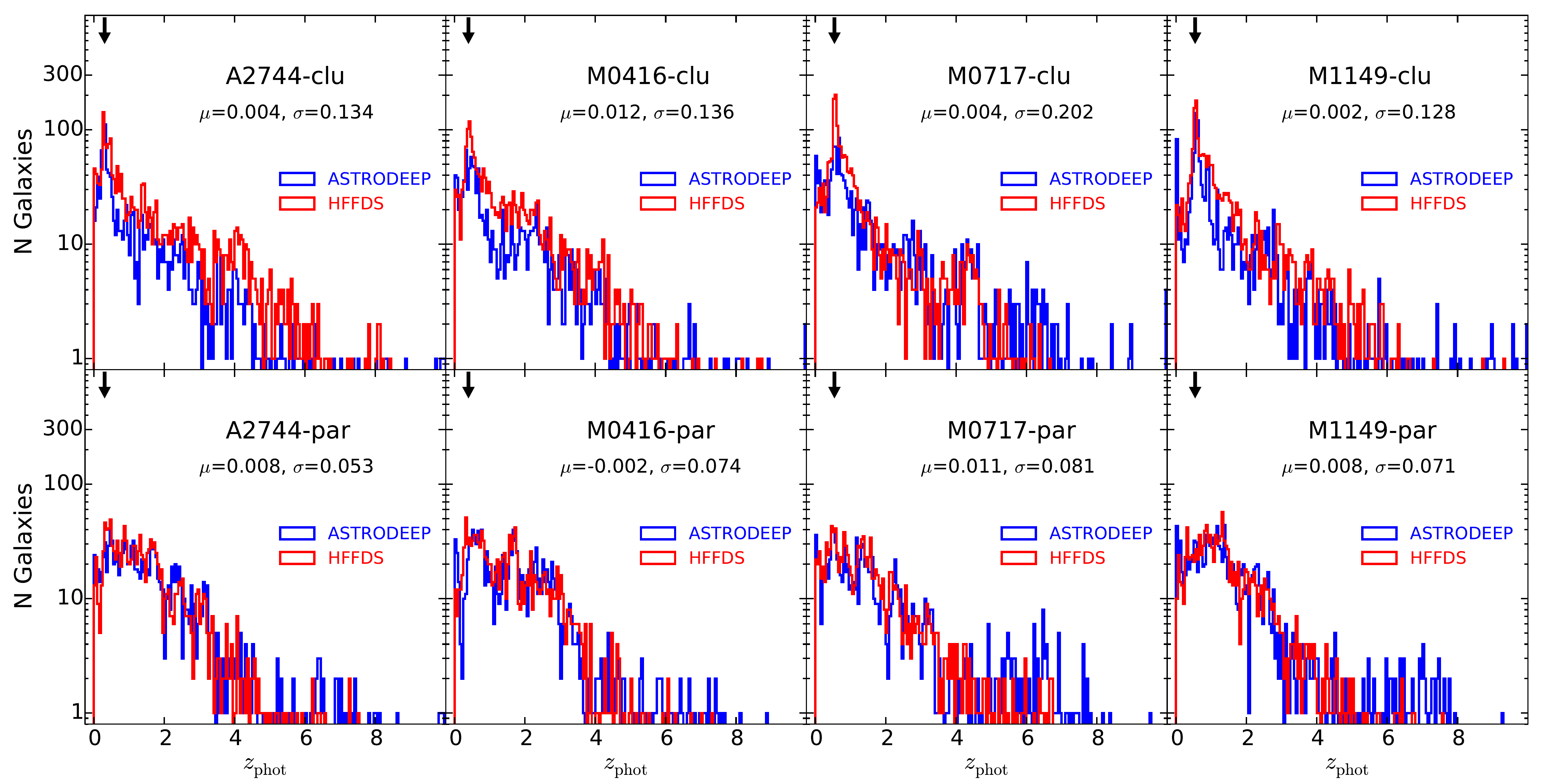}
\caption{Histogram comparison of the \hff-DeepSpace photometric redshifts ($z_\mathrm{phot}$, red) to ASTRODEEP \citep[blue,][]{Merlin2016,Castellano2016,DC2017}.  Galaxies were selected to be brighter than AB magnitude of 27 in the $F160W$ band (corresponding roughly to the 90\% completeness limit for both catalogs) over comparable source detection areas (i.e., WFC3 footprint of the \hff).  The bi-weight mean offset ($\mu$) and scatter ($\sigma$) are given for each field in their respective panel.  The black arrow in each panel represents the redshift of the associated cluster.}
\label{astrohist}
\end{figure*}		%%%%%%%%%%%%%%%  PLOT  %%%%%%%%%%%%%%%%%

\clearpage
\bibliography{myrefs}

\end{document}